\documentclass[fleqn,usenatbib]{mnras}

\usepackage{newtxtext,newtxmath}

\usepackage[T1]{fontenc}

\DeclareRobustCommand{\VAN}[3]{#2}
\let\VANthebibliography\thebibliography
\def\thebibliography{\DeclareRobustCommand{\VAN}[3]{##3}\VANthebibliography}

\newcommand{\Msol}{M_{\odot}}
\newcommand{\ang}{\textup{\AA}}

\usepackage{graphicx}	
\usepackage{amsmath}	
\usepackage{tabularx}
\usepackage[flushleft]{threeparttable}
\usepackage{xfrac}
\usepackage{hyperref}

\title[Lanthanides in Nebular Phase KNe]{Lanthanide Impact on the Infra-Red Spectra of Nebular Phase Kilonovae}

\author[Q. Pognan et al.]{
Quentin Pognan,$^{1}$\thanks{E-mail: quentin.pognan@aei.mpg.de}
Kyohei Kawaguchi,$^{1,2}$
Shinya Wanajo,$^{2,1}$
Sho Fujibayashi$^{4,3,1}$
Anders Jerkstrand,$^{5}$ \newauthor
and Jon Grumer$^{6}$
\\
$^{1}$Max Planck Institute for Gravitational Physics (Albert Einstein Institute), Am M\"{u}hlenberg 1, Potsdam-Golm,14476, Germany\\
$^{2}$Center for Gravitational Physics and Quantum Information, Yukawa Institute for Theoretical Physics, Kyoto University, Kyoto, 606-8502, Japan \\
$^{3}$ Astronomical Institute, Graduate School of Science, Tohoku University, Sendai 980-8578, Japan \\
$^{4}$ Frontier Research Institute of Interdisciplinary Sciences (FRIS), Tohoku University, Sendai 980-9578 \\
$^{5}$ Oskar Klein Center, Department of Astronomy, Stockholm University Albanova, SE-10691, Stockholm, Sweden \\
$^{6}$ Theoretical Astrophysics, Department of Physics and Astronomy, Uppsala University, Box 516, SE-751 20 Uppsala Sweden
}

\date{Accepted XXX. Received YYY; in original form ZZZ}

\pubyear{\the\year{}}

\begin{document}
\label{firstpage}
\pagerange{\pageref{firstpage}--\pageref{lastpage}}
\maketitle

\begin{abstract}
Nebular phase kilonovae (KNe) have significant infra-red (IR) emission thought to be mostly forbidden emission lines from rapid neutron capture (r-process) species in neutron star merger ejecta. Lanthanide elements in particular have complex, open f-shell atomic structures with many IR transitions. Using non-local thermodynamic equilibrium (NLTE) radiative transfer simulations, we explore the impact of lanthanides on the IR spectra of KNe in the nebular phase, exploring a parameter space of ejecta mass and lanthanide fraction. We find that lanthanide impact is greater at higher densities, corresponding to earlier epochs and greater ejecta masses. The wavelengths most affected are found to be $\lambda \lesssim 4~\mu$m, with the species Ce\,\textsc{iii} and Nd \textsc{ii} being the most important contributors to spectral formation. We also find significant emission from species proposed in observations, notably Te\,\textsc{iii} at 2.1 $\mu$m, and Se\,\textsc{iii} at 4.5 and 5.7 $\mu$m, while W\,\textsc{iii} is subdominant at 4.5 $\mu$m. The Te\,\textsc{iii} feature at 2.1 $\mu$m is always blended, particularly with Zr\,\textsc{ii}, Ce\,\textsc{iii}, and Nd\,\textsc{ii}.  We do not reproduce the smooth blackbody-like continua observed in AT2023vfi. Based on our results, we argue that line opacity alone is likely insufficient to produce optically thick continua in the nebular phase, even in the case of lanthanide/actinide-rich ejecta, as our models are optically thin in the IR at these epochs. Given that lanthanide contributions are dominant below 4 $\mu$m, we suggest that NIR observations best probe these elements, while MIR spectroscopy with \textit{JWST} can reliably probe non-lanthanide emission even in relatively lanthanide-rich cases. 
\end{abstract}



\section{Introduction}
\label{sec:intro}

Binary neutron star (BNS) mergers are known to produce ultra-violet (UV)/optical/infra-red (IR) transients referred to as kilonovae (KNe) \citep[][]{Li.Paczynski:98,Metzger.etal:10,Rosswog.etal:13}. These transients are powered by the radioactive decay of unstable isotopes in the merger ejecta, created primarily by rapid neutron capture (r-process) nucleosynthesis \citep[][]{Lattimer.Schramm:74,Lattimer.Schramm:76,Symbalisty.Schramm:82,Eichler.etal:89,Freiburghaus.etal:99,Rosswog.etal:99}. Since the origin of r-process elements in the Universe is a long-standing, open question, observations of KNe may provide some answers as to the importance of NS mergers as astrophysical r-process sites. 

Thus far, a complete series of photometric and spectral observations only exists for one KN, AT2017gfo \citep[e.g.][]{Abbott.etal:17,Cowperthwaite.etal:17,Kasen.etal:17,Smartt.etal:17,Villar.etal:17}. The initial emission, on a time-scale of a week, was dominated by photospheric emission, with various absorption and P-Cygni features being interpreted as signatures of diverse r-process species \citep[e.g.][]{Watson.etal:2019,Domoto.etal:21,Gillanders.etal:21,Domoto.etal:22,Gillanders.etal:22,Sneppen.Watson:23,Tarumi.etal:23}. Thus far, the majority of species (potentially) identified in observed spectra are found by the features of a single or a few strong lines \citep[e.g. Sr\,\textsc{ii}, Y\,\textsc{ii}, La\,\textsc{iii}, Ce\,\textsc{iii}][]{Watson.etal:2019,Domoto.etal:22,Sneppen.Watson:23,Sneppen.etal:24b}. Lanthanide species with open f-shell structures, however, typically produce forests of lines that can lead to broad absorption/emission features, as well as strong line blanketing in at bluer wavelengths ($\lambda \lesssim 7000~\ang$) leading to redder colours in the emergent KN \citep[e.g.][]{Perego.etal:17,Smartt.etal:17,Tanaka.etal:20,Gillanders.etal:22}. Since many lanthanide lines have relatively small energies, and therefore yield transition wavelengths in the IR, their spectral signature at these wavelengths may be quite different from the aforementioned empty f-shell species.

Evolving with time, AT2017gfo became redder and the emission began to appear non-thermal, with the spectra past 7 days being dominated by a prominent emission feature at 2.1\,$\mu$m, interpreted as forbidden emission from Te\,\textsc{III} \citep[][]{Hotokezaka.etal:23,Gillanders.etal:24}. Photometry from \textit{Spitzer} in the 3.6$\mu$m and 4.5$\mu$m bands taken at 43 and 74 days after merger, represent the latest (in time) observations of AT2017gfo, with a consistent non-detection in the bluer band, and detections of 21.9 and 23.9 AB magnitudes in the redder band respectively \citep[][]{Villar.etal:18,Kasliwal.etal:22}, the origin of which is potentially attributed to Se\,\textsc{III} and/or W\,\textsc{III} \citep[][]{Hotokezaka.etal:22}.

The only other spectral data of potentially r-process powered emission to exist, is that of the KN candidate AT2023vfi, which was detected by excess IR emission in the afterglow of the long gamma-ray burst (GRB)230307A \citep[][]{Levan.etal:24,Gillanders.Smartt:25}. For this object, two spectra were taken by \textit{JWST} at 29 and 61 days after the GRB trigger.The spectral features of AT2023vfi are found to be well fit by a $\sim 600$ -- 700K Blackbody continuum at 29 days, with Gaussian emission lines overlain at 2.1 and 4.5~$\mu$m that have been tentatively attributed to Te\,\textsc{iii} and Se\,\textsc{III} and/or W\,\textsc{III} respectively, with the origin of the continuum being suggested as arising from significant lanthanide opacity \citep[][]{Levan.etal:24,Gillanders.Smartt:25}. At 61 days, the spectrum is relatively flat and featureless aside from some emission remaining at 2.1~$\mu$m.

While the UV and optical emission of KNe has been studied in great depth by virtue of the high quality spectra of AT2017gfo \citep[e.g.][]{Cowperthwaite.etal:17,Villar.etal:17,Smartt.etal:17,Kawaguchi.etal:18,Domoto.etal:21,Banerjee.etal:24}, both the NIR and mid-IR (MIR) have a great potential for providing insight into the composition of NS merger ejecta powering the KN. Generally, IR wavelengths are expected to be less optically thick than the UV/optical wavebands due to the presence of fewer absorption lines, in the context that bound-bound transitions provide the vast majority of optical depth in KNe \citep[e.g.][]{Tanaka.etal:20,Banerjee.etal:22,Carvajal.etal:22,Carvajal.etal:23,Kato.etal:24,Deprince.etal:25}. This signifies that observations of KNe in the IR reach deeper into the ejecta than corresponding UV/optical observations for a given epoch, and therefore probe more of the ejecta. Furthermore, while optical depth remains high at blue optical and UV wavelengths, fluorescence to redder IR wavelengths will occur, such that IR emission also indirectly probes the optically thick regions of the ejecta. 

As time progresses, the density decreases such that the ejecta conditions transition from local thermodynamic equilibrium (LTE) to non-local thermodynamic equilibrium (NLTE) \citep[e.g.][]{Hotokezaka.etal:21,Pognan.etal:22b,Pognan.etal:23,Gillanders.etal:24}, and the ejecta become optically thin to most wavelengths, with this regime being called the nebular phase. The efficiency of collisional processes decreases with time, with high-lying states becoming inaccessible by thermal excitation. The lowest-lying states are often fine-split energy levels of the ground state, where strong allowed transitions (E1) are not available for radiative deexcitation. Instead, the strongest transitions are forbidden magnetic dipoles (M1) \citep[e.g.][]{Hotokezaka.etal:22}, and the small energies typically involved in these bound-bound transitions yield wavelengths in the NIR and/or MIR wavebands. 

In KNe, the nebular phase is believed to be reached on time-scales of $t \gtrsim 10$d \citep[e.g.][]{Waxman.etal:19,Hotokezaka.etal:21,Pognan.etal:22b}. Correspondingly, forbidden IR emission lines go through a transition regime from optically thick LTE to full NLTE around the same time, with the exact moment depending not only on ejecta properties, but also on the atomic structure of the emitting species \citep[see e.g.][Jerkstrand et al. (in prep.) for an in-depth analysis in supernovae and KNe respectively]{Jerkstrand:17}. Once fully in the nebular phase, the luminosity of forbidden lines is proportional to the abundance of the emitting species, such that observation of forbidden emission provides a direct method of assessing the composition of NS merger ejecta.

As in the photospheric phase, some species are expected to be identifiable by strong, single emission line features from M1 transitions within the fine splitting of their ground state, such as Te\,\textsc{iii}, Se\,\textsc{iii}, and W\,\textsc{iii}. However, lanthanides may break this trend due to their level-rich structure, which can yield many blended IR emission lines of similar strength and wavelength. Notably, many levels are low-lying and energetically accessible even for the small energies of the nebular phase, e.g. the first 5 levels of Ce\,\textsc{iii} all lie below $5500~\rm{cm^{-1}}$, including levels of opposite parities necessary for E1 transitions to occur, cf. the first excited level of Kr\,\textsc{ii} with $\rm{E} = 5370~\rm{cm^{-1}}$. As such, the IR signatures of lanthanides in nebular phase KNe may be significant and complex, particularly at early ($t\sim 10~$days) times where such E1 transitions may still be optically thick, and therefore affect forbidden emission from non-lanthanide species. Given that AT2017gfo was assessed to have a lanthanide fraction in the range of $X_{\rm{La}} \sim 10^{-4} - 10^{-2}$ \citep[e.g.][]{Domoto.etal:22,Gillanders.etal:22}, investigating their impact on nebular phase emission will be useful for the analysis of current and future IR observations. 

The analysis of KN emission in the IR has been previously approached using semi-analytic methodology, by fixing parameters such as temperature and ionisation structure, and conducting detailed calculations of line emissivity for key species \citep[e.g.][]{Hotokezaka.etal:22,Hotokezaka.etal:23,Gillanders.etal:24}. Ultimately, however, these studies are to some extent inconsistent, as the temperature and ionisation structure of the ejecta is constantly evolving in time \citep[e.g.][]{Hotokezaka.etal:20,Pognan.etal:22a}. Furthermore, these studies have omitted potential interactions between lines, such as possible absorption of forbidden emission by stronger, optically thick E1 transitions, which may be important role in the early nebular phase, particularly in the presence of lanthanides. 

We conduct a parameter space study of IR KN emission with a focus on the impact of lanthanide species, providing a complementary investigation to the work on lanthanide-free ejecta by Jerkstrand et al. (in prep.). We employ the NLTE spectral synthesis code \textsc{sumo} \citep{Jerkstrand.etal:11,Jerkstrand.etal:12}, in its modified KN version \citep{Pognan.etal:22a,Pognan.etal:25}, to produce a grid of synthetic spectra covering a wavelength range of $1.2 - 30\, \mu$m, roughly corresponding to the operational range of \textit{JWST}. We cover epochs from 10 to 75 days after merger, corresponding to the late-time spectra of AT2017gfo, and covering all currently relevant IR observations, including the \textit{Spitzer} photometry of AT2017gfo, as well as the \textit{JWST} observations of AT2023vfi. We describe the ejecta models and radiative transfer (RT) simulation in Section \ref{sec:Models}. We begin by examining the evolution of temperature and ionisation structure of the models in Section \ref{sec:thermo}, and follow with an in-depth analysis of the emergent spectra in Section \ref{sec:spectra}. We also consider the broad-band evolution of the model lightcurves (LCs) in Section \ref{sec:LCs}, and end with our conclusions in Section \ref{sec:discussion}.

\section{Ejecta Models and radiative transfer simulation}
\label{sec:Models}

\subsection{Hydrodynamical simulations and ejecta models}
\label{subsec:hydro}

The ejecta profile, composition, and radioactive power which are used as input to the RT simulation are based on the DD2-135 model of \citet{Kawaguchi.etal:22}. The merger scenario is that of a symmetric BNS merger with component masses of $M_{\mathrm{NS}} = 1.35 \, \Msol$ and the DD2 equation of state \citep[][]{Banik.etal:14}. This is initially simulated by a numerical relativity (NR) simulation \citep[][]{Fujibayashi.etal:20b}, the outflow data of which is used as the inner boundary conditions for the 2-D axisymmetric hydrodynamic (HD) simulation following the methodology described in \citet{Shibata.Hotokezaka:19,Kawaguchi.etal:21,Kawaguchi.etal:22}. The elemental abundances, as well as associated radioactive power are calculated for tracer particles in a post-processing step of the NR simulation \citep{Fujibayashi.etal:20b,Fujibayashi.etal:23}, using the \texttt{rNET} nuclear network code \citep{Wanajo:18}. The radioactive energy is thermalised according to decay product following \citet{Barnes.etal:16}, aside from fission fragments which are neglected, and adopted in the HD simulation. The energy deposition from the HD simulation is then used directly as the energy input to the RT simulation, taking into account the expansion of the ejecta and time-evolving thermalisation efficiencies.

The ejecta in the HD simulation are separated into dynamical and post-merger components depending on their entrance time into the computational domain, with a cut-off of $t_{\rm{cut}} = 0.15\,$s, such that the dynamical ejecta component mass approximately agrees with that reported in the NR simulation for consistency \citep[see][]{Fujibayashi.etal:20b,Kawaguchi.etal:21}. The two ejecta components are then averaged over all angles and radii in order to recover the total component masses as well as mass-weighted average abundances and radioactive heating rates.

The dynamical component contains tidally disrupted material, shock heated ejecta at the moment of contact between the two neutron stars, as well as some of the faster moving neutrino-driven ejecta, and is found to have a relatively high lanthanide mass-fraction of $X_{\rm{La}} = 0.054$. We note that the inclusion of some neutrino-driven ejecta in our `dynamical' ejecta component is due to the use of the 0.15\,s cut-off time; physically this latter part is usually associated to the post-merger component, and we clarify that the $f_{\rm{dyn}}$ parameter we use (see below) is with respect to the dynamical ejecta as defined in this study, i.e. $t_{\rm{eject}} < t_{\rm{cut}}$. The post-merger ejecta contain neutrino-driven and viscosity driven components, the majority of which comes from the remnant accretion disc. The post-merger component has a markedly lower lanthanide mass-fraction of $X_{\rm{La}} = 0.002$.

The ejecta models given as input to the RT simulation all have the same density profiles of $\rho \propto v^{-4}$, with 5 radial zones spanning $v_{\rm{ej}} = 0.05 - 0.3$c in steps of $0.05$c. Three total ejecta masses are taken: $0.05, 0.01, 0.005\, \Msol$, with homogeneous elemental compositions parametrized according to the fraction of dynamical ejecta $f_{\rm{dyn}}$, such that the elemental mass fractions are given by:

\begin{equation}
    X_{\rm{tot}} = f_{\rm{dyn}}X_{\rm{dyn}} + (1-f_{\rm{dyn}})X_{\rm{pm}}
    \label{eq:composition}
\end{equation}

\noindent where $X_{\rm{dyn}}$ and $X_{\rm{pm}}$ are the mass fractions of elements in the dynamical and post-merger components respectively. 

From the above formula, the original masses of the components are not taken into account. For instance, a model with $M_{\rm{ej}} = 0.05\, \Msol$ and $f_{\rm{dyn}} = 0.5$ has a composition that is taken evenly from the post-merger and dynamical components, e.g. there is $0.025\, \Msol$ mass of dynamical ejecta in the model. Additionally, we remove all elements heavier than $_{56}$Ba from the post-merger component such that it is lanthanide-free. This is done to allow the consideration of a lanthanide-free model in the context of a fiducial $f_{\rm{dyn}}=0$ value. While this truncation may seem artificial, we note that the abundances of lanthanide and third-peak elements in the dynamical ejecta component are greater by over an order of magnitude by mass-fraction than in the post-merger ejecta, such that for any value of $f_{\rm{dyn}} \geq 0.01$, their abundances will be largely dominated by the dynamical component.

We adopt a uniform ejecta composition for simplicity, and to reduce the complexity of analysing the emergent spectra in this initial study. In reality, the composition will depend on velocity coordinate (and angle), with post-merger ejecta typically having slower velocities. In this sense, our post-merger component dominated models may have an excess of lighter r-process elements ($Z \lesssim 40$) in the outer ejecta layers. However, we expect their contribution to the emergent spectra to be minor, considering the fact that most of the mass is located in the low-velocity layers as a result of the steep density profile ($\rho\propto v^{-4}$). We note that the stratification of ejecta composition would be highly non-trivial in the presence of the long-lived remnant: the post-merger activity is more violent, yielding a high velocity wind driven by the magnetic field \citep[e.g.][]{Kiuchi.etal:24} and neutrino heating \citep[e.g.][]{Just.etal:23}.

\begin{figure}
    \centering
    \includegraphics[trim={0.4cm 0.cm 0.4cm 0.3cm},width=1.1\linewidth]{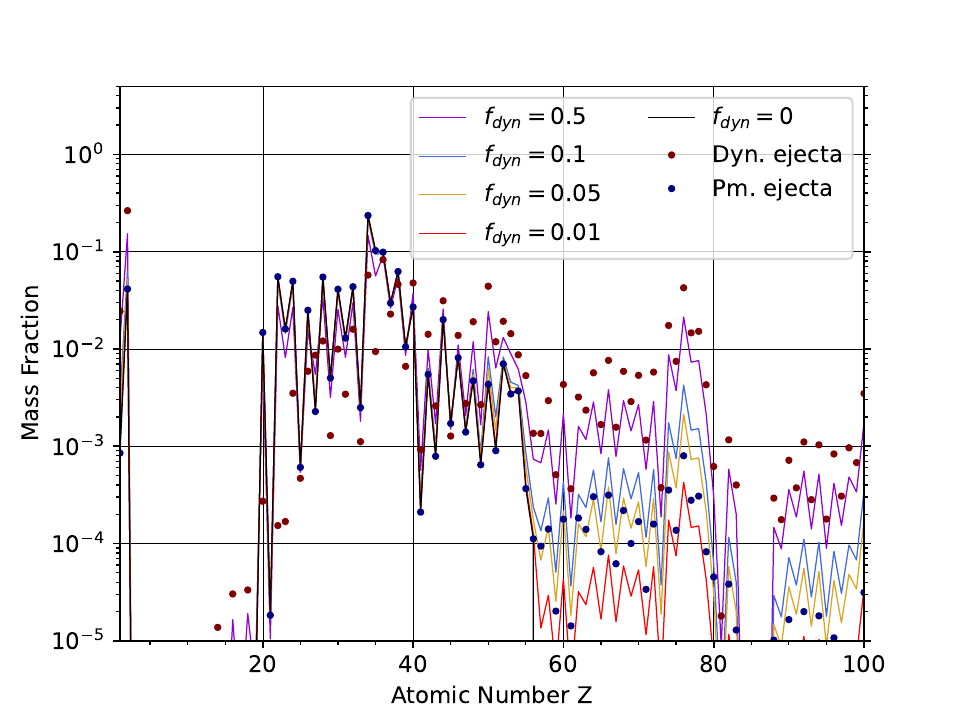} \\
    \includegraphics[trim={0.4cm 0.cm 0.4cm 0.3cm},width=1.1\linewidth]{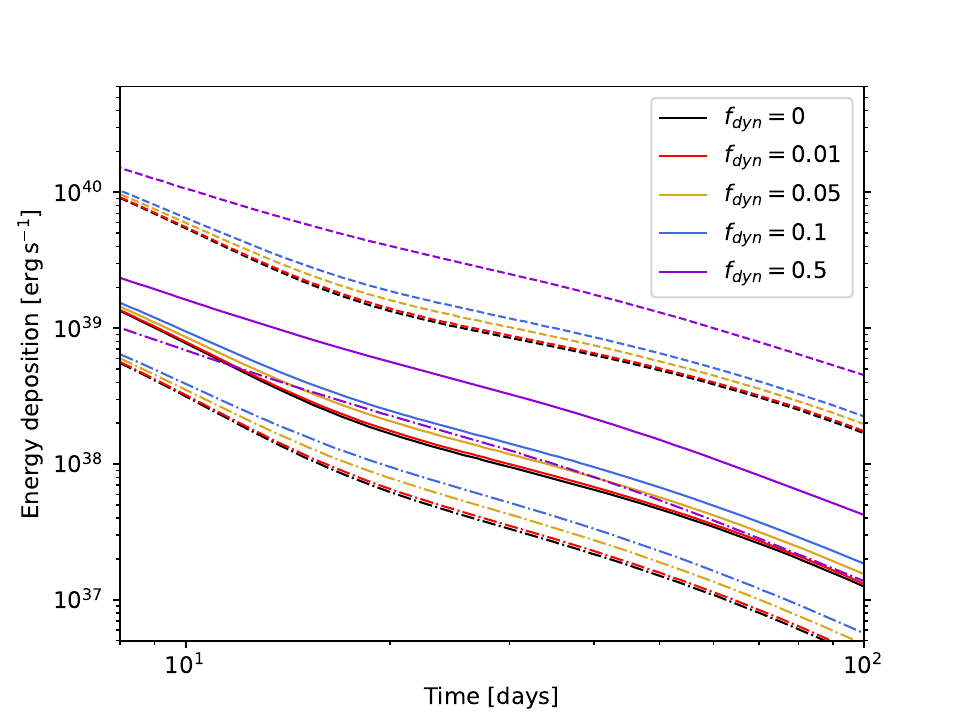}
    \caption{Model abundances (top panel) and associated energy depositions (bottom panel) including thermalisation efficiency, where the solid lines have $M_{\rm{ej}} = 0.01\Msol$, the dashed lines have $M_{\rm{ej}} = 0.05\Msol$, and the dashed-dotted lines have $M_{\rm{ej}} = 0.005\Msol$. In the top panel, we show the abundance pattern of the post-merger and dynamical components in dark blue and red points respectively. Note that the model compositions limited to 30 elements is not shown here, but are explicitly written in Table \ref{tab:compositions}.}
    \label{fig:dep_compare}
\end{figure}

The inner boundary of $v_{\rm{in}} = 0.05$c is chosen for consistency with previous KN studies which have used this value and due to convergence issues and memory constraints when solving line-by-line transfer in media with many optically thick lines \citep[e.g.][]{Pognan.etal:23}. In \textsc{sumo}, lines with a Sobolev optical depth of $\tau_S \geq 10^{-5}$ are explicitly treated in the transfer. The total of such lines may exceed the maximum allowed within current memory constraints ($\sim$350000) by a combination of dense ejecta (e.g. early times, slow velocities, and/or high masses) and the inclusion of many line-rich species, particularly lanthanides as in this study. For the lanthanide-rich models studied here with a starting epoch of 10~d and maximum ejecta mass of $0.05\,\Msol$, an inner boundary of $v_{\rm{in}} = 0.05$c was found to provide the best balance between including as many elements as possible (see below and Table \ref{tab:compositions}), while still being slow enough to cover the velocities of features expected from late time observations (e.g. $v \gtrsim 0.05c$, from the \textit{JWST} spectra of AT2023vfi, see Section \ref{subsubsec:AT2023vfi}) and \citet{Levan.etal:24,Gillanders.Smartt:25}.

Nevertheless, it is expected that more massive, slower moving inner ejecta be significant to spectral formation at late times. In comparison to a slower inner boundary, we expect more blending of features from Doppler broadening, as well as hotter temperatures and a higher degree of ionisation for a given epoch. It is difficult to gauge how temperature will change, as both heating and cooling are density dependent, with cooling additionally depending on the ionisation state, since different ions have different line-cooling capacities. The radioactive power and bolometric luminosity of the models will be slightly lower, as thermalisation of decay products will be less efficient at a given time. We note that this inner-boundary velocity lies in between those of other nebular phase studies that take $v_{\rm{in}} = 0.02, 0.08$c respectively \citep[][]{Jerkstrand.etal:25}. 

We select five values of $f_{\rm{dyn}} = (0,0.01,0.05,0.1,0.5)$, resulting in 15 individual models when considering the 3 ejecta masses, and find elemental abundances following equation \eqref{eq:composition}. The original NR and HD simulations used to create the ejecta models in this study, i.e. the symmetric BNS merger yielding a long-lived remnant, most closely corresponds to $M_{\rm{ej}} = 0.05\, \Msol$ with $f_{\rm{dyn}} = 0.01 - 0.05$ in our parameter space \citep[][]{Fujibayashi.etal:18,Kawaguchi.etal:21}. BNS mergers leading to short-lived remnants typically conserve a similar amount of dynamical ejecta, but much less post-merger ejecta, such that our model with $f_{\rm{dyn}} = 0.5$ and $M_{\rm{ej}} = 0.01\, \Msol$ is representative of this case \citep[e.g.][]{Fujibayashi.etal:23,Kawaguchi.etal:23}. Generally, black hole (BH)NS mergers may cover a broad range of total ejecta masses and component ratios, but are also able to reach parts of parameter space inaccessible to BNS mergers, for instance $f_{\rm{dyn}} = 0.5$ and $M_{\rm{ej}} = 0.05\, \Msol$ \citep[e.g.][]{Hayashi.etal:22,Kawaguchi.etal:24}. The $f_{\rm{dyn}} = 0$ model is a fiducial case of pure-post merger ejecta, and is useful for comparison to the $f_{\rm{dyn}} = 0.01$ model in order to gauge the minimal lanthanide mass-fraction required to yield significant impacts in the emergent spectra. 

\begin{table*}
    \centering
    \setlength\tabcolsep{0.4cm}
    \begin{tabular}{cccccc}
    \hline \hline
    Element & $f_{\rm{dyn}}=0$ & $f_{\rm{dyn}}=0.01$ & $f_{\rm{dyn}}=0.05$ & $f_{\rm{dyn}}=0.1$ & $f_{\rm{dyn}}=0.5$ \\
    \hline 
    $_{2}$He & 0.0521 & 0.0549  & 0.0660 & 0.0799 & 0.1898 \\
    $_{22}$Ti & 0.0698 & 0.0691 & 0.0662 & 0.0627 & 0.0346 \\
    $_{26}$Fe & 0.0315 & 0.0312 & 0.0303 & 0.0290 & 0.0192  \\
    $_{28}$Ni & 0.0691 & 0.0686 & 0.0663 & 0.0636 & 0.0417 \\
    $_{34}$Se & 0.2970 & 0.2942 & 0.2855 & 0.2739 & 0.1826  \\
    $_{35}$Br & 0.1295 & 0.1283 & 0.1235 & 0.1175 & 0.0699  \\
    $_{36}$Kr & 0.1247 & 0.1245 & 0.1236 & 0.1224 & 0.1131 \\
    $_{37}$Rb & 0.0374 & 0.0373 & 0.0369 & 0.0364 & 0.0327  \\
    $_{38}$Sr & 0.0788 & 0.0786 & 0.0777 & 0.0766 & 0.0677 \\
    $_{39}$Y & 0.0133 & 0.0132 & 0.0130 & 0.0127 & 0.0107 \\
    $_{40}$Zr & 0.0341 & 0.0343 & 0.0353 & 0.0366 & 0.0465 \\
    $_{42}$Mo & 0.0069 & 0.0070 & 0.0074 & 0.0080 & 0.0122 \\
    $_{44}$Ru & 0.0253 & 0.0254 & 0.0259 & 0.0266 & 0.0319 \\
    $_{46}$Pd & 0.0102 & 0.0103 & 0.0106 & 0.0109 & 0.0136 \\
    $_{48}$Cd & 0.0059 & 0.0061 & 0.0068 & 0.0077 & 0.0148 \\
    $_{50}$Sn & 0.0054 & 0.0059 & 0.0079 & 0.0104 & 0.0301 \\
    $_{52}$Te & 0.0089 & 0.0090 & 0.0096 & 0.0104 & 0.0163 \\
    $_{58}$Ce & - & 0.0001 & 0.0002 & 0.0004 & 0.0018 \\
    $_{60}$Nd & - & 0.0001 & 0.0003 & 0.0005 & 0.0027 \\
    $_{62}$Sm & - & 0.0001 & 0.0002 & 0.0004 & 0.0020 \\
    $_{63}$Eu & - & 0.0001 & 0.0001 & 0.0003 & 0.0015 \\
    $_{64}$Gd & - & 0.0001 & 0.0004 & 0.0007 & 0.0035 \\
    $_{65}$Tb &  - & 0.0001 & 0.0001 & 0.0002 & 0.0010 \\
    $_{66}$Dy & - & 0.0001 & 0.0005 & 0.0010 & 0.0047 \\
    $_{68}$Er & - & 0.0001 & 0.0004 & 0.0007 & 0.0037 \\
    $_{69}$Tm & - & 0.0001 & 0.0002 & 0.0004 & 0.0018 \\
    $_{70}$Yb & - & 0.0001 & 0.0003 & 0.0007 & 0.0033 \\
    $_{74}$W & - & 0.0002 & 0.0011 & 0.0022 & 0.0108 \\
    $_{76}$Os & - & 0.0005 & 0.0027 & 0.0053 & 0.0265 \\
    $_{78}$Pt & - & 0.0002 & 0.0010 & 0.0019 & 0.0094 \\ \hline
    $X_{\rm{La}}$ & 0 & 0.0010 & 0.0027 & 0.0053 & 0.0260 \\
    \hline \hline
    \end{tabular}
    \caption{Model compositions by elemental mass fraction, with total lanthanide mass fractions at the bottom. Note the minimum values of 0.0001 for the lanthanides in the $f_{\rm{dyn}} = 0.01$ model as described in the text.} 
    \label{tab:compositions}
\end{table*}

As previously mentioned, current computational limitations prevent us from conducting detailed line-by-line transfer in NLTE for all elements of the periodic table, and so we must limit the species included in the models. We select 30 elements with a minimum mass fraction of $10^{-4}$ \citep[][]{Pognan.etal:23}. We choose 10 elements with lines known or predicted to be relevant to KNe: He, Se, Br, Kr, Rb, Sr, Y, Zr, Te, and W. We consider lines both in the optical and IR; while this study focuses on the IR wavelengths, strongly absorbing or emitting lines in the optical still play an important role in determining the temperature of the ejecta, which can affect the emergent IR spectrum, or more directly by fluorescence. As this study focusses on the impact of lanthanides, we always include the 10 most abundant lanthanides even if their individual abundances lie below the minimum threshold of $10^{-4}$, in which case this minimum value is taken; this occurs for the $f_{\rm{dyn}} = 0.01$ model. Finally, the remaining 10 elements are chosen by taking into account both their overall abundance in the composition and their contribution to total opacity across all values of $f_{\rm{dyn}}$. For the values studied here, this tends to favour 1st and 2nd peak species. However previous NLTE studies have found that 3rd peak species typically contribute little spectrally or in terms of cooling when present, as lanthanides and actinides dominate instead \citep[][]{Pognan.etal:23,Pognan.etal:25}. As such, the omission of most 3rd peak elements aside from W, Os and Pt is not expected to significantly impact our results. 

The abundance of the chosen elements may vary with time due to continued radioactive decay of certain isotopes, however this variation is not currently implemented in \textsc{sumo} when solving the time-dependent temperature and ionisation structure. As such, we consider for every model the evolution of elemental abundances across the 10 - 75 day timespan studied here, finding that the abundances at 25 days represent a good average, and most abundances do not vary significantly in this timespan. The largest variation is found for $_{28}$Ni, of which the radioactive isotopes with $A = 56,66$ continue to decay with time \citep{Wanajo:18,Fujibayashi.etal:20b}, though the stable isotopes with $A = 60,62,64$ dominate the abundance. The change in $_{28}$Ni is most significant in the dynamical component, where the $_{28}$Ni mass fraction drops from $1.49 \times 10^{-2}$ at 10~d, to $1.15 \times 10^{-2}$ at 75~d. However, most of the $_{28}$Ni comes from the post-merger component, where the variation is only of $\sim 1 \times 10^{-3}$. 

In the case of long-lived ($t \gtrsim 100\,$ms) BNS merger remnants, it has been found that a substantial amount of radioactive $^{56}$Ni may be formed by a proton-rich wind \citep[][]{Jacobi.etal:25} (though see \citet{Cheong.etal:24}). Such a wind is not found in the simulations used here, potentially due to usage of a simpler neutrino leakage with absorption scheme as opposed to an M1 transport scheme \citep[see][for details]{Fujibayashi.etal:20b}, and the $_{28}$Ni isotopes produced are mainly the aforementioned stable ones. However, the NR simulations used as a basis for this study continue much further into the post-merger phase, 6\,s c.f. $\sim$100\,ms, such that it is not entirely clear whether unstable $^{56}$Ni isotopes would still play a large role in the case where most of the ejecta mass comes from the later post-merger components \citep[see also e.g.][]{Just.etal:23}.

Given the significant uncertainty in other areas of NLTE modelling, from the nuclear network calculations of abundances and radioactive power, to the formulae used to model the diverse NLTE processes, we expect that the inconsistency introduced by taking a fixed composition at each epoch to be minor in comparison. The resulting final compositions by mass fraction as well as associated total energy deposition, that is the radioactive power with thermalisation efficiency taken into account \citep[][]{Barnes.etal:16,Kasen.Barnes:19,Waxman.etal:19}, for each model may be visualised in Fig. \ref{fig:dep_compare}. We also provide the exact values of the mass-fractions for each element in Table \ref{tab:compositions}.

\subsection{Radiative transfer simulation}
\label{subsec:RTsim}

The NLTE radiative transfer (RT) code \textsc{sumo} \citep[][]{Jerkstrand.etal:11,Jerkstrand.etal:12} is used to perform the spectral synthesis simulations, with the most up to date modifications for KN models described in \citet{Pognan.etal:25}. To briefly summarise, \textsc{sumo} is a spherically symmetric RT code that solves NLTE rate equations for the ejecta temperature, ionisation, and excitation structures iteratively with the radiation field using a Monte Carlo photon propagation method. As this code is designed to model the nebular phases of explosive transients, the stationarity approximation is used, such that light travel time effects are neglected. More details on the processes included in the modelling and their calculation can be found in \citet{Pognan.etal:23}. The ejecta models described above are evolved over a timespan of 10 - 75 days in time-dependent mode \citep[][]{Pognan.etal:22a}, placed at a distance of 40 Mpc, and with a spectral range of $500\ang - 30\mu\rm{m}$. Since this study is focussed on the IR spectra of KNe, we show the spectra for $ \lambda = 1.2 - 30 \, \mu$m. 

The changes made to the RT simulations concern the usage of atomic data for r-process species. Previously, the thermal collision strengths, calculated using the van Regemorter \citep[][]{Regemorter:62} and Axelrod \citep{Axelrod:80} formulae for allowed and forbidden transitions respectively, were adjusted using a simple scaling factor based on ionisation state and ejecta temperature. While this treatment remains for most r-process species, collision strengths taken directly from atomic data files are now also included where available, e.g. in standard ADF04 files from the OPEN-ADAS database\footnote{https://open.adas.ac.uk/}. In these files, Maxwellian-averaged collision strengths $\Upsilon_{i,j}(T_e)$, from lower level \textit{i} to upper level \textit{j}, are given for a set of electron temperatures $T_e$. These values are fit by a simple power law in order to be incorporated into \textsc{sumo} at any given ejecta temperature. In order to prioritise better fits for temperatures relevant to those found in KN models, data points above $6 \times 10^{4}\,$K are excluded. 

Alongside the aforementioned changes to atomic data, some of the original atomic data calculated using the Flexible Atomic Code \citep[\textsc{fac},][]{Gu:08} have been modified. The low-lying energy levels of several species have been calibrated to data available from the National Institute for Standards and Technology (NIST) Atomic Spectra Database \citep[NIST ASD][]{NIST_ASD}. For a given energy level in NIST, the closest (in energy difference) available \textsc{fac} level with the same labelling, parity and total angular momentum \textit{J} is scaled to the value given in NIST \citep[see same method in e.g.][]{Domoto.etal:22}. Once all the desired energy levels are adjusted in this way, the Einstein A-coefficients for the radiative transitions associated to shifted energy levels are also scaled according to the following relation:

\begin{equation}
    A_{\rm{new}} = \left(\lambda_{\rm{FAC}}/\lambda_{\rm{new}}\right)^n A_{\rm{FAC}}
    \label{eq:A_vals}
\end{equation}

\noindent where $\lambda_{\rm{FAC}}$ is the original transition wavelength as calculated by \textsc{fac}, $\lambda_{\rm{new}}$ is the wavelength obtained when using the newly calibrated energy level(s), and $n = 3, 5$ for dipole and quadrupole transitions respectively \citep[][]{Cowan_atom81}. 

\begin{figure*}
    \centering
    \includegraphics[trim={0.4cm 0.cm 0.4cm 0.3cm},width = 0.47\textwidth]{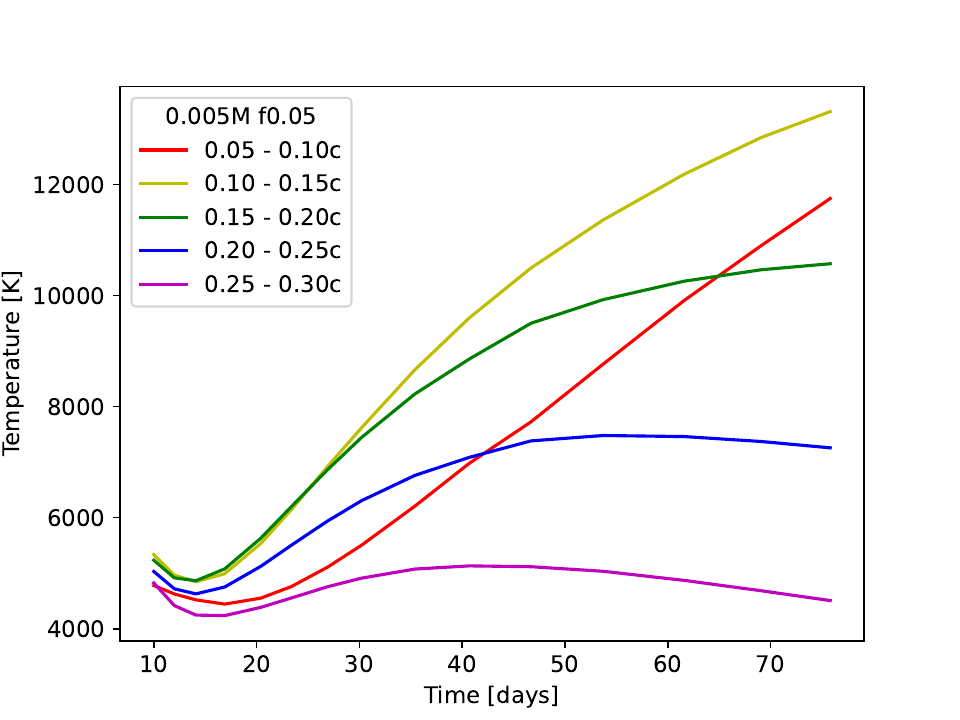} 
    \includegraphics[trim={0.4cm 0.cm 0.4cm 0.3cm},width = 0.47\textwidth]{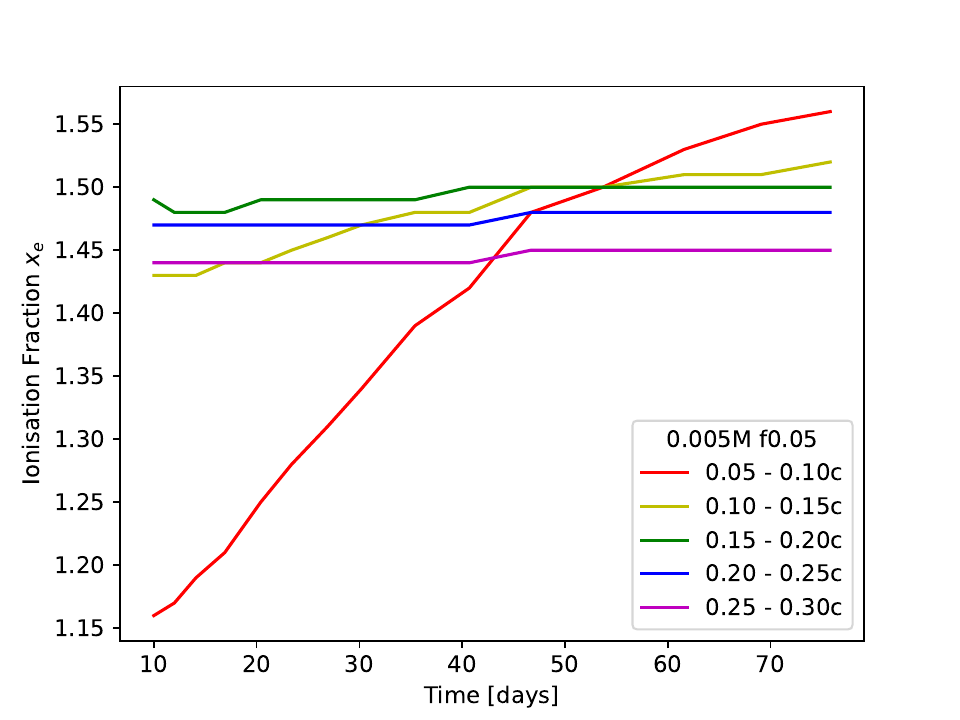}
    \includegraphics[trim={0.4cm 0.cm 0.4cm 0.3cm},width = 0.47\textwidth]{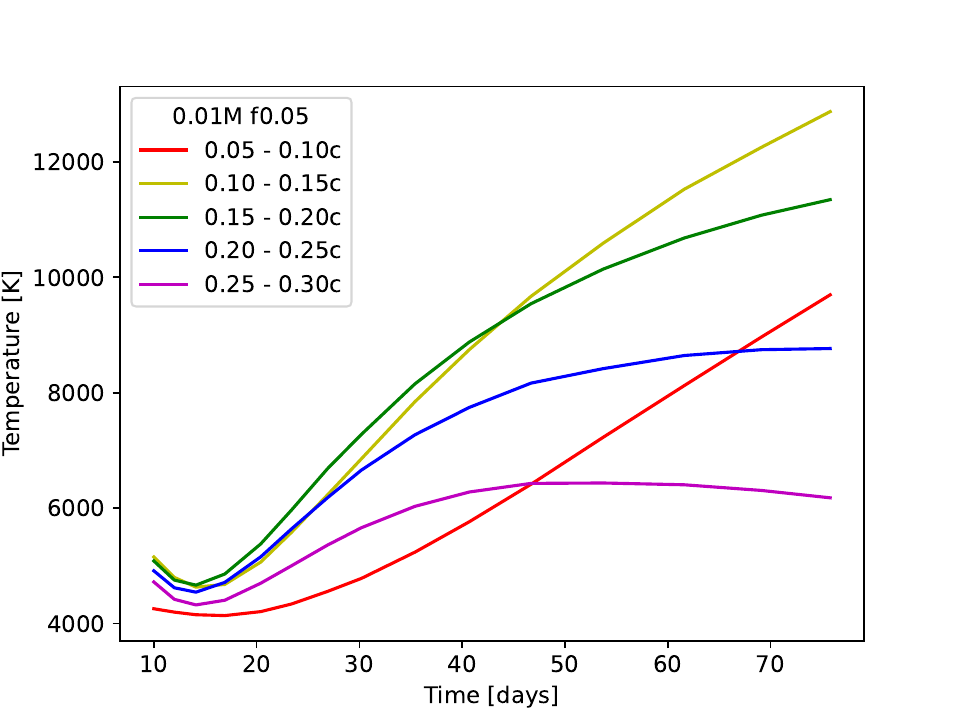}
    \includegraphics[trim={0.4cm 0.cm 0.4cm 0.3cm},width = 0.47\textwidth]{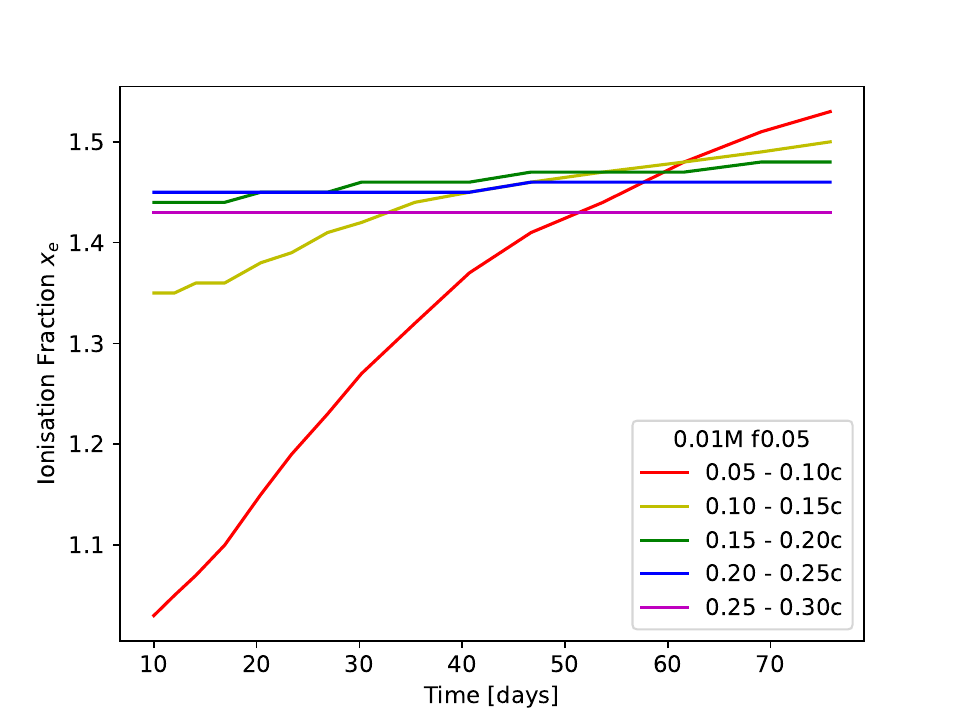}
    \includegraphics[trim={0.4cm 0.cm 0.4cm 0.3cm},width = 0.47\textwidth]{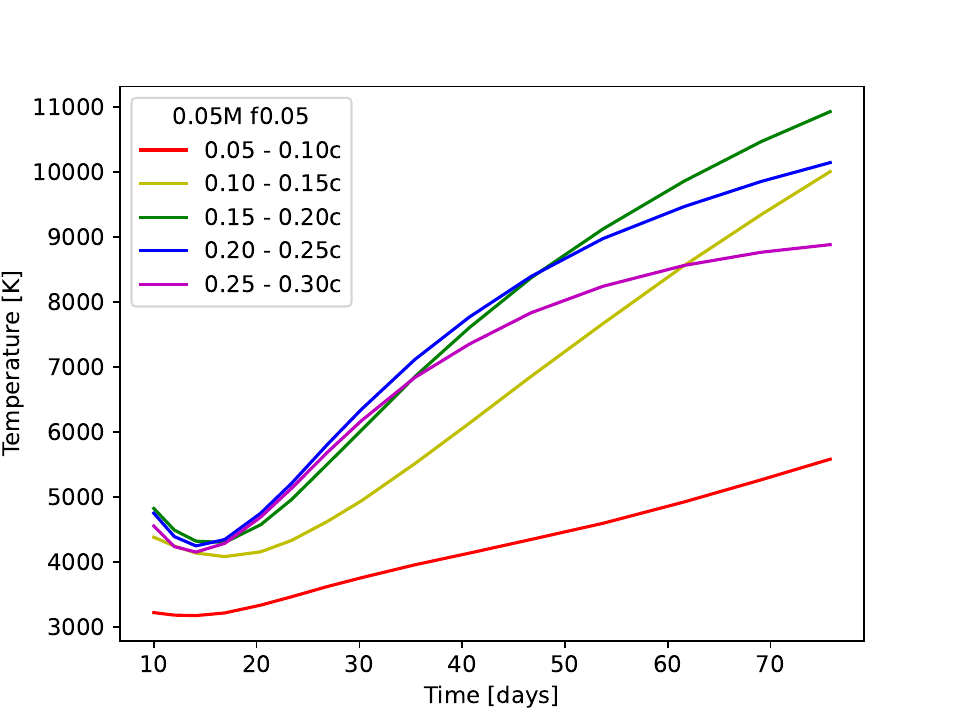}
    \includegraphics[trim={0.4cm 0.cm 0.4cm 0.3cm},width = 0.47\textwidth]{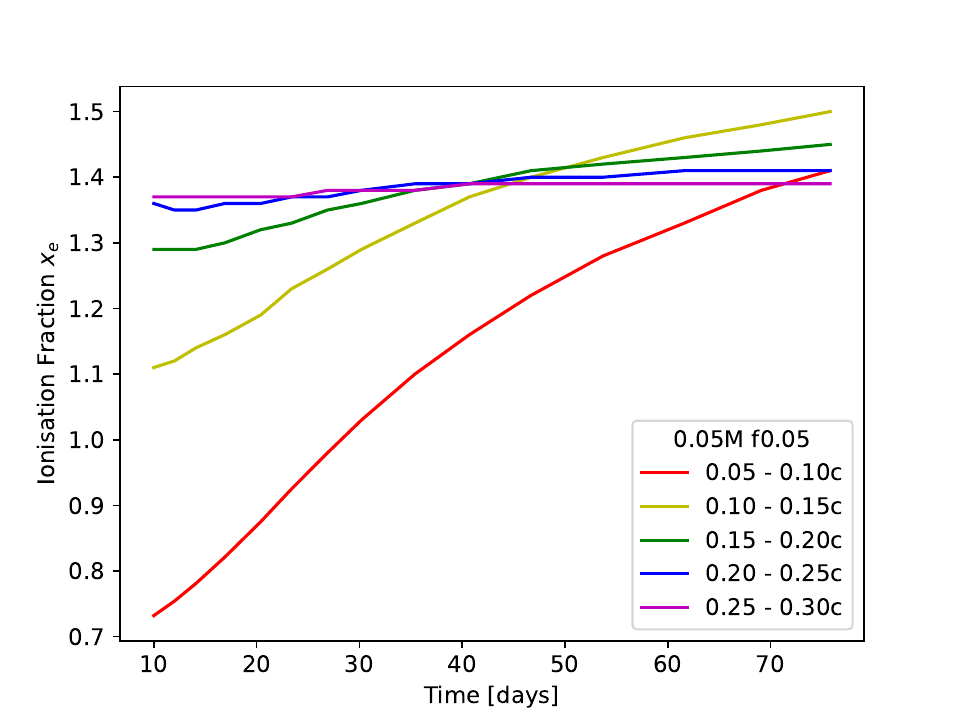}
    \caption{The evolution of temperature (left panels) and ionisation degree (electron fraction $x_e$, right panels) by zone of the $f_{\rm{dyn}} = 0.05$ model, with masses $M_{\rm{ej}} = 0.005, 0.01, 0.05 \, \Msol$ in the top, middle and bottom rows respectively}
    \label{fig:thermo_evolution}
\end{figure*}

Light elements ($Z \leq 28$) do not have atomic data originating from \textsc{fac}, but instead from well verified sources such as NIST \citep[][]{NIST_ASD} and/or the Kurucz database \citep[][]{Kurucz:18}. Additionally, they are not subject to the constant total recombination rate of $10^{-11} \, \mathrm{cm^3 \, s^{-1}}$ that is applied to r-process species, but either have tabulated data or use fitting formulae from various sources. Importantly, data only up to the doubly ionized stage was used for these light elements, such that Ti\,\textsc{iv}, Fe\,\textsc{iv} and Ni\,\textsc{iv} are not actively modelled in the emergent spectra, and are only considered for the calculation of ionisation structure. 

While some recombination rate data now exists for certain r-process species \citep[][]{Banerjee.etal:25,Singh.etal:25}, this study continues to use the previous constant rate of $\alpha_{\rm{rec}} = 10^{-11}\, \mathrm{cm^3 \, s^{-1}}$ for r-process ions, as these new data were not available at the initiation of this study (see Appendix \ref{app:atomic_data} for a discussion on this point). The impact of these new data on NLTE spectra of KNe is explored in \citet{Banerjee.etal:25}, as well as in the companion paper focussing on light ($Z\leq 40$) r-process elements in the nebular phase \citep[][]{Jerkstrand.etal:25}. When data for a given species are available, we discuss the possible impact on our spectral results (see Section \ref{sec:spectra}).

We describe details pertaining to the treatment of light elements, as well as changes to the \textsc{fac} r-process data in Appendix \ref{app:atomic_data}. The impact of not including triply ionized states for the previously mentioned light elements on the emergent spectra is discussed in Section \ref{sec:spectra}.

\section{Temperature and ionization structure in the nebular phase}
\label{sec:thermo}

While the evolution of temperature and ionisation structure of KNe in the nebular phase has been studied before, it remains useful to explicitly examine these quantities in order to better understand the emergent spectra. The models evolve in qualitatively similar manners, and so we present here the evolution of the $f_{\rm{dyn}} = 0.05$ model in Fig. \ref{fig:thermo_evolution}, while the others may be viewed in the supplementary data repository.

Looking first at the left-hand panels of Fig. \ref{fig:thermo_evolution}, we see how the temperatures of the 5 radial zones evolve. The layers initially cool until about 15\,d, which is contrary to previous NLTE predictions \citep[e.g.][]{Hotokezaka.etal:21,Pognan.etal:22a}. This likely arises from the radioactive power calculated by the nuclear network used for these particular models, as we see in Fig. \ref{fig:dep_compare} that the energy deposition has a steeper slope around these epochs. The reduced energy deposition impacts the heating rate per volume, which then scales more steeply with time than the line cooling rate per volume ($c_{\rm{line}} \propto t^{-6}$), therefore leading to ejecta cooling. 

Assuming that $\beta$-decay electrons dominate heating, and that atomic line emission dominates cooling, previous studies in NLTE have predicted an initial rapid heating of $T \sim t^{1.7}$, followed by a flatter evolution of $T \sim t^{0.2}$, after the thermalisation break of the $\beta$-decay electrons \citep[e.g.][]{Hotokezaka.etal:21}. At low enough densities, time-dependent effects begin to occur, degrading the heating and providing additional cooling such that the temperature flattens then turns around and decreases \citep[][]{Pognan.etal:22a}. We see this general evolution, of initial rapid heating, followed by shallower heating and eventual cooling, occurring the lower density outer-layers of the models. The variation in slopes between the different zones and models are due to deviations from the above assumptions, e.g. the raw heating rate not following exactly $t^{-1.3}$ at all times, the density-dependent onset and magnitude of time-dependent effects, the complex dependency of line-cooling on ionisation state and temperature etc. The time-dependent effects arise earlier in the low-density outer ejecta layers, which leads to a temperature structure inversion, where the innermost layer eventually becomes the hottest. The time at which this inversion is density dependent and varies by ejecta mass; greater densities lead to time-dependent effects becoming significant at later times. 

Turning now to the right hand panels, we examine the degree of ionisation in each ejecta layer,defined by the ionisation (electron) fraction $x_e = n_e/\sum_i n_i$, where $n_e$ is the number density of free electrons, and $n_i$ is the number density of a given ion, summed across all ions of all elements in the ejecta. A key result here is that time-dependent effects lead to ionisation structure freeze-out in the outermost three layers for every ejecta mass, while the inner two layers have a slowly increasing degree of ionisation which flattens with time. As for temperature, this eventually leads to an inversion in the ionisation structure when compared to early times. 

The two innermost layers are less susceptible to time-dependent effects, and slowly become more ionised with time, though the change remains relatively minor, at most from $x_e = 0.72$ to $x_e = 1.4$ in the innermost layer of the model with $M_{\rm{ej}} = 0.05\,\Msol$. Since these inner layers contain 80 per-cent of the ejecta mass, and therefore drive spectral formation at late times, it is expected that the species dominating the emergent spectra at early times may remain significant until late times. As such, the spectral shape of KNe in the nebular phase may be slowly changing, in contrast to the photospheric phase, where rapidly cooling ejecta lead to quick and drastic ionisation structure changes in a matter of hours \citep[e.g.][]{Banerjee.etal:20,Sneppen.etal:24a,Sneppen.etal:24b,Brethauer.etal:26}. 

The degree to which time-dependent effects play a role depends on several factors, particularly ejecta density. They will be reduced for a higher density at a given epoch, which could arise from a slower inner boundary or more centrally concentrated mass-profile. The ionisation freeze-out additionally depends on the recombination rate of a given species; a faster rate will delay this, while a slower one will hasten the effect. Combination of fast recombination rates with high densities can lead to steady-state conditions holding for a significant duration, while slower rates may lead to an earlier onset of time-dependent effects. Ultimately, a fully accurate treatment of NLTE physical conditions requires both usage of time-dependent physics alongside accurate recombination rates for all species.

\section{Spectra}
\label{sec:spectra}

\begin{figure*}
    \centering
    \includegraphics[trim={0.4cm 0.cm 0.4cm 0.3cm},width = 0.47\textwidth]{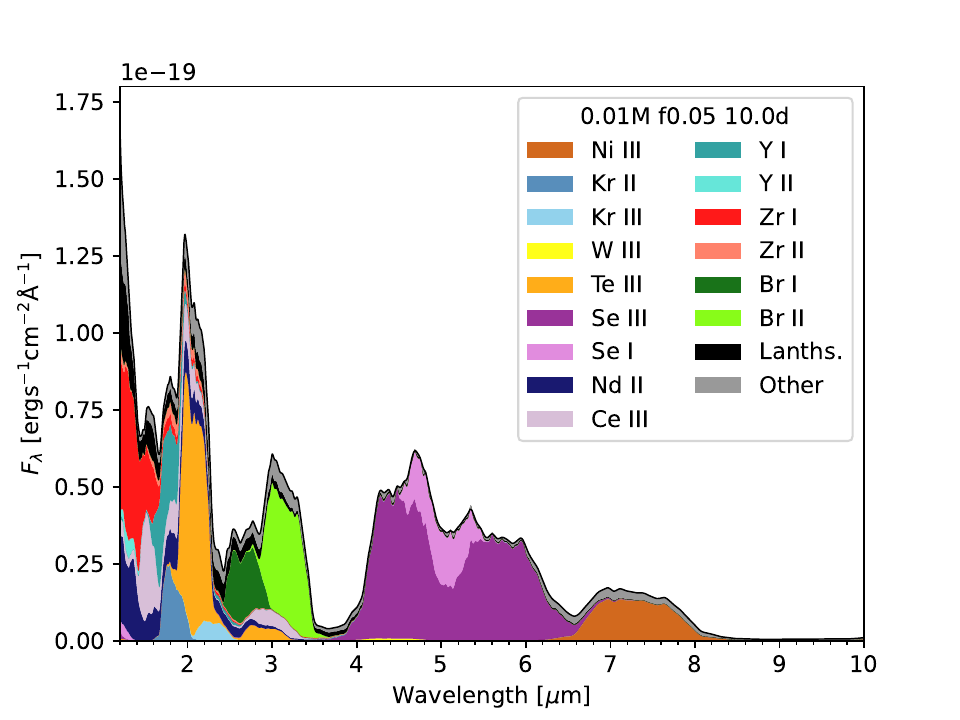} 
    \includegraphics[trim={0.4cm 0.cm 0.4cm 0.3cm},width = 0.47\textwidth]{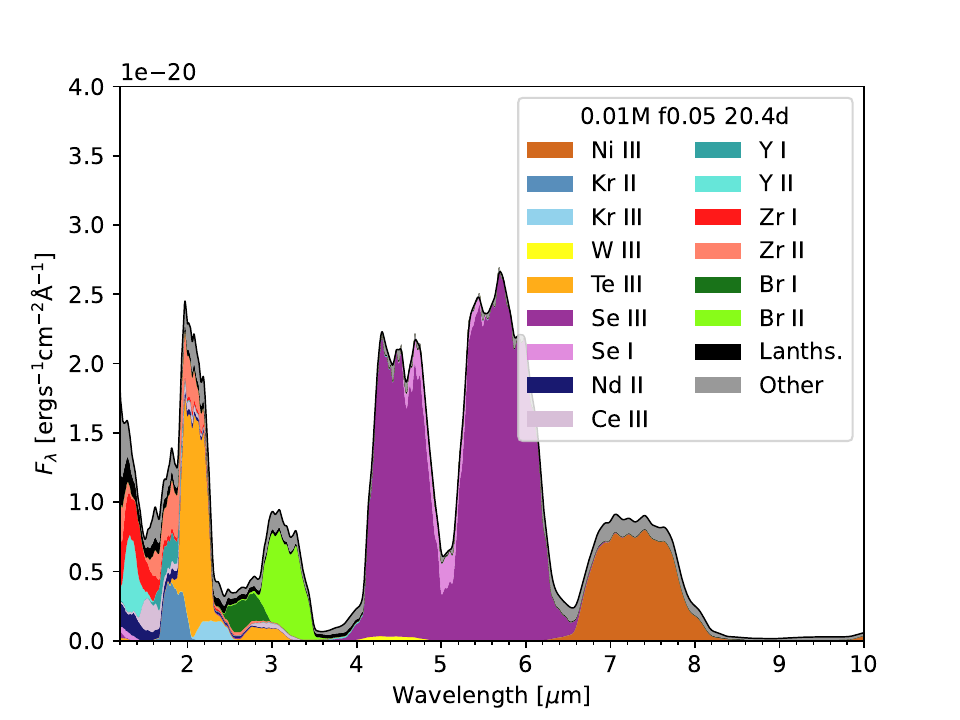} 
    \includegraphics[trim={0.4cm 0.cm 0.4cm 0.3cm},width = 0.47\textwidth]{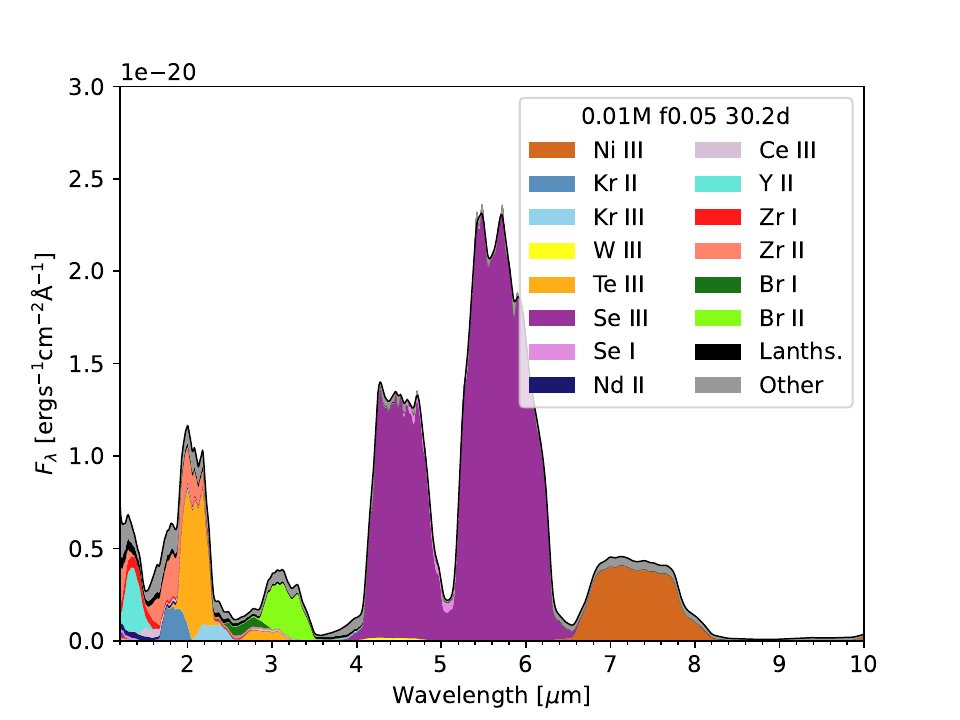} 
    \includegraphics[trim={0.4cm 0.cm 0.4cm 0.3cm},width = 0.47\textwidth]{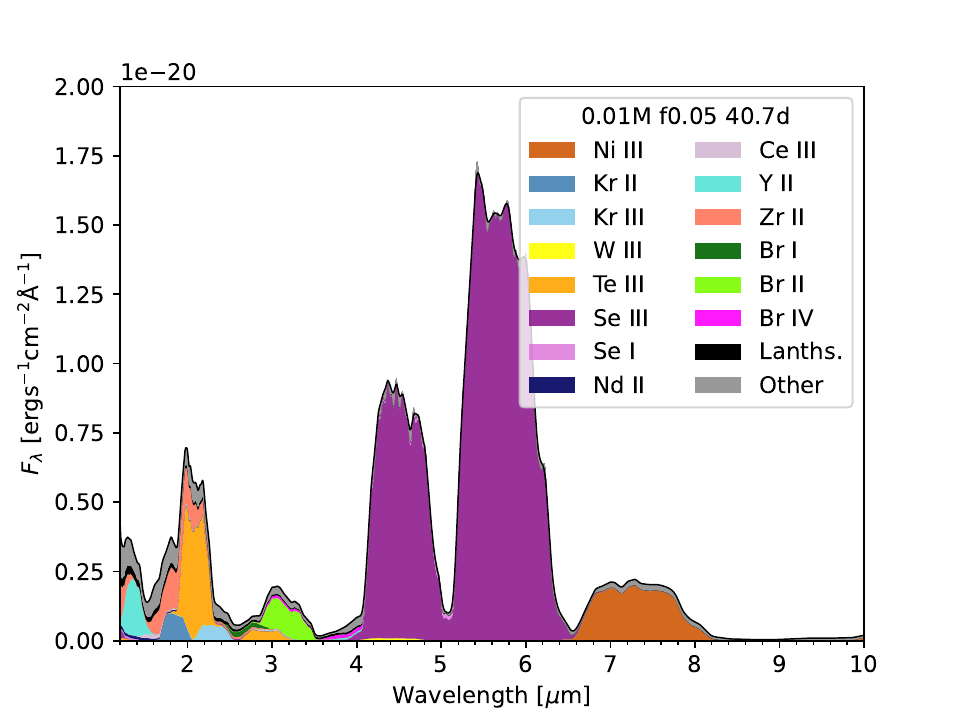} 
    \includegraphics[trim={0.4cm 0.cm 0.4cm 0.3cm},width = 0.47\textwidth]{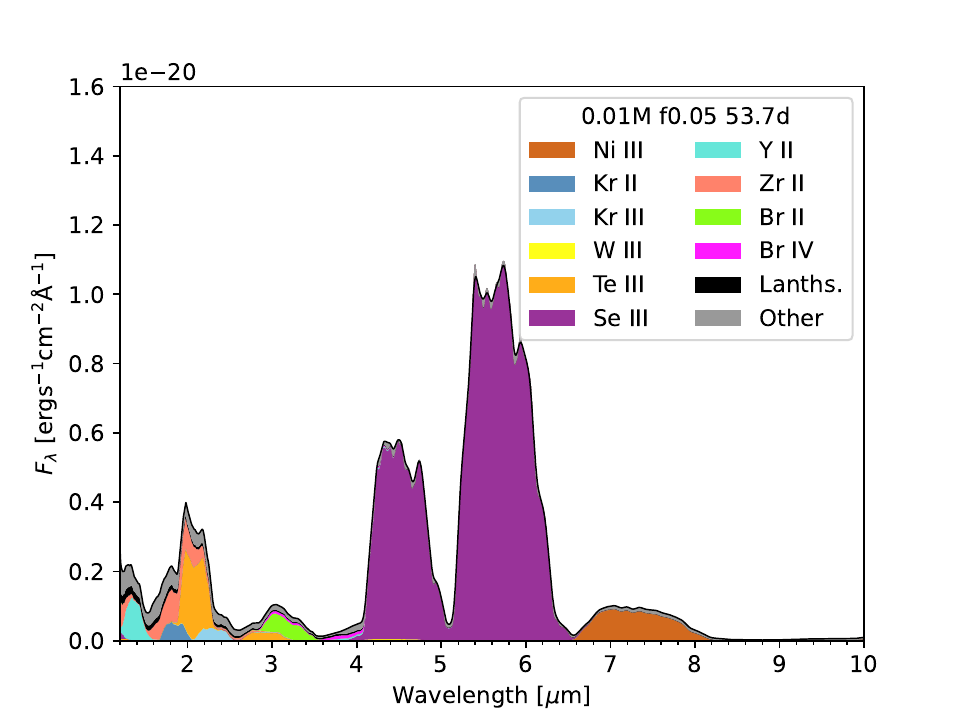} 
    \includegraphics[trim={0.4cm 0.cm 0.4cm 0.3cm},width = 0.47\textwidth]{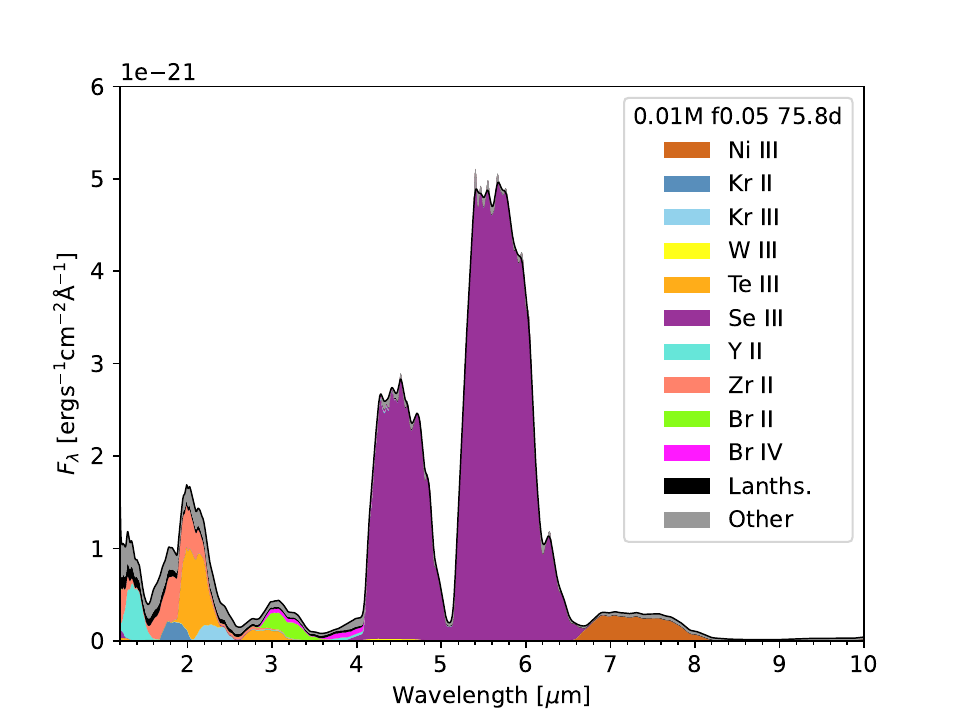} 
    \caption{Spectra of the $f_{\rm{dyn}} = 0.05$, $M_{\rm{ej}} = 0.01\, \Msol$ model from 10 to 75 days post-merger. Elemental compositions are shaded. Note that the contributions are summed (i.e. stacked), and that the black-coloured lanthanide contribution excludes lanthanide species already present otherwise.}
    \label{fig:f005M001_NIRspec}
\end{figure*}

We synthesise spectra for each of the 15 models from 10 to 75 days, resulting in a total of 105 individual spectra. Since the chosen elements for every value of $f_{\rm{dyn}} \neq 0$ are identical and only vary in abundances, many of the spectra evolve in a similar manner across the timespan studied here. We take the $f_{\rm{dyn}} = 0.05$, $M_{\rm{ej}} = 0.01\, \Msol$ model as representative of the average evolution, and refer to it as our `standard' model. We first present the general spectral trends found ubiquitously across the model parameter space, and then explore the impact of mass at various epochs on the `standard' model. After this, we examine the edges of our parameter space: the $f_{\rm{dyn}} = 0.01$ and $f_{\rm{dyn}} = 0.5$ models. 

Since the spectral range presented is relatively large ($1.2 - 30\, \mu$m), we split the spectra into two parts: $1.2 - 10 \, \mu$m and $10 - 30\, \mu$m, for presentation purposes. We focus first on the shorter wavelengths which are more spectrally active, and then cover the longer wavelengths which are fainter and highly similar between all models at most epochs. We then compare our model spectra to existing spectral observations of KNe in the nebular phase, that is the 10.4 day spectrum of AT2017gfo \citep[e.g.][]{Abbott.etal:17,Smartt.etal:17,Pian.etal:17}, and the 29 and 61 day \textit{JWST} spectra of AT2023vfi \citep[][]{Levan.etal:24,Gillanders.Smartt:25}. The spectra for every model at every epoch may be viewed in their entirety in the supplementary material.

\subsection{General trends in 1.2 to 10 micron range}
\label{subsec:general_NIRspec}

The spectral evolution of the `standard' $f_{\rm{dyn}} = 0.05$, $M_{\rm{ej}} = 0.01\, \Msol$ model is shown in Fig. \ref{fig:f005M001_NIRspec}, with key individual species contributions shaded. Starting with the top left panel which shows the model at 10 days, we see that the landscape of contributions at $\lambda \leq 2.4\, \mu$m is quite complex, with diverse emitting species. We find significant emission from the Te\,\textsc{iii} 2.1 $\mu$m line, with additional contributions from the lanthanides Nd\,\textsc{ii} and Ce\,\textsc{iii}, as well as some blending with emission from Kr\,\textsc{ii} at 1.86 $\mu$m and Kr\,\textsc{iii} at 2.32 $\mu$m. We note that both of these latter are M1 transitions from the first excited state to the ground state. and also that the Kr\,\textsc{iii} transition is not calibrated, having a true wavelength of $\lambda = 2.20\,\mu$m. 

\begin{figure}
    \centering
    \includegraphics[trim={0.4cm 0.cm 0.4cm 0.3cm},width=1\linewidth]{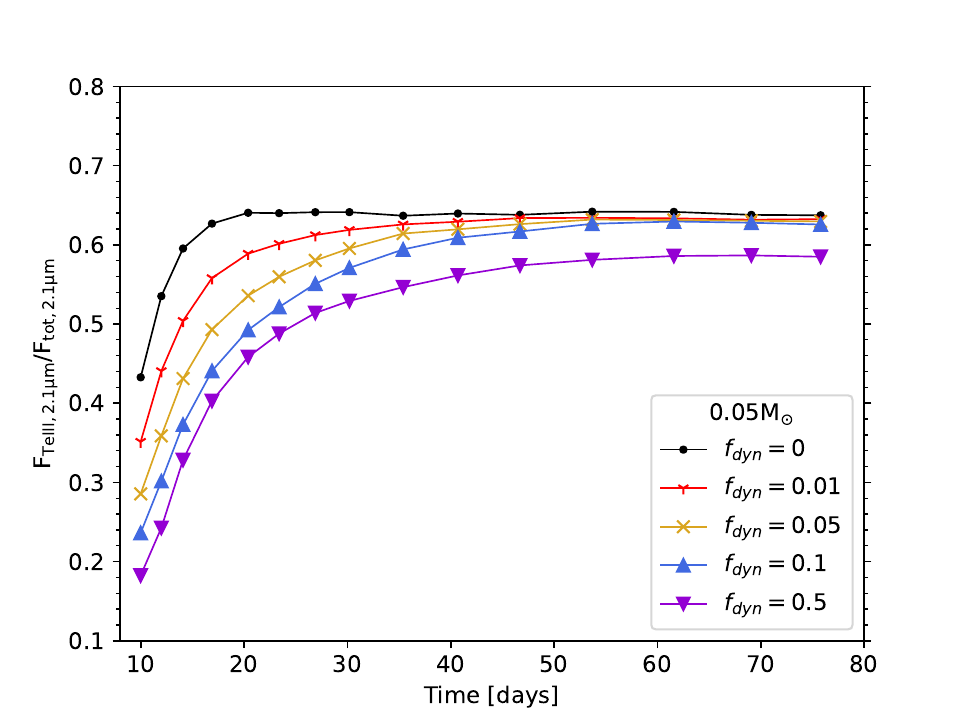}\\
    \includegraphics[trim={0.4cm 0.cm 0.4cm 0.3cm},width=1\linewidth]{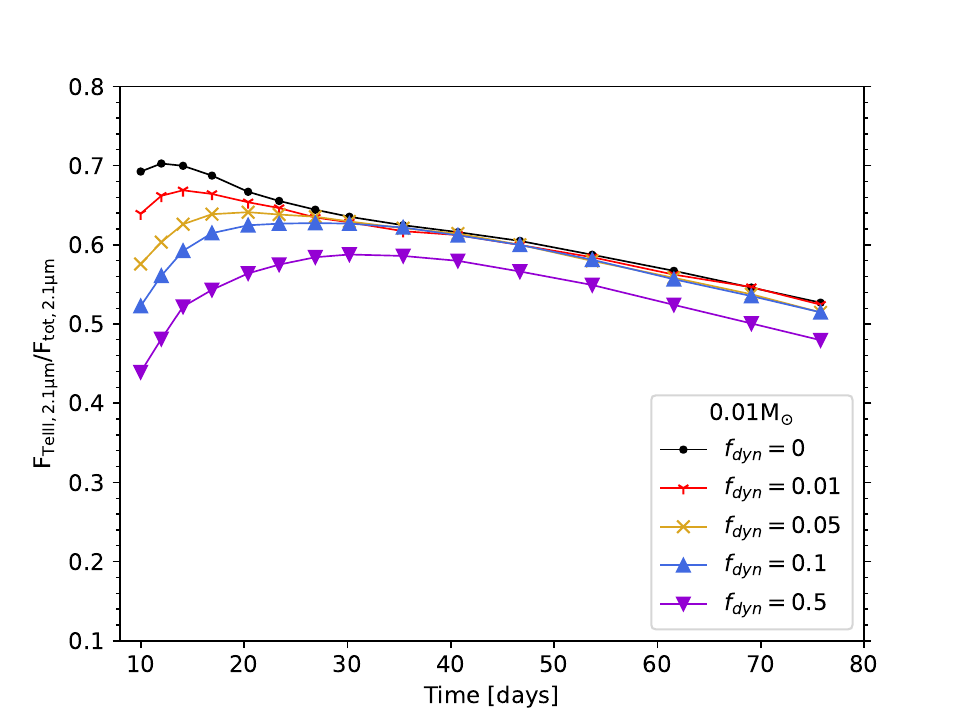} \\
    \includegraphics[trim={0.4cm 0.cm 0.4cm 0.3cm},width=1\linewidth]{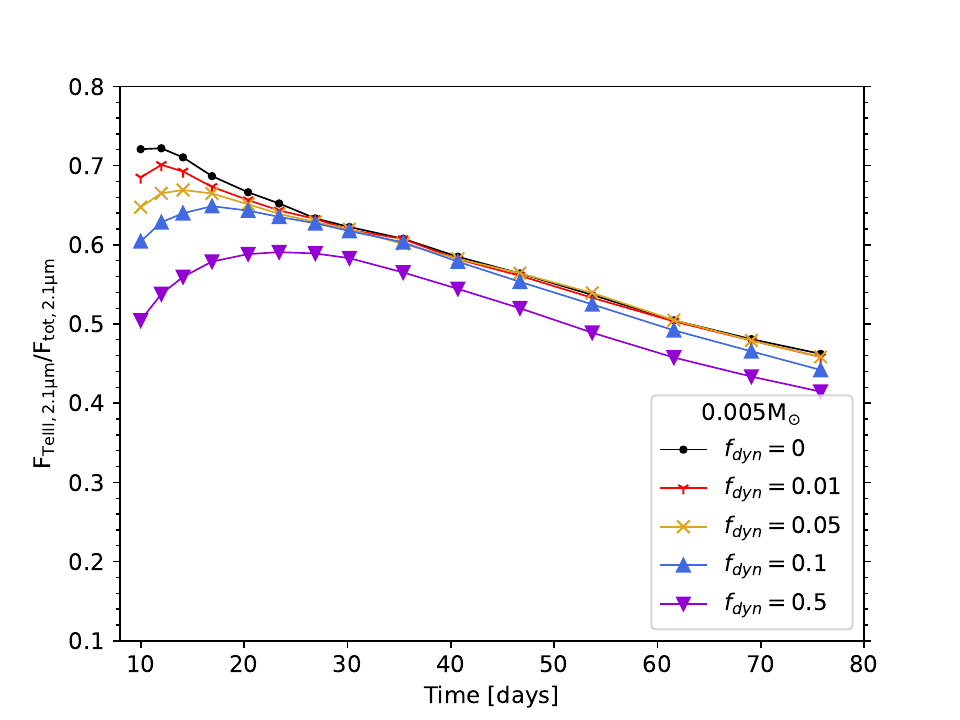}
    \caption{Fraction of Te\,\textsc{iii} flux in the range of 1.9 -- 2.3 $\mu$m.}
    \label{fig:TeIII_fracs}
\end{figure}

We examine the significance of the Te\,\textsc{iii} emission at 2.1 $\mu$m more quantitatively by calculating what fraction of the total emitted flux it is responsible for in a range of 1.9 -- 2.3 $\mu$m, the results of which are shown for all models and epochs in Fig. \ref{fig:TeIII_fracs}. We find that its contribution at 10 days varies significantly by model composition and ejecta mass, ranging from $\sim$0.18 -- 0.72, with our `standard' model (yellow crosses in the centre panel) at $\sim$ 0.58.

Below 2 $\mu$m we find a significant blending of species, with important contributions from Y\,\textsc{i}, Zr\,\textsc{i}, and various lanthanides, of which Nd\,\textsc{ii} and Ce\,\textsc{iii} are the most dominant. The emission of these species arises from various E1 transitions, contrary to the low-lying M1 transitions of the other species found in the emergent spectrum. The vast majority of these E1 lines are, however, not optically thick at these wavelengths, i.e. they have the Sobolev optical depth $\tau_{\rm{sob}} < 1$, such that they are likely not absorbing emission from other species, and merely blending their emission with key species like Te\,\textsc{iii}. Regardless, their non-negligible contribution to the IR spectrum at 10 days highlights the importance of taking E1 dipole emission into account in the early nebular phase ($t \lesssim 10$ days). 

Moving redwards, we find important M1 forbidden emission from Br\,\textsc{i} at 2.71 $\mu$m, and Br\,\textsc{ii} at 3.18 $\mu$m, both transitions from the first excited state to the ground state, and with some minor blending from other species, notably the lanthanides and the Te\,\textsc{iii} $n = 3 \rightarrow 2$ forbidden transition at 2.9 $\mu$m. We then find a significant drop in emission at $\lambda \sim$ 3.5 -- 4.0 $\mu$m, followed by a double-peaked emission feature from Se\,\textsc{iii} at 4.5 and 5.7 $\mu$m, blended with Se\,\textsc{i} at 5.0 $\mu$m. The Se emission lines are M1 forbidden transitions between low lying states, the same as those initially predicted in previous semi-analytical studies and also suggested within observational contexts \citep[][]{Hotokezaka.etal:22,Gillanders.etal:24,Gillanders.Smartt:25}. Centred on 7.35 $\mu$m, we then have a broad emission feature from Ni\,\textsc{iii} $n = 2 \rightarrow 1$, after which the emission drops to negligible levels.  

As we move forwards in time to 20 and 30 days, we see the spectra evolve in various ways. Notably, the influence of neutral species is decreased, as they are ionized out of the inner two layers (see Fig. \ref{fig:thermo_evolution}). At 30 days, we find that the emission from Se\,\textsc{i}, Br\,\textsc{i}, Y\,\textsc{i} and Zr\,\textsc{i} has become minor, if not negligible, while Y\,\textsc{ii} and Zr\,\textsc{ii} have become more prominent. Now, the Y\,\textsc{ii} emission at 1.39 $\mu$m, and the Zr\,\textsc{ii} around $\sim 1.7\, \mu$m are blends of several low-lying M1 emission lines. Both of these species have low-lying states all of even parity, which precludes them from having E1 dipole transitions \citep[see e.g.][]{NIST_ASD}. The emission from doubly ionized species such as Se\,\textsc{iii} and Ni\,\textsc{iii} remains significant, even gaining in relative importance to the rest of the species as the ionisation structure favours doubly ionized states. This also occurs for Te\,\textsc{iii}, and we find that it becomes slightly more dominant at 2.1 $\mu$m, as is well shown in Fig. \ref{fig:TeIII_fracs}.

The lanthanide emission is significantly reduced in this time-span, even though the ionisation balance still includes significant quantities of Nd\,\textsc{ii} and Ce\,\textsc{iii} at 30 days, e.g. within the innermost layer, Nd\,\textsc{ii} is 36.3 per-cent of $_{60}$Nd, and Ce\,\textsc{iii} is 44.8 per-cent of $_{58}$Ce. The decrease in their emission is due to the way the levels responsible for the IR emission are populated. In the case of Nd\,\textsc{ii}, we find that a broad range of levels with energies ranging from $\sim 10 000 - 20 000 \, \rm{cm^{-1}}$, corresponding to temperatures of $\sim 14 000 - 29 000\,$K, are initially responsible for the IR emission. This is hotter than the temperature of the ejecta at this time (see Fig. \ref{fig:thermo_evolution}), such that their population comes mostly from recombination of Nd\,\textsc{iii}. As time progresses, recombination becomes inefficient due to decreasing density, despite an increase in available Nd\,\textsc{iii}, such that these states become less populated. Additionally, with optical depth at bluer wavelengths also decreasing, previously optically thick optical transitions become available escape channels for emitted photons, and are therefore energetically favoured deexcitation channels over those emitting in the IR.

The Ce\,\textsc{iii} levels on the other hand are more low-lying and thermally accessible, suggesting they are populated by thermal collisions and photoexcitation (PE). It was found in previous studies that PE may continue to yield significant impact on the excitation structure of species in NLTE after thermal collisions become inefficient \citep[][]{Pognan.etal:22b}, so we focus on PE rates in the following. We find significant PE occurring from the first 20 levels ($E \leq 9900\, \mathrm{cm^{-1}}$) of Ce\,\textsc{iii} at 10 days for our standard model, ranging from $\sim 10^{-3} - 20\, \rm{s^{-1}}$, and reaching higher levels of up to $\lesssim 20 000\, \rm{cm^{-1}}$. At 40 days however, the rates drop to a range of $\sim 10^{-4} - 10^{-1}\, \rm{s^{-1}}$, with only the first 8 levels ($E \leq 5000\, \rm{cm^{-1}}$) having any significant PE, and only reaching levels up to $\sim 10 000 \, \rm{cm^{-1}}$. As both PE and collisional excitation become inefficient with time, only the lowest lying levels of Ce\textsc{iii} remain accessible by either process, thus reducing the species' spectral impact. 

Given the similar structure of lanthanides, similar explanations or a combination of the aforementioned reasons may also explain the general lack of lanthanide contribution at later times. One must also keep in mind the relatively low elemental abundance of these species to begin with; this model only has $X_{\rm{La}} = 0.0027$, which does remain, however, well within the estimates for AT2017gfo.

Looking now at the spectral evolution in the 40 - 75 day range, we see that the spectral shape evolves very slowly. One should still note the decreasing flux levels, corresponding to the reduction in deposited energy at these later epochs. The reason for this effectively frozen spectral shape lies with the slowly changing temperature and ionisation states of the ejecta. This highlights the critical importance of taking into account time-dependent effects in NLTE calculations of nebular phase KNe, without which the degree of ionisation would steadily continue to increase, and the emergent spectra would continue to noticeably change in this timespan. We also find that lanthanide emission in this time range is negligible compared to that of other species, particularly first peak species and Te\,\textsc{iii}. 

In terms of relative contributions, the Te\,\textsc{iii} emission at 2.1 $\mu$m decreases steadily in our fiducial model (see Fig. \ref{fig:TeIII_fracs}), going from a flux fraction of $\sim$0.61 at 40 days, and reaching $\sim$0.52 at 75 days, while the Se\,\textsc{iii} double-peaked emission at 4.5 and 5.7 $\mu$m remains the dominant spectral feature in this wavelength range. The Ni\,\textsc{iii} feature at 7 $\mu$m slowly reduces with time however, as it is ionized out to Ni\,\textsc{iv}. Though the ionisation threshold of Ni\,\textsc{iii} is quite high at $\sim 35$~eV, the prevalence of Ni\,\textsc{iv} at late times is explainable. 

Firstly, both Ni\,\textsc{i} and Ni\,\textsc{ii} have relatively low recombination rates of $\sim 10^{-13}$ and $10^{-12}\, \mathrm{cm^3 \, s^{-1}}$ respectively \citep{Shull.Steenberg:82}, leading to a large amount of Ni\,\textsc{iii} at early times: 65 per-cent of all $_{28}$Ni in the innermost layer of the `standard' model at 10 days. Additionally, Ni\,\textsc{iii} is found to have high non-thermal (NT) ionisation rates, roughly 2 -- 4 times higher than doubly ionized r-process elements at the same epoch. The large amount of Ni\,\textsc{iii} present at early times, combined with slightly higher NT ionisation rates for this species thus leads to an increasing amount of Ni\,\textsc{iv} with time, maximally 71 per-cent at 75 days in the innermost layer of the standard model.

As was mentioned in Section \ref{subsec:RTsim} however, Ni\,\textsc{iv} is not explicitly included in the RT modelling since no atomic data exists within \textsc{sumo} for this ion. Given the dominance of this ion in the ionisation balance of $_{28}$Ni at late times, it is likely that this species would produce some spectral features. Examining the level structure of Ni\,\textsc{iv} as given by NIST, we find a ground multiplet structure potentially allowing for three IR transitions of relevance, at wavelengths of 8.405, 11.726 and 17.271 $\mu$m. Transition probabilities are not available on NIST for these lines, but taking them to be of similar strength as the ground multiplet transitions of Fe\,\textsc{ii} for which we have data, we estimate that their A-values would be on the order of $10^{-3}$ --  $10^{-4} \, \mathrm{s^{-1}}$. Therefore, our model spectra may be missing some additional Ni\,\textsc{iv} features at the aforementioned wavelengths. Conducting a similar analysis for Ti\,\textsc{iv} and Fe\,\textsc{iv}, which are found to be maximally 66 and 86 per-cent of their ionisation balance respectively at 75 days, we find a potential Ti\,\textsc{iv} transition at 26.28 $\mu$m, while the structure of Fe\,\textsc{iv} does not permit any low-lying M1 transitions.

\begin{figure*}
    \centering
    \includegraphics[trim={0.4cm 0.cm 0.4cm 0.3cm},width=0.47\linewidth]{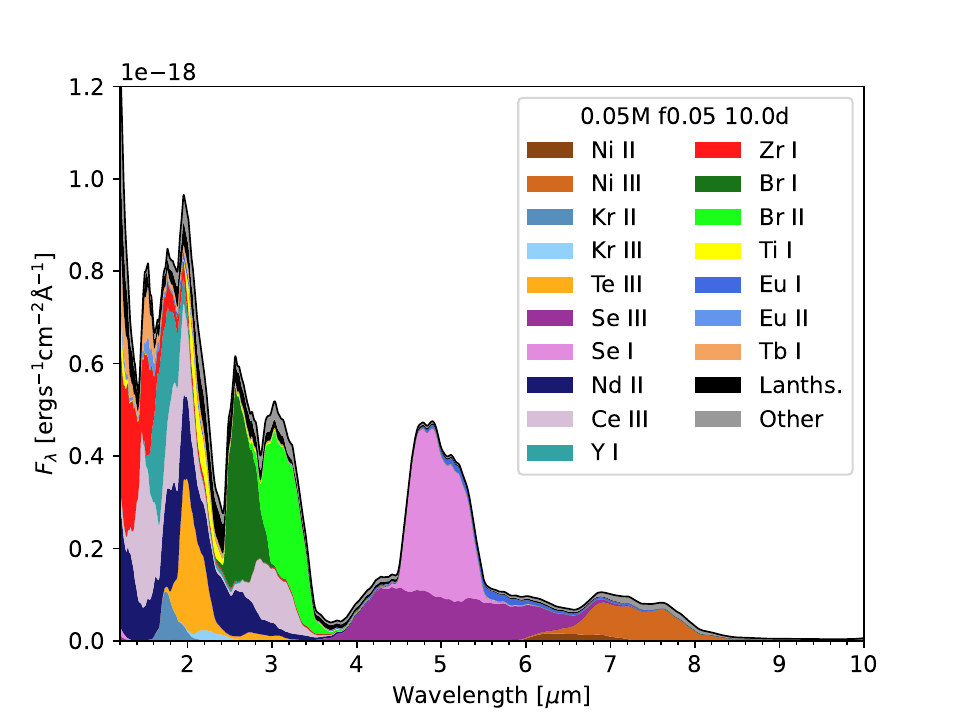}
    \includegraphics[trim={0.4cm 0.cm 0.4cm 0.3cm},width=0.47\linewidth]{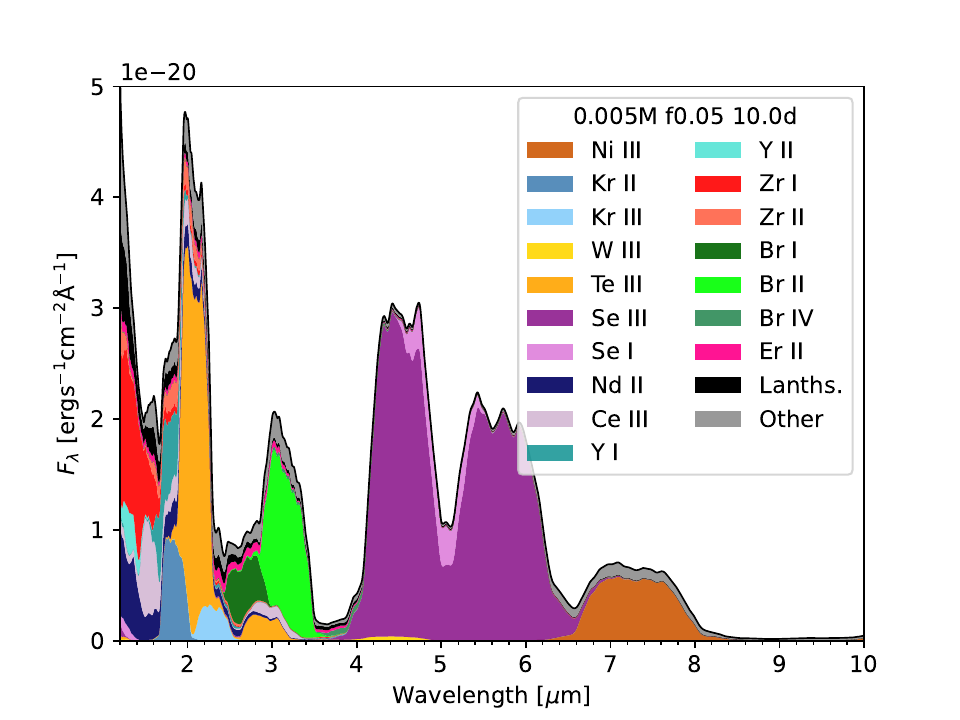}
    \includegraphics[trim={0.4cm 0.cm 0.4cm 0.3cm},width=0.47\linewidth]{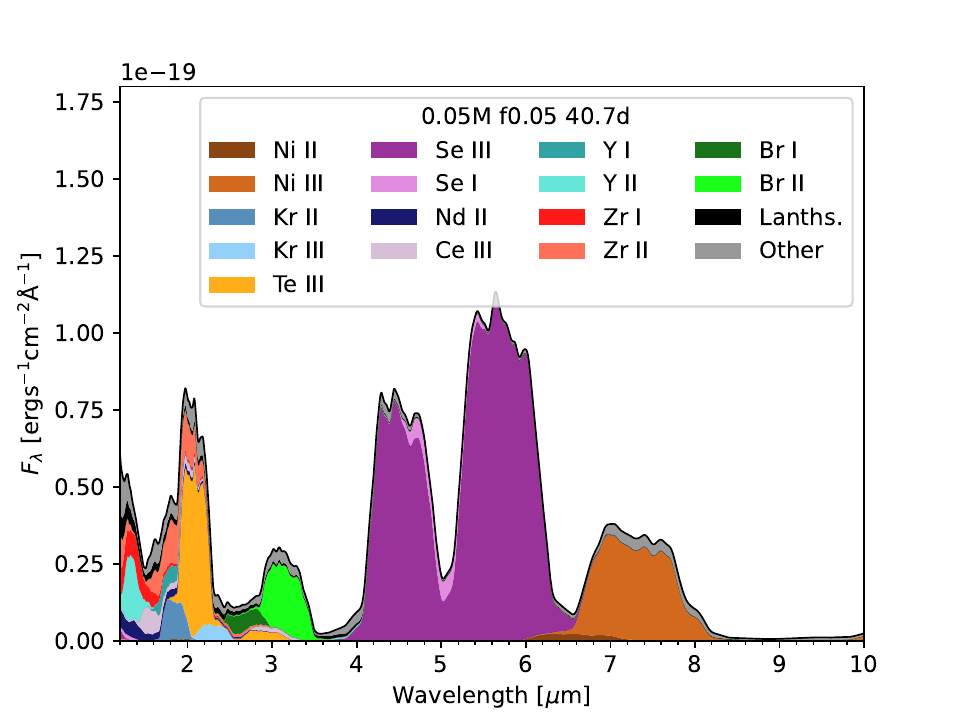}
    \includegraphics[trim={0.4cm 0.cm 0.4cm 0.3cm},width=0.47\linewidth]{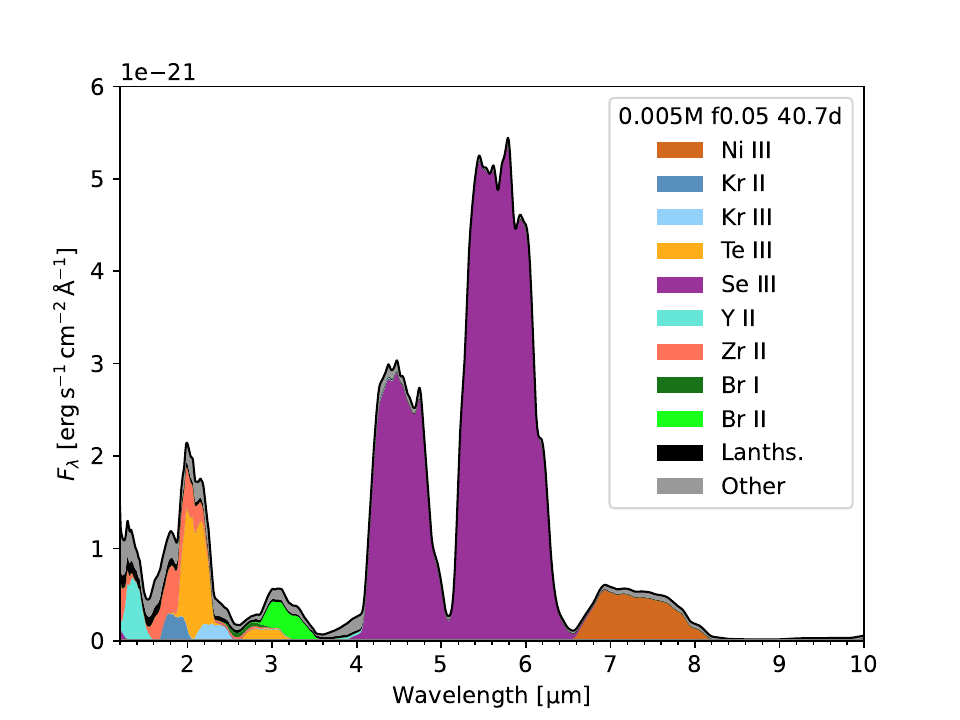}
    \includegraphics[trim={0.4cm 0.cm 0.4cm 0.3cm},width=0.47\linewidth]{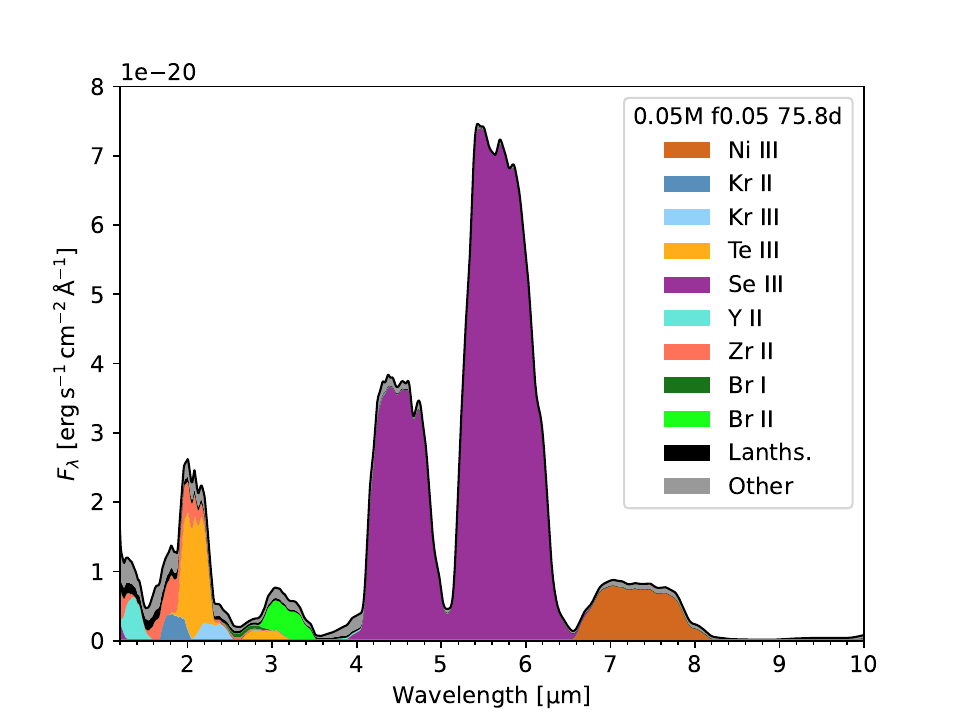}
    \includegraphics[trim={0.4cm 0.cm 0.4cm 0.3cm},width=0.47\linewidth]{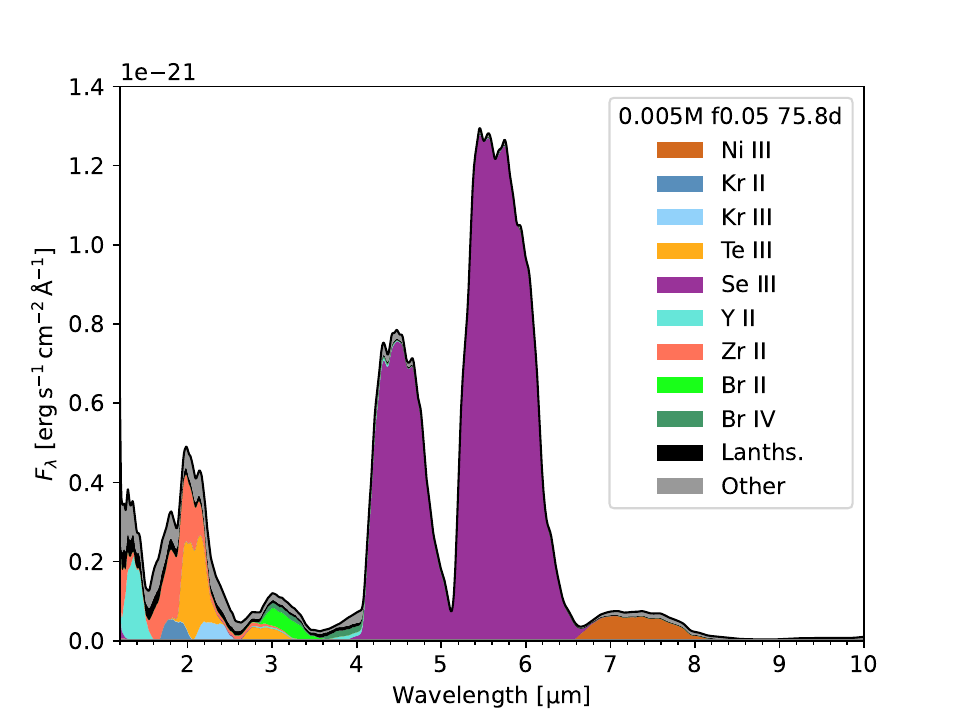}
    \caption{Spectra of the $f_{\rm{dyn}} = 0.05$ model at select epochs with $M_{\rm{ej}} = 0.05\, \Msol$ in the left-hand panels, and $M_{\rm{ej}} = 0.005\, \Msol$ in the right-hand panels.}
    \label{fig:f005spec_massvar}
\end{figure*}

The persistence of the Ni\,\textsc{iii} line at 7.3\,$\mu$m and its relative lack of blending with any other species makes it a potentially interesting observational target \citep[see also][]{Jerkstrand.etal:25}. As mentioned in Section \ref{subsec:hydro}, the $_{28}$Ni in our models is mostly in the form of stable isotopes, such that the overall abundance of Ni\,\textsc{iii} is negligibly affected by the decay of $^{56}$Ni. However, if this unstable isotope is produced in greater quantities, as may be the case for long-lived ($t \gtrsim 100\,$ms) NS remnants \citep[][]{Jacobi.etal:25}, then we expect the Ni\,\textsc{iii} line, and any emission from $_{28}$Ni in general, to decrease more significantly with time as it decays to $^{56}$Co. Further work is needed to fully establish how the 7.3\,$\mu$m feature evolves due to ejecta conditions as opposed to isotope decay, but a markedly different evolution as to the one found here, combined with the predicted lightcurve impact of $^{56}$Ni when it is the dominant isotope \citep[][]{Jacobi.etal:25}, could potentially establish the nature of the $_{28}$Ni produced in BNS mergers.

\subsubsection{Features from elements with known recombination rates}
\label{subsubsec:recomb_impact}

Of the diverse features described above, some arise from elements for which recombination rates have recently been calculated \citep[e.g.][]{Banerjee.etal:25,Singh.etal:25}. We discuss here the potential impact of these new rates on important features in our emergent spectra in comparison to our employed flat rate of $\alpha = 10^{-11}\, \rm{cm^3 \, s^{-1}}$.

We consider first the Te\,\textsc{iii} feature, which plays a central role in our emergent spectra. Dielectronic recombination rates are available for Te\,\textsc{iii} to Te\,\textsc{ii}, in the range of $\sim 3\times10^{-11} - 3\times 10^{-12}\, \rm{cm^3 s^{-1}}$ for $T = 1000 - 10000$K \citep[][]{Singh.etal:25}. As such, our fiducial value of $10^{-11}\, \rm{cm^3 s^{-1}}$ is found to be slightly higher than the calculated rate for our temperature solutions, implying that we should potentially have more Te\,\textsc{iii} than our current amount. However, without knowledge of rates for the other ions of $_{52}$Te, it is difficult to establish in more detail the impact of our assumed rate. 

Looking at the recombination rates for Se\,\textsc{i} - Se\,\textsc{iii} from \citet{Banerjee.etal:25}, we have that recombination to these species differs by factors of $\lesssim 2$, $\sim 2 - 10$, and $\sim 0.5 - 2$ respectively compared to our fiducial rate. It is therefore likely that our current model may slightly overproduce Se\,\textsc{i} and Se\,\textsc{iii}, and underproduce Se\,\textsc{ii}. Since Se\,\textsc{ii} does not have any IR lines from low-lying states, the impact is potentially that the overall emission of $_{34}$Se in our model is slightly decreased at all epochs. It is difficult to gauge however, to what extent this would affect the dominance of a single-peak Se\,\textsc{i} feature vs. a double-peaked Se\,\textsc{iii} for a given epoch, or how this feature would evolve in time. 

Both $_{39}$Y and $_{40}$Zr have some recent recombination rate data available \citep[][]{Banerjee.etal:25,Singh.etal:25}. For recombination to the neutral states of these elements, we have that Zr\,\textsc{ii} recombines with a rate of $\sim 10^{-10} - 10^{11} \, \rm{cm^3 s^{-1}}$ for $T = 10^3 - 10^4$K, implying that the impact of Zr\,\textsc{i} may be in fact be somewhat underestimated in our model. The values for recombination of Y\,\textsc{ii} are different depending on the source, but in the range of $\sim 10^{-11} - 10^{-10} \, \rm{cm^3 s^{-1}}$. In either case, the values are greater than our fiducial choice, and so we expect a greater Y\,\textsc{i} presence at the behest of Y\,\textsc{ii}. The exact degree of variation expected is difficult to gauge, though the general spectral shape is likely to remain mostly similar given the limited impact of $_{39}$Y in general. The greatest change in lanthanide-bearing models may be a reduced 1.4 $\mu$m feature from Y\,\textsc{ii} at later times ($t \gtrsim 30$ days), as shown in Figure \ref{fig:f005M001_NIRspec}, while more lanthanide-poor models may find a greater blending of Y\,\textsc{i} with Te\,\textsc{iii} at early times (see Section \ref{subsec:f001_spec}).

Dielectronic recombination rates for Ce\,\textsc{iii} to Ce\,\textsc{ii} are also available, with values of $\sim 10^{-10} - 10^{-9} \, \rm{cm^3 s^{-1}}$ for our temperature range \citep[][]{Singh.etal:25}. These values are substantially higher than our fiducial value, implying that we have too much Ce\,\textsc{iii} in terms of ion fraction, again within the uncertainty of not knowing the rates for the other ions of $_{58}$Ce. Assuming that we should instead have a much greater abundance of Ce\,\textsc{ii}, we predict an even greater spectral impact from this ion due to its enhanced amount of low-lying (e.g. $E \leq 10 000 \, \rm{cm^{-1}}$) states compared to Ce\,\textsc{iii}. While the latter has 20 levels under this limit, the former has 96, including 21 of opposite parity allowing E1 dipole transitions to occur. Therefore, we expect Ce\,\textsc{ii} to provide significantly more M1 emission, as well as more E1 emission and potentially absorption in the IR, if the transitions become optically thick as the fraction of this ion is significantly increased. 

Overall, we find that our fiducial recombination rate is a relatively good average value for the species and temperatures considered here. The biggest discrepancy is for the lanthanide $_{58}$Ce, specifically recombination from the doubly to singly ionized state, where our fiducial value significantly underestimates the calculated rates.  Generally, it is difficult to gauge the exact impact on the ionisation structure and the emergent spectra without knowing not only recombination rates for all ions of an element, but also the ionisation cross-sections for all relevant ionising processes, which for KNe are primarily non-thermal collisional ionisation and photoionisation.

\subsubsection{Impact of ejecta mass}
\label{subsubsec:mass_impact}

We now examine the impact of total ejecta mass on the emergent spectra of our standard model at 10, 40 and 75 days, shown in Fig. \ref{fig:f005spec_massvar}, with the heavier $M_{\rm{ej}} = 0.05\, \Msol$ variant in the left-hand panels, and the lighter $M_{\rm{ej}} = 0.005\, \Msol$ model in the right-hand panels. Looking first at the heavier mass model at 10 days in the top left panel, we see that the 1 - 2 $\mu$m part of the spectrum is particularly affected. There, we find that neutral species are emitting much more prominently, which follows from a more neutral ionisation structure as shown in the bottom panel of Fig. \ref{fig:thermo_evolution}, particularly in the innermost layer which has an electron fraction of 0.73 cf. 1.03 in the `standard' model. We see much more emission from Br\,\textsc{i} at 2.71 $\mu$m, and the 4 - 6.4 $\mu$m range is now dominated by a single peak arising from forbidden Se\,\textsc{i} emission. Some features are still present as in the `standard' model, such as the emission from Kr\,\textsc{ii} and Kr\,\textsc{iii}, the peak from Br\,\textsc{ii}, and the broad emission from Ni\,\textsc{iii}.  

Alongside the enhanced emission from neutral species, we see a greater impact from the key lanthanides Nd\,\textsc{ii} and Ce\,\textsc{iii}, to the point where the Te\,\textsc{iii} line at 2.1 $\mu$m is no longer dominating the emission at this wavelength, only contributing $\sim$28 per-cent of the flux, as shown in Fig. \ref{fig:TeIII_fracs}. We additionally find some enhanced Tb\,\textsc{i} emission below 2~$\mu$m, as well as some Eu\,\textsc{i} in the 5 -- 6~$\mu$m range, though neither species is calibrated and so the features are not wavelength accurate. Since the increased mass at the same epoch leads to greater densities, we find a non-negligible number of E1 lines below 2 $\mu$m to be optically thick in the innermost layer, while many lines are partially optically thick ($\tau \lesssim 0.8$) in the second layer, such that some line absorption is likely occurring. Alongside the greater density which enhances collisional deexcitation, the absorption from these E1 lines contributes to the suppression of forbidden emission from Kr\,\textsc{ii}, Kr \textsc{iii}, and Te\,\textsc{iii} at these wavelengths. 

Turning to the lighter model at 10 days in the top right panel, we see a relatively similar landscape as the `standard' model, with some key differences. The neutral species are still present, yet the singly ionized species like Zr\,\textsc{ii} and Y\,\textsc{ii} also play a minor role, while the lanthanide contribution is almost identical to that in the $M_{\rm{ej}} = 0.01\, \Msol$ model. Compared to the `standard' model, the Te\,\textsc{iii} 2.1 $\mu$m emission is much more significant at $\sim$65 per-cent of emitted flux (top panel of Fig. \ref{fig:TeIII_fracs}), due to reduced collisional deexcitation efficiency allowing more forbidden emission to occur, as well as a slightly greater amount of Te\,\textsc{iii} in the ionisation structure: 49 per-cent cf. 44 per-cent in the `standard' model. We also have the double peaked Se\,\textsc{iii} structure in the 4 - 6.4 $\mu$m range, as Se\,\textsc{i} has been mostly ionized out. We recover slightly more Ni\,\textsc{iii} emission at 7 $\mu$m, for the same reason as the boosted Te\,\textsc{iii} emission. In this sense, the 10 day spectrum of the low mass model somewhat resembles the spectrum of the `standard' model at 20 days, highlighting the impact of ejecta density on the emergent spectral shape. 

Moving forward in time to study the evolution of these models, we find that the general trends found in the `standard' model are reproduced. Neutral species are ionized out, lanthanide contributions decrease, while Se\,\textsc{iii} dominates the spectrum. A key difference however, is the time-scale on which these changes occur. A greater ejecta mass leads to higher densities at a given epoch, such that the time-scales relevant to important physical processes like recombination and ionisation remain short compared to the evolutionary time until later. Time-dependent effects which occur when those time-scales become significant compared to the evolutionary therefore also occur later. As such, we see that the spectral shape of the heavy model continues to evolve significantly from 40 to 75 days. Conversely, the low mass model has more minor changes from 40 to 75 days, the most noticeable of which is the slow decrease of Ni\,\textsc{iii} emission.

The overall spectral shape, however, remains essentially identical. This difference in evolution links back to the NLTE calculations of temperature and ionisation structure using the time-dependent equations; the heavy model stays closer to a steady-state evolution for a longer period of time. At late enough times however, the impact of mass is almost entirely limited to the brightness of the emergent spectrum, and looking at the bottom row of Fig. \ref{fig:f005spec_massvar}, we see that the emergent spectra are practically identical aside from the flux levels. Notably, in both cases we find the impact of lanthanides at these late times to be minimal on the emergent IR spectra in this model with $X_{\rm{La}} = 0.0027$.

\begin{figure}
    \centering
    \includegraphics[trim={0.4cm 0.cm 0.4cm 0.3cm},width = 0.5\textwidth]{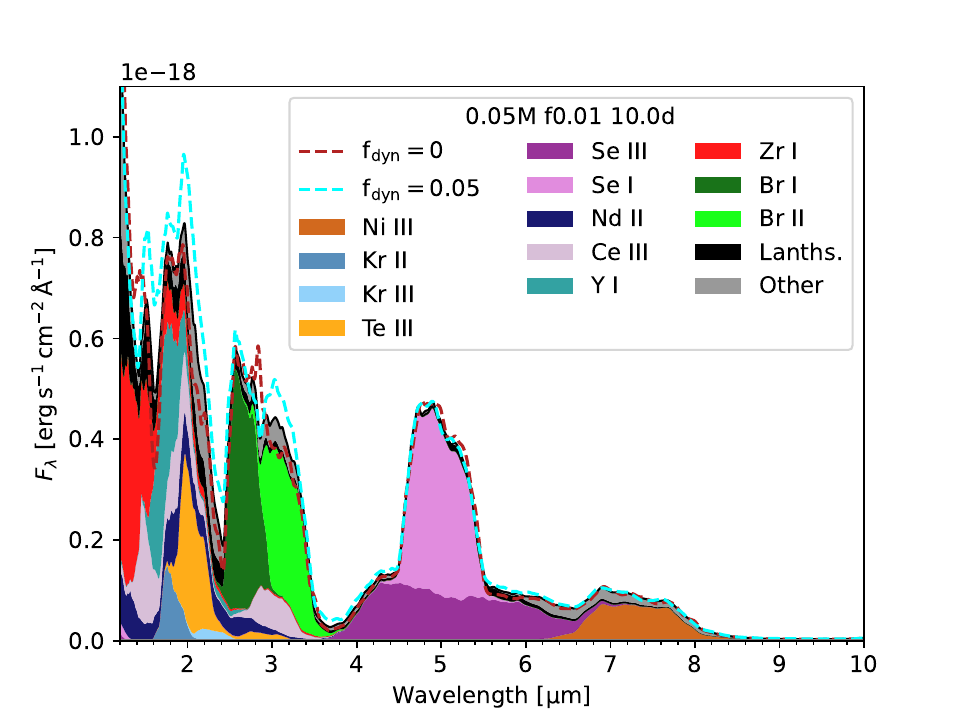} 
    \caption{Spectrum of the $f_{\rm{dyn}} = 0.01$ model at 10 days, with the $f_{\rm{dyn}} = 0,0.05$ models plotted in the dashed lines for comparison. All models have $M_{\rm{ej}} = 0.05 \, \Msol$.}
    \label{fig:f001_NIRspec}
\end{figure}

\subsection{Lanthanide-poor model}
\label{subsec:f001_spec}

In this section, we explore the $f_{\rm{dyn}} = 0.01$ model as a representative case for relatively lanthanide-poor ejecta with $X_{\rm{La}} = 0.001$, corresponding most closely to the physical case of a BNS merger with a long-lived remnant \citep[e.g.][]{Fujibayashi.etal:18,Kawaguchi.etal:21}. Despite being a lanthanide-poor case in our parameter space, we do note that this fraction remains within the range estimated for AT2017gfo. From the analysis of lanthanide trends in Section \ref{subsec:general_NIRspec}, we see that lanthanide impact is greatest at early times and for greater masses (higher densities). We therefore show the 10 day spectra of the $f_{\rm{dyn}} = 0.01$ model compared to the lanthanide-free $f_{\rm{dyn}} = 0$ model, both with $M_{\rm{ej}} = 0.05\, \Msol$, in Fig. \ref{fig:f001_NIRspec}, in order to gauge the impact of a relatively small lanthanide mass-fraction with respect to a completely lanthanide-free case. 

Looking at Fig. \ref{fig:f001_NIRspec}, we see that Ce\,\textsc{iii} and Nd\,\textsc{ii} still play a non-negligible role at 10~d, notably blending with Te\,\textsc{iii} at 2.1\,$\mu$m. The first peak species Zr\,\textsc{i} and Y\textsc{i} are also significant at the same wavelength, such that the Te\,\textsc{iii} flux contribution is only of $\sim$ 35 per-cent. Contrary to the $f_{\rm{dyn}} = 0.05$ model with the same ejecta mass however, the lanthanide E1 lines are not optically thick in this case, likely due to species' low abundances in the model composition. As we decrease the total ejecta mass, we find the same reduction of lanthanide impact as in the `standard' model, albeit to a greater extent, such that lanthanide emission for the $f_{\rm{dyn}} = 0.01$ model with $M_{\rm{ej}} \leq 0.01 \, \Msol$ is essentially negligible. 

Ignoring the shaded ion contributions, but instead comparing the total emergent spectrum to the $f_{\rm{dyn}} = 0$ model plotted in cyan in Fig. \ref{fig:f001_NIRspec}, we see that the $f_{\rm{dyn}} = 0.01$ model is almost identical, aside from minor differences around 1.4\,$\mu$m, such that the two models would likely not be distinguishable if observed. The $f_{\rm{dyn}} = 0.05$ model has some more differences from the lanthanide-free model, though it may still be difficult to distinguish it observationally. All three models are practically identical for $\lambda \gtrsim 3.5~\mu$m however, as expected from the lack of lanthanide emission found there in the `standard' model. 

As time goes on, we find that the impact of lanthanides continues to decrease, such that the $f_{\rm{dyn}} = 0.01$ model becomes identical to the $f_{\rm{dyn}} = 0$ case at $t \gtrsim 20$ days (spectra may be viewed in the supplementary material). These models suggest that spectrally distinguishing mildly and poor lanthanide-bearing ejecta ($X_{\rm{La}} \lesssim 0.0027$) at IR wavelengths in the NLTE regime may be observationally difficult. Since we generally find that lanthanides play a stronger role in denser ejecta, more centrally concentrated ejecta profiles may yield stronger signals at these epochs.

Further studies exploring different ejecta models would be required in order to establish more rigorous constraints on the minimal lanthanide mass-fraction necessary to produce a significant spectral IR impact in the nebular phase. Nevertheless, these results imply that the IR regime past $\lambda \sim 4\, \mu$m are likely to be dominated by first r-process peak species even for moderately lanthanide-bearing ejecta \citep[see][for an in depth analysis of lanthanide-free models]{Jerkstrand.etal:25}.

\subsection{Lanthanide-rich model}
\label{subsec:f05_spec}

\begin{figure}
    \centering
    \includegraphics[trim={0.4cm 0.cm 0.4cm 0.3cm},width=0.99\linewidth]{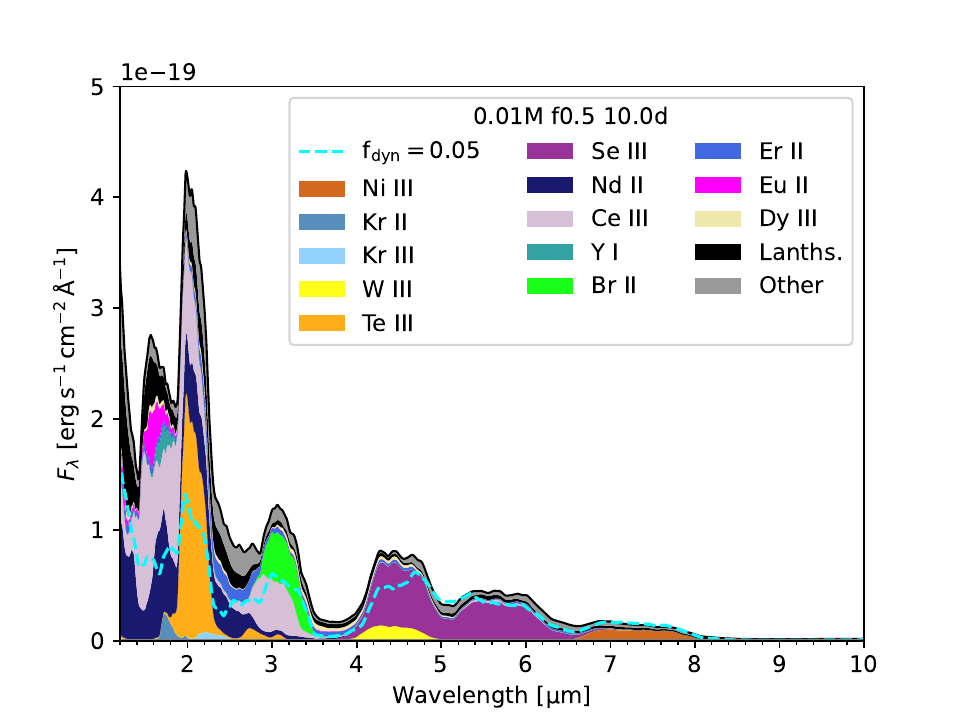} \\
    \includegraphics[trim={0.4cm 0.cm 0.4cm 0.3cm},width=0.99\linewidth]{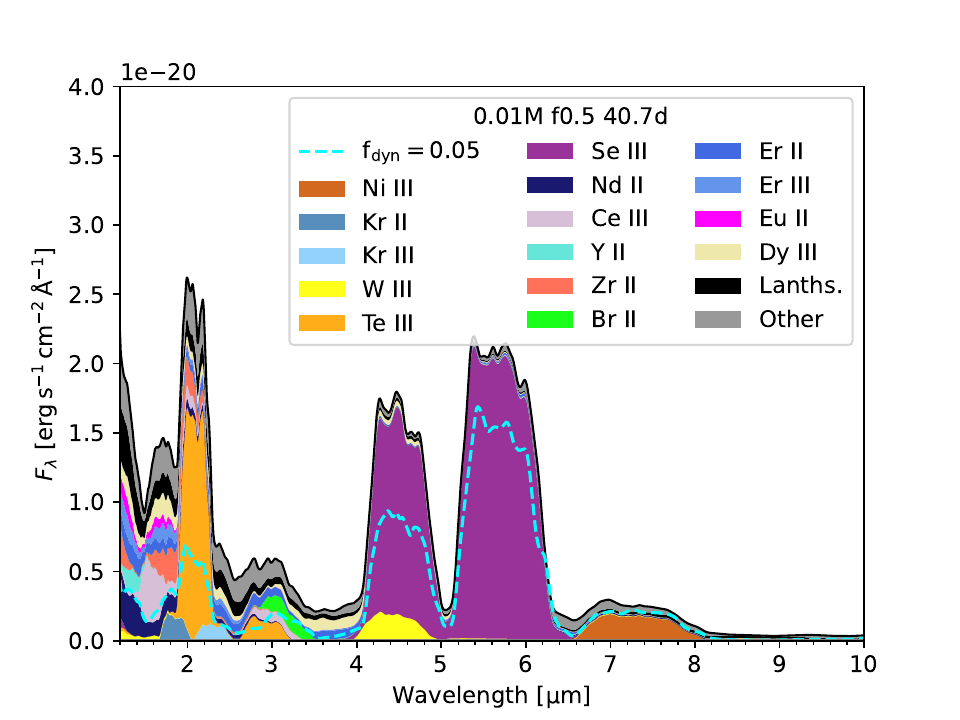} \\
    \includegraphics[trim={0.4cm 0.cm 0.4cm 0.3cm},width=0.99\linewidth]{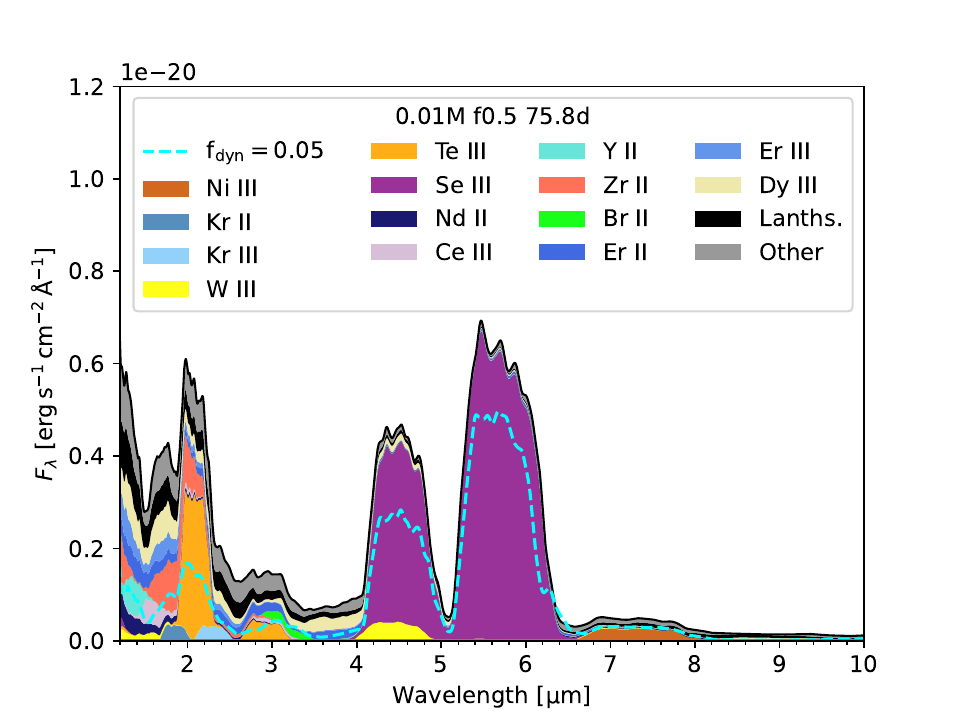}
    \caption{Spectra of the $f_{\rm{dyn}} = 0.5$ model with $M_{\rm{ej}} = 0.01\, \Msol$ at select epochs, compared to the $f_{\rm{dyn}} = 0.05$ model shown by the dashed cyan line.}
    \label{fig:f05M001_NIRspec}
\end{figure}

We now turn our attention to the opposite edge of our parameter space by examining the lanthanide-rich $f_{\rm{dyn}} = 0.5$ model, with $X_{\rm{La}} = 0.026$, approximately 10 times higher than that of the $f_{\rm{dyn}} = 0.05$ model. Physically, this model corresponds most closely to a BNS merger with a short lived remnant, or certain cases of BHNS mergers depending on the total ejecta mass \citep[e.g.][]{Hayashi.etal:22,Fujibayashi.etal:23,Kawaguchi.etal:23,Kawaguchi.etal:24}. We first examine the evolution of the $M_{\rm{ej}} = 0.01\, \Msol$ model in Fig. \ref{fig:f05M001_NIRspec} at select epochs. 

\begin{figure*}
    \centering
    \includegraphics[trim={0.4cm 0.cm 0.4cm 0.3cm},width=0.47\linewidth]{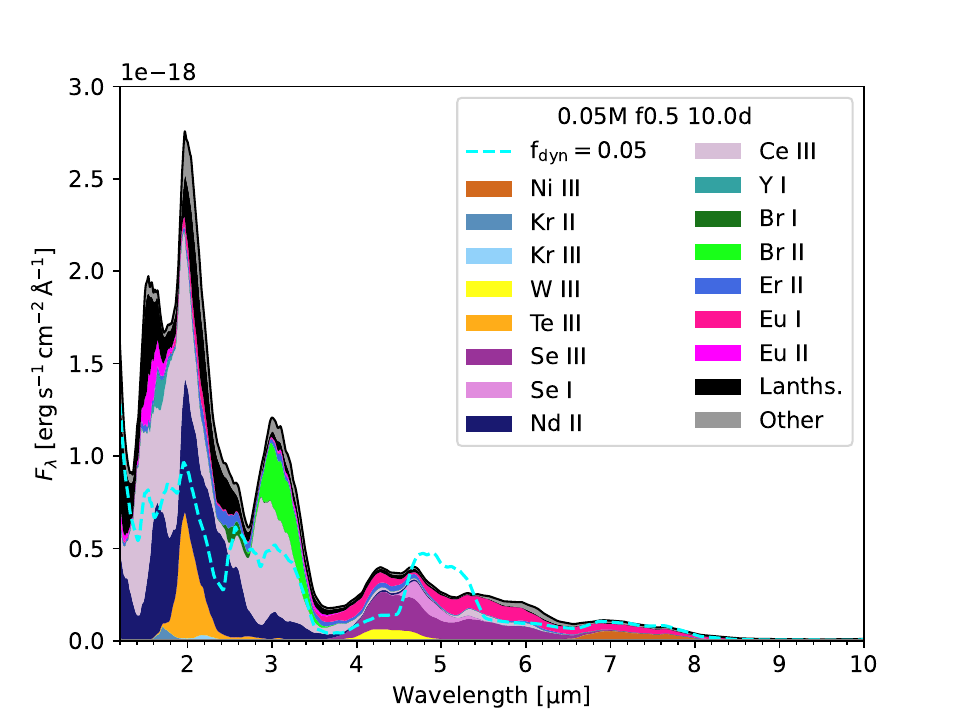}
    \includegraphics[trim={0.4cm 0.cm 0.4cm 0.3cm},width=0.47\linewidth]{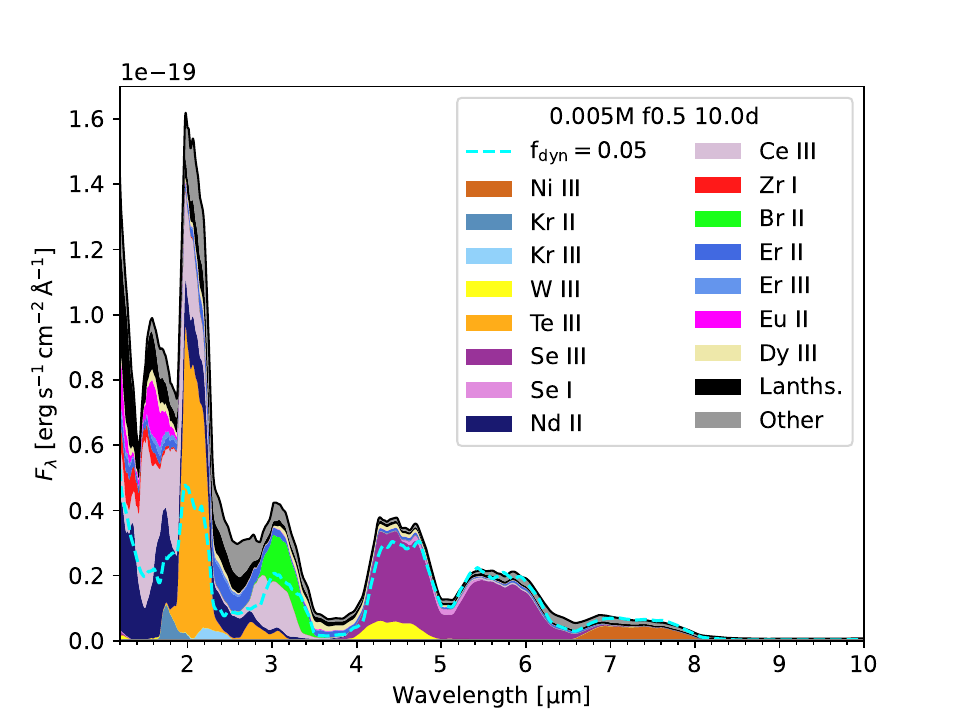} 
    \caption{The $f_{\rm{dyn}} = 0.5$ model at 10 days with $M_{\rm{ej}} = 0.05\, \Msol$ and $M_{\rm{ej}} = 0.005\, \Msol$ in the left and right panels respectively, compared to the $f_{\rm{dyn}} = 0.05$ model at the same masses shown by the dashed cyan line}.
    \label{fig:f05_massvar10d}
\end{figure*}

Considering first the top panel showing the model at 10 days, we immediately see that the presence of many lanthanides and heavier species drastically changes the spectrum compared to the `standard' model at the same epochs. Notably, we have significant, dominant lanthanide emission below 2 $\mu$m, such that other species known to emit there are suppressed, and we also do not recover Br\,\textsc{i} emission at 2.7 $\mu$m. The Te\,\textsc{iii} feature is heavily blended with Nd\,\textsc{ii} and Ce\,\textsc{iii}, only contributing $\sim$44 per-cent of the flux at 2.1 $\mu$m (Fig. \ref{fig:TeIII_fracs}). We also find more emission from other lanthanides, specifically Er\,\textsc{ii}, Er\,\textsc{iii} and Eu\,\textsc{ii}, though we note that these species are not calibrated, and so their features are not wavelength accurate. The lanthanide lines are not found to be optically thick however, with only a few of the strongest E1 transitions reaching $\tau_{\rm{sob}} \lesssim 1$ in the innermost layer. We also find the greatest W\,\textsc{iii} contribution in this model, particularly at 4.5 $\mu$m, but this remains subdominant with respect to Se\,\textsc{iii}. 

Both $_{34}$Se and $_{74}$W have been proposed as origins for the 4.5 $\mu$m emission seen in the \textit{Spitzer} photometry of AT2017gfo \citep[e.g.][]{Hotokezaka.etal:22}, as well as for the emission in the 29 day spectrum of AT2023vfi \citep[][]{Levan.etal:24,Gillanders.Smartt:25}. From the models in this study, however, we find that Se\,\textsc{iii} consistently dominates in every spectrum, with W\,\textsc{iii} maximally contributing $\sim 1/6$ of the flux in the $f_{\rm{dyn}} = 0.5$ model. Considering that all lines are firmly emitting in the NLTE regime at 40 days \citep[see][]{Jerkstrand.etal:25}, the line luminosities scale as $L_{\rm{line}} \propto M_{\rm{ion}} \Upsilon$, where $M_{\rm{ion}}$ is the mass of the ion in the ejecta, and $\Upsilon$ is the collision strength between the lower and upper level. For the solution of the innermost zone of our $f_{\rm{dyn}} = 0.5$, $M_{\rm{ej}} = 0.05\, \Msol$ model at 40 days, we have $\Upsilon_{\rm{Se_{III}}} = 4.5$, and $\Upsilon_{\rm{W_{III}}} = 1.45,3.65$ for the lines with $\lambda = 4.43,4.54\, \mu$m respectively. For equal ion masses, both lines are therefore expected to be comparable, if slightly brighter for W\,\textsc{iii}. However, our model has a much lower abundance of $_{74}$W than $_{34}$Se: $\sim$ 1 per-cent cf. $\sim$ 18 per-cent. Taking into account the fraction of each species in their doubly ionized state, of 39 per-cent and 18 per-cent for $_{74}$W and $_{34}$Se respectively, we find that the ion mass of Se\,\textsc{iii} is greater by an order of magnitude, thus leading to its dominant emission. 

Calculations of the mass of W\,\textsc{iii} required to produce the 4.5~$\mu$m emission in both AT2017gfo and AT2023vfi have been carried out based on detailed R-matrix atomic data (the same used in this study) in \citet{McCann.etal:25}. For AT2017gfo, a mass of $1.65\times 10^{-4}\,\Msol$ is estimated, similar to that in our $f_{\rm{dyn}} = 0.5$ model with $M_{\rm{ej}} = 0.05\Msol$ which has $M_{\rm{WIII}} = 2.12\times 10^{-4} \, \Msol$. Though the exact electron density and temperature solutions found in our model are slightly different than those used to infer the requisite W\,\textsc{iii} mass ($n_e \approx 5\times 10^{-5}$ c.f. $10^6 \, \rm{cm^{-3}}$ and $T \approx 4700$K c.f. $3500$K), similar line luminosities are expected. However, the stronger emission from Se\,\textsc{iii} at this same wavelength leads to this model not being consistent with the measured photometry (see Section \ref{sec:LCs}).

The merger scenario underlying the ejecta models in this work is one that yields a long-lived remnant \citep[see][]{Fujibayashi.etal:20a,Kawaguchi.etal:21}, and so it is not surprising that we find a dominance of first-peak elements in terms of composition. Even taking solely the dynamical ejecta component of our hydrodynamical model, we find that the abundance of $_{34}$Se is $\sim$ 5 per-cent, greater than that of $_{74}$W at $\sim$ 1.7 per-cent. Third peak elements may be more abundant in cases where significant amounts of low $\rm{Y_e} \lesssim 0.2$ ejecta are produced, as may be the case in NSBH mergers or BNS mergers with short-lived remnants ($t_{\rm{BH}} \lesssim 10$~ms) \citep[e.g.][]{Fujibayashi.etal:23}. While the lifetime of the remnant for AT2017gfo is still currently debated, it is generally believed to be longer than $\sim 10$ms \citep[e.g.][but see also \citet{Sneppen.etal:25}]{Just.etal:23,Kawaguchi.etal:23,Curtis.etal:24,Sippens.etal:25,Vieira.etal:26}. Therefore, we suggest that Se\,\textsc{iii} is the more plausible candidate for the emission at 4.5 $\mu$m in AT2017gfo, though it is possible that W\,\textsc{iii} contributes in a sub-dominant, yet non-negligible fashion (see Section \ref{subsubsec:AT2023vfi} for a discussion of this point related to AT2023vfi).

Moving forwards in time, the spectral landscape of the $f_{\rm{dyn}} = 0.5$ model remains complex, particularly for wavelengths $\lambda \lesssim 4\, \mu$m. There, we find a large amount of blending between diverse species, where we have made the effort to individually highlight some of the more important lanthanide contributions. While the Te\,\textsc{iii} emission at 2.1 $\mu$m continues to yield a significant peak in the overall spectrum, it is blended with various species and only provides $\sim$ 50 per-cent of the total flux there at 75 days. The landscape bluewards of this feature is a complex blend of many species, with lanthanides largely dominating. However, some first r-process peak species like Zr\,\textsc{ii} and Y\,\textsc{ii} do appear with their contributions of blended, low-lying M1 emission lines. Past 4 $\mu$m, we recover the usual double peaked Se\,\textsc{iii} and Ni\,\textsc{iii} features, with some smaller contribution from W\,\textsc{iii} at 4.5 $\mu$m. The enduring impact of lanthanides even at late times in this model is in contrast to the evolution of the standard model, and some key features are significantly different, such as the absence of important Br\,\textsc{ii} emission at late times. As such, this lanthanide-case should be observationally differentiable from the more mild composition of the $f_{\rm{dyn}} = 0.05$ model with $X_{\rm{La}} = 0.0027$.

Since it was shown in Section \ref{subsubsec:mass_impact} that ejecta mass plays an important role in the impact of lanthanides on the emergent spectra, we further examine the $f_{\rm{dyn}} = 0.5$ model with $M_{\rm{ej}} = 0.05, 0.005\, \Msol$ at 10 days in Fig. \ref{fig:f05_massvar10d}. Similarly to the other models, we find stronger lanthanide signatures for the heavier model. While the $M_{\rm{ej}} = 0.005\, \Msol$ case does retain significant spectral lanthanide presence, we continue to recover typical features such as the Te\textsc{iii} 2.1 $\mu$m peak, as well as the redder Se\,\textsc{iii} and Ni\,\textsc{iii} features. Conversely, the $M_{\rm{ej}} = 0.5\, \Msol$ case shows a significantly altered spectral shape, where the aforementioned MIR features are smaller and blended with significant emission from Eu\,\textsc{i}. While Eu\,\textsc{i} has not been calibrated, we find that the emission in the $\lambda \sim 3.7$ -- 6~$\mu$m range arises from several low-lying E1 transitions should be located at somewhat redder wavelengths of $\lambda \sim 4.2$ -- 9~$\mu$m \citep[e.g.][]{NIST_ASD}. Additionally, the landscape below $\lambda \sim 4~\mu$m is heavily dominated by Nd\,\textsc{ii} and Ce\,\textsc{iii}, to the extent that Te\,\textsc{iii} produces only 18 per-cent of the flux at 2.1~$\mu$m. 

\begin{figure}
    \centering
    \includegraphics[trim={0.4cm 0.cm 0.4cm 0.3cm},width=1\linewidth]{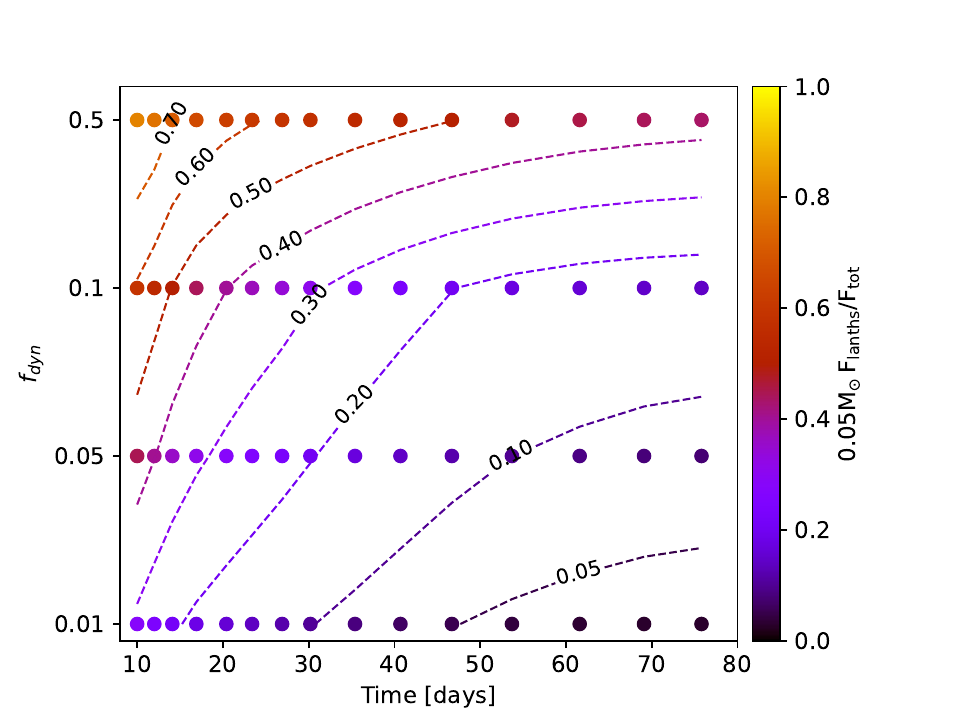}\\
    \includegraphics[trim={0.4cm 0.cm 0.4cm 0.3cm},width=1\linewidth]{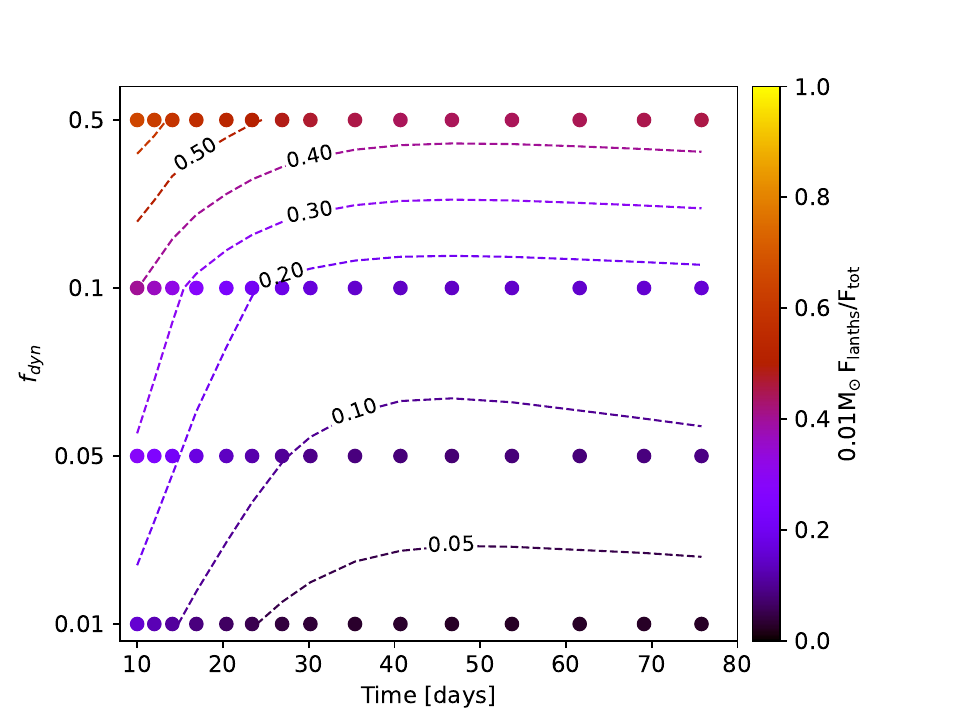} \\
    \includegraphics[trim={0.4cm 0.cm 0.4cm 0.3cm},width=1\linewidth]{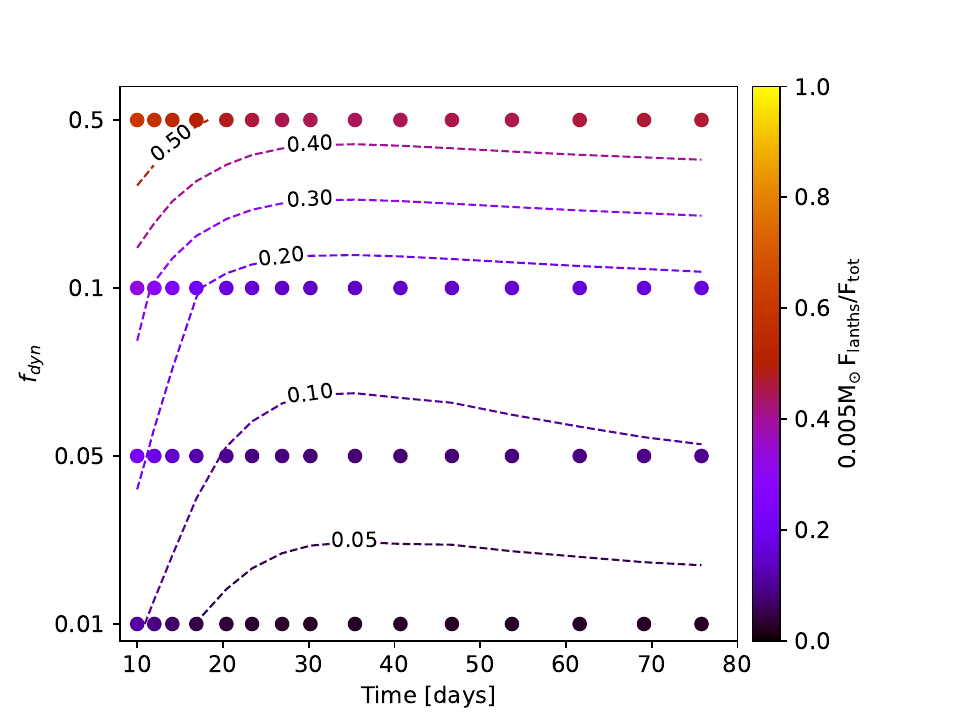}
    \caption{Evolution of the lanthanide flux fraction in the range of 1.2 -- 3.6$\mu$m for every model, with the heaviest ejecta mass in the top panel, and the lightest ejecta mass in the bottom panel respectively. The dashed lines are contours for the flux fraction of lanthanides, showing the trends across parameter space. Note that the same colour scheme is applied to both the points and contours, and that the $f_{\rm{dyn}} = 0$ model is absent as no lanthanides are included in the composition. }
    \label{fig:lanth_fracs}
\end{figure}

Generally, we find that the impact of lanthanides on the emergent spectra is greater at earlier times and for larger ejecta masses, i.e. greater densities, with the largest impact usually occurring below $\sim 3.6\, \mu$m. Correspondingly, we find that the majority of models show few lanthanide features at late times e.g. $t \gtrsim 40$ days, or at wavelengths redder than 3.6 $\mu$m. In order to better quantify how important the lanthanide species are in models across all epochs and masses, we show their contribution to the total emerging flux in the 1.2 -- 3.6 $\mu$m range in Fig. \ref{fig:lanth_fracs}. There, we clearly see how the impact of lanthanides in terms of emission is distributed over our parameter space, and how their significance tends to decrease with time. Extrapolating these trends to higher ejecta masses or greater densities for more centrally concentrated ejecta profiles, as well as greater values of $f_{\rm{dyn}}$ that may correspond to different merger scenarios, we forecast that the impact of lanthanides will be greater, and their effects longer lasting in the evolution of the KN. 

\subsection{Emission in the 10 to 30 micron range}
\label{subsec:MIRspec}

We now consider the emission further into the MIR, in the 10 - 30 $\mu$m range, corresponding to reddest spectral range of \textit{JWST}. As we find that the vast majority of the models evolve similarly in this range, we focus on our $f_{\rm{dyn}} = 0.05,0.5$ models with $M_{\rm{ej}} = 0.01\,\Msol$. The complete set of 10 -- 30 $\mu$m spectra for all models may be viewed in the supplementary material. 

In Fig. \ref{fig:MIRspec}, we present the `standard' model at 10, 40 and 75 days in the left-hand panels, and the lanthanide-rich $f_{\rm{dyn}} = 0.5$ model in the right-hand panels. Considering first the top-left panel at 10 days, we find a notable emission peak at 11 $\mu$m, consisting mostly of Ni\,\textsc{iii} with contributions from Ru\,\textsc{ii} and Ru\,\textsc{iii}, as well as trace amounts of lanthanides and other diverse species. The lanthanide-rich model has more emission from Ce\,\textsc{iii}, leading to a slightly brighter overall feature.

At $\sim$ 14 $\mu$m, we have a blended emission feature consisting of mainly Br\,\textsc{ii}, with some contribution from Zr\,\textsc{i} and Zr\,\textsc{iii}. This is followed redwards by another heavily blended feature consisting mainly of Ru\,\textsc{ii}, Ru\,\textsc{iii}, Zr\,\textsc{ii} and Se\,\textsc{i}. Finally, we have some Fe\,\textsc{iii} and Zr\,\textsc{ii} emission at 22.5 $\mu$m, with the emission dropping off past of this final feature. The $f_{\rm{dyn}} = 0.5$ model has a similar SED shape, the main difference being more emission between 12.5 -- 20.0\,$\mu$m due to contributions from the lanthanides Ce\,\textsc{iii} and Er\,\textsc{ii}.

As time progresses, we see that the main Ni\,\textsc{iii} feature at 11 $\mu$m remains significant, with some additional contribution from Mo\,\textsc{iii} coming in at later times. The Br\,\textsc{ii} emission at 14 $\mu$m is reduced, but the overall emission feature increases in relative importance due to more significant emission from Zr\,\textsc{iii} as well as some Mo\,\textsc{iv}. There is also relatively more Zr\,\textsc{ii} emission redwards of this, as well as additional Mo\,\textsc{iii} contributions, such that there is significant blending of species at late times. In the $f_{\rm{dyn}} = 0.05$ model, we therefore find that for the entire timespan covered here, the lanthanides play an extremely minor, if not negligible role in the emergent spectra at 10 -- 30 $\mu$m. 

\begin{figure*}
    \centering
    \includegraphics[trim={0.4cm 0.cm 0.4cm 0.3cm},width=0.49\linewidth]{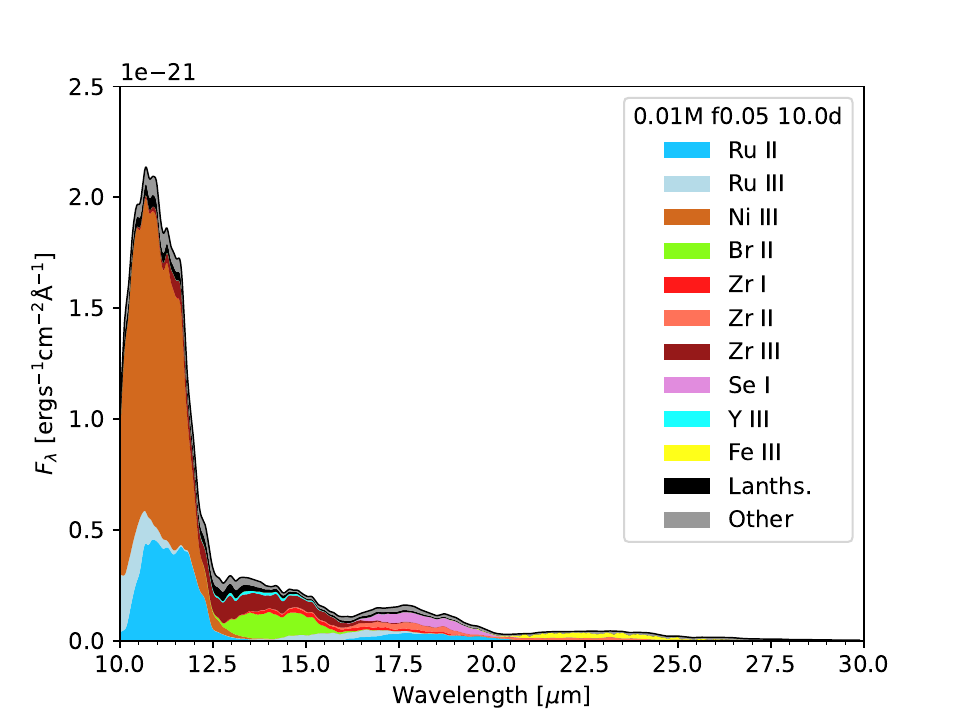}
    \includegraphics[trim={0.4cm 0.cm 0.4cm 0.3cm},width=0.49\linewidth]{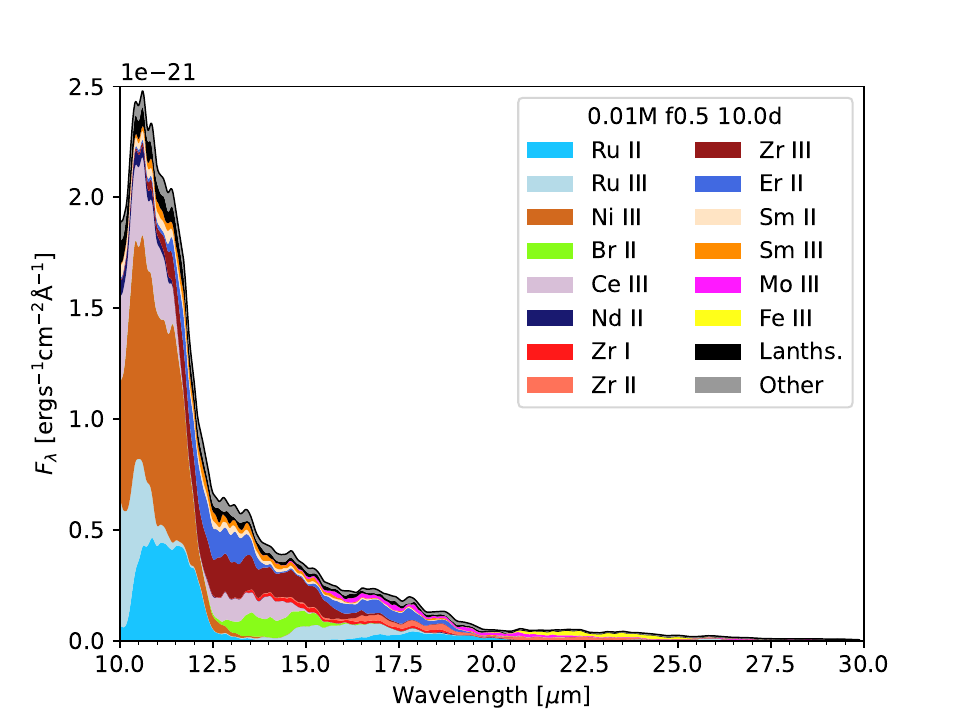} \\
    \includegraphics[trim={0.4cm 0.cm 0.4cm 0.3cm},width=0.49\linewidth]{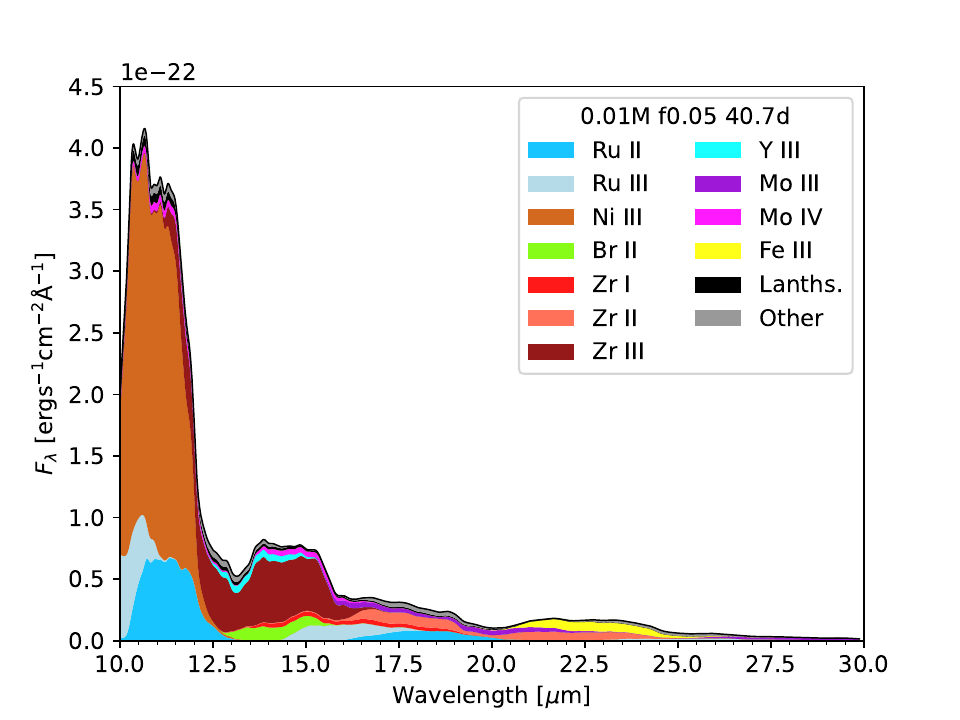}
    \includegraphics[trim={0.4cm 0.cm 0.4cm 0.3cm},width=0.49\linewidth]{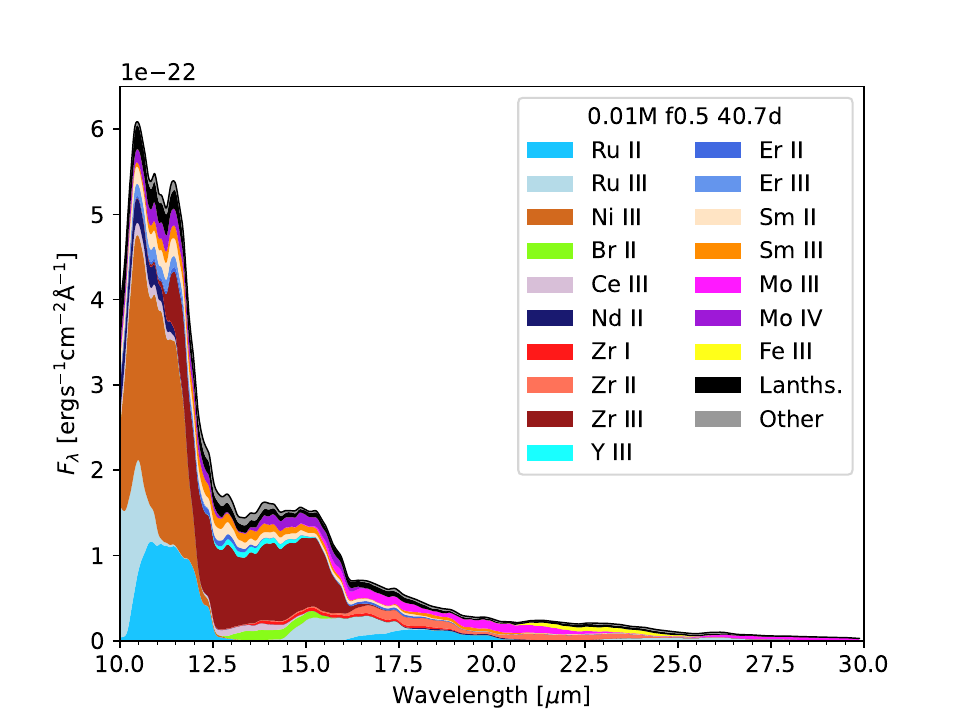} \\
    \includegraphics[trim={0.4cm 0.cm 0.4cm 0.3cm},width=0.49\linewidth]{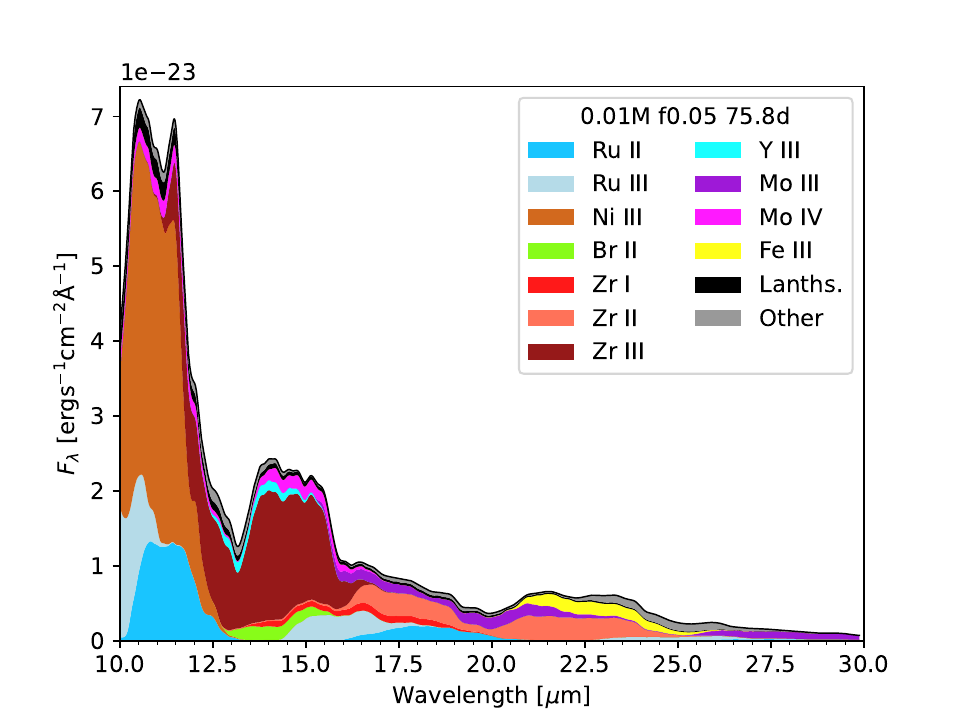}
    \includegraphics[trim={0.4cm 0.cm 0.4cm 0.3cm},width=0.49\linewidth]{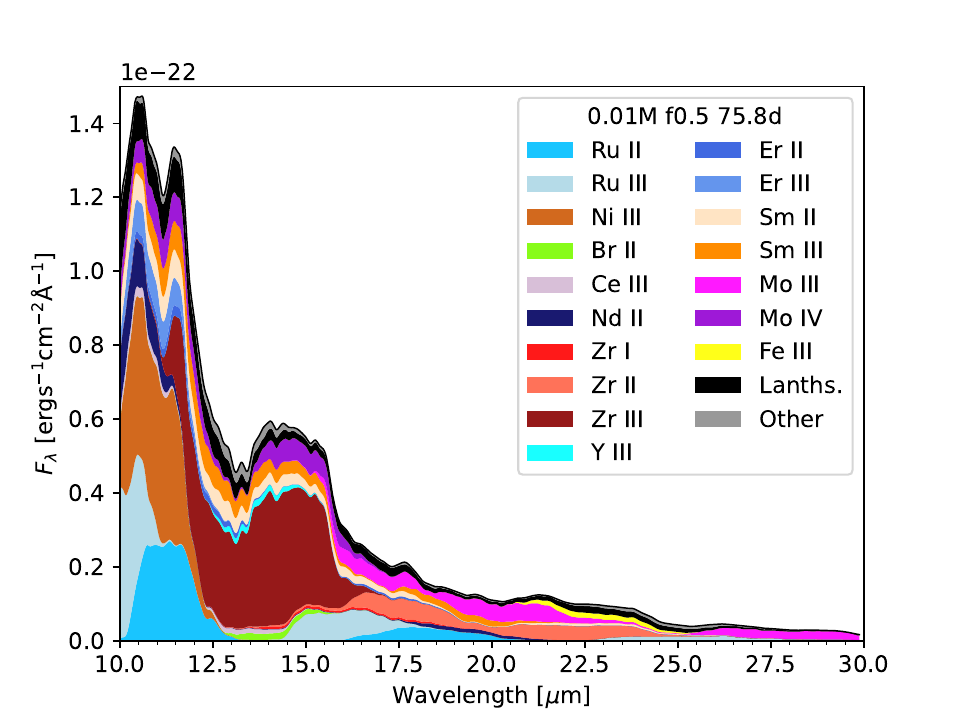}
    \caption{Spectra of the $f_{\rm{dyn}} = 0.05,0.5$ models with $M_{\rm{ej}} = 0.01\, \Msol$ in the 10 to 30\,$\mu$m range in the left and right-hand panels respectively.}
    \label{fig:MIRspec}
\end{figure*}

Looking at the evolution of the lanthanide-rich model in the left-hand panels of Fig. \ref{fig:MIRspec}, we see that the general trends from the `standard' model are reproduced, with some lanthanide species provide additional, highly blended emission. At 75.4~d, this model has approximately two times more flux at these wavelengths than the $f_{\rm{dyn}} = 0.05$ model, but we caution that this is an effect of greater energy deposition (see Fig. \ref{fig:dep_compare}) and not due to lanthanide emission.

Generally, we find that lanthanide impact in the 10 -- 30\,$\mu$m range to be very limited, such that spectrally distinguishing lanthanide-rich or poor ejecta from observations at these wavelengths seems unfeasible. Furthermore, even in our most lanthanide-rich model, the impact of these species does not change the overall SED shape, generally yielding broad, blended features. Quantitatively, we find the biggest difference at late times between 10 -- 15~$\mu$m for our $f_{\rm{dyn}} = 0.01,0.5$ models. Taking the broadband \textit{JWST} filters F1500W - F1130W colour, such that we gauge the relative flux of the peaks at 11 and 14.5 $\mu$m and eliminate the impact of different bolometric luminosity between the models, we find for the $f_{\rm{dyn}} = 0.5$ and 0.01 models at 75 days colours of 0.47 and 0.79 magnitudes respectively (see also Section \ref{sec:LCs} and the supplementary material).

It may be initially surprising to find so little lanthanide emission in the MIR, given that these elements are known for their extremely dense level structures, which would presumably yield many MIR transitions. Considering first the species that do dominate past 10~$\mu$m, we see that they share a similar level structure: a ground state that is fine-split such that the energy level spacing is $\lesssim 1000~\rm{cm^{-1}}$, followed by a first excited state that is much higher energy, potentially to the point that it is not accessible by thermal excitations (collisional or PE). This implies that the level population of these ions is concentrated within the fine-split ground state, therefore leading to significant MIR emission. Ni\,\textsc{iii} and Ru\,\textsc{iii} are excellent examples of this atomic structure, where the first excited level above the fine-split ground state has an energy of $E > 10 000~\rm{cm^{-1}}$.

Considering now the low-lying energy levels of various lanthanide species, we find a broader variety in structure. Some simply have no low-lying states, such as Eu\,\textsc{iii}, which has a singlet ground state followed by an excited state at 28000~$\rm{cm^{-1}}$. Some follow the `ideal' structure of the non-lanthanide emitters, such as Dy\,\textsc{ii}, but are likely prevented from significantly emitting due to their lower abundances, e.g. $X_{\rm{Dy}} = 0.0047$ c.f. $X_{\rm{Ru}} = 0.0319$ in the $f_{\rm{dyn}} = 0.5$ model, as line luminosity scales with ion mass in the nebular regime. Instead, species with a dense level structure at small energies, which yields many MIR lines emit more significantly. Nd\,\textsc{ii} is a good example, having 10 even parity states of similar angular momentum \textit{J} and $E \leq 5000~\rm{cm^{-1}}$, therefore allowing many M1 transitions, with energy spacing between consecutive states of $< 1000~\rm{cm^{-1}}$. A dense level structure is therefore necessary but not sufficient for lanthanide emission in the MIR; these levels must additionally be of low enough energy so as to be accessible by inefficient, nebular phase excitation processes.

As for the 1 - 10 $\mu$m spectra, it is important to verify the accuracy of emergent features, particularly at such red wavelengths where the slightest deviation of energy levels in theoretical atomic data from their `true' values can lead to large changes in wavelength. Many of the main emitting species, notably Ni\,\textsc{iii}, Zr\,\textsc{i} -- \textsc{iii}, Fe\,\textsc{iii}, Br\,\textsc{ii}, Y\,\textsc{iii}, Se\,\textsc{i} and Sm\,\textsc{iii}, have had at least their lowest lying energy levels calibrated to NIST, or are taken from highly precise data sets (see Table \ref{tab:atomic_data}), and so their features in the 10 -- 30 $\mu$m range are wavelength accurate. Therefore, the predicted emission from these species is taken to be robust within the intrinsic uncertainty of the NLTE modelling. 

The $_{42}$Mo and $_{44}$Ru features are generally found to be slightly too red, with the exact values shown in Table \ref{tab:MIR_transitions}. The \textsc{fac} atomic data for the lanthanide Sm\,\textsc{ii} is jumbled relative to that available in NIST, such that the location of Sm\,\textsc{ii} features in our models are not generally accurate. However, given the energy levels found in NIST, we predict potential Sm\,\textsc{ii} features from low-lying transitions at 12.735, 14.725 and 19.227 $\mu$m, with emission on the same order of magnitude as that found in our models. Unfortunately, we find that our computer \textsc{fac} data for $_{68}$Er is inaccurate. When comparing to NIST, we find that Er\,\textsc{iii} should not be emitting anything in the 10 -- 30 micron range \citep[see also][]{Gaigalas.etal:20}, while Er\,\textsc{ii} has only one potential transition emitting at 22.705 $\mu$m. As such, the $_{68}$Er emission in our models should be broadly disregarded, though we note that the impact of this element is relatively minor across the entire 1.2 -- 30 $\mu$m range studied here. 

\begin{table}
    \centering
    \setlength\tabcolsep{0.4cm}
    \begin{tabular}{cccc}
    \hline \hline
    Species & Transition & $\lambda_{\textsc{fac}}$ [$\mu$m] & $\lambda_{\rm{NIST}}$ [$\mu$m] \\
    \hline
                        & $3 \rightarrow 2$ & 28.17 & 23.46 \\
    Mo\,\textsc{iii}     & $4 \rightarrow 3$ & 20.58 & 18.00 \\
                        & $5 \rightarrow 4$ & 16.95 & 15.42 \\
    \hline
                        & $2 \rightarrow 1$ & 15.00 & 12.85 \\
    Mo\,\textsc{iv}      & $3 \rightarrow 2$ & 11.56 & 10.19 \\
                        & $4 \rightarrow 3$ & 9.86 & 9.05 \\
    \hline
    Ru\,\textsc{ii}      & $3 \rightarrow 2$ & 11.35 & 10.30 \\
                        & $4 \rightarrow 3$ & 17.80 & 16.39 \\
    \hline
    Ru\,\textsc{iii}     & $2 \rightarrow 1$ & 9.97 & 8.63 \\
                        & $3 \rightarrow 2$ & 16.01 & 14.98 \\
    \hline \hline
    \end{tabular}
    \caption{Key non-lanthanide transitions in the 10 -- 30 $\mu$m range that were not calibrated in the \textsc{fac} data. The levels in the `Transition' column refer to those in the ground multiplet as found in the NIST database.}
    \label{tab:MIR_transitions}
\end{table}

Despite the inaccurate predictions for several emitting species in our models, the most significantly contributing species remain those that have been calibrated or have accurate atomic data to begin with. Notably, we always find significant emission at 11 $\mu$m, driven mostly by Ni\,\textsc{iii} emission. Combined with the Ni\,\textsc{iii} feature predicted at 7 $\mu$m, simultaneous detection of both of these emission peaks by \textit{JWST}'s MIRI would provide strong observational evidence for the presence of $_{28}$Ni in KN ejecta, despite the potential blending of the 11 $\mu$m feature with other species. 

A key nuance of this latter point is that our model composition does not include every element that may be created in KN ejecta, such that it is possible that certain key species have been omitted, e.g. Ge\,\textsc{i} which is not included here, is found to have a strong emission line at 11.7 $\mu$m \citep[][]{Jerkstrand.etal:25} that would blend heavily with the Ni\,\textsc{iii} predicted here. More generally, it is also worth noting that the emission in this range remains rather faint, and the scale of flux in this wavelength range is approximately two orders of magnitude smaller than in the 1 - 10 $\mu$m range for a given model (see e.g. Figs. \ref{fig:f005M001_NIRspec} and \ref{fig:MIRspec}). For instance, in the `standard' model at 10~d, we find that 78.4 per-cent of the total flux is emitted in the 1 -- 10~$\mu$m range, compared to only 1.3 per-cent in the 10 -- 30~$\mu$m range, while the rest is bluer. The model becomes slightly redder in time, with 81.9 per-cent and 3.3 per-cent flux emitted in the 1 -- 10 and 10 -- 30~$\mu$m ranges respectively at 30.2~d, but the overall luminosity drops to a greater extent, such that it remains to be seen whether this spectral range could be well observed even for nearby events at any epoch.

\subsection{Comparison to existing spectral observations}
\label{subsec:specobs_compare}

Thus far, spectral observations of only two (potential) KNe exist, AT2017gfo \citep[e.g.][]{Abbott.etal:17,Smartt.etal:17,Pian.etal:17} and AT2023vfi \citep[][]{Levan.etal:24,Gillanders.Smartt:25}, the former of which had spectra up to 10 days, and the latter spectra at 29 and 61 days. Since these epochs are covered by our models, it is interesting to examine whether any features in our synthetic spectra match up to those found in previous analyses. We note that in the following analyses, we do not fit the model spectra to observations, and flux scaling of the model spectra is for visualisation purposes, in order to compare SED and feature shapes.

\subsubsection{AT2017gfo}
\label{subsec:AT2017gfo}

\begin{figure}
    \centering
    \includegraphics[trim={0.4cm 0.cm 0.4cm 0.3cm},width=1\linewidth]{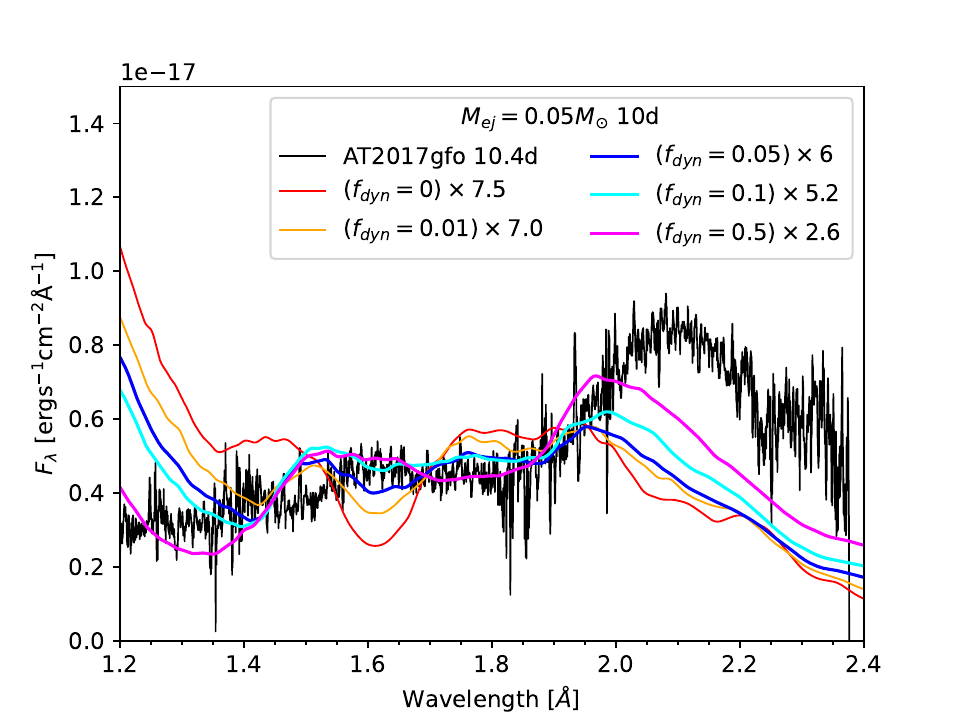}\\
    \caption{Comparison of model spectra with $M_{\rm{ej}} = 0.05\, M_{\odot}$ to AT2017gfo at 10 days. Note that the model spectra have had their flux scaled by the values indicated in the legends.}
    \label{fig:AT2017gfo}
\end{figure}

Starting with AT2017gfo, we compare our model spectra with $M_{\rm{ej}} = 0.05\, \Msol$, closest to the inferred ejecta mass of 0.04 -- 0.08 $\Msol$ \citep[e.g.][]{Perego.etal:17,Smartt.etal:17,Waxman.etal:18}, at 10 days to the observed spectrum\footnote{ENGRAVE data release: \url{http://www.engrave-eso.org/AT2017gfo-Data-Release/}} at 10.4 days in Fig. \ref{fig:AT2017gfo}. We find that these models are generally lacking flux in the 1.2 -- 2.4 $\mu$m range, as even the most energetic ($f_{\rm{dyn}}=0.5$) requires some scaling up in order to match the observation. However, checking the models' bolometric luminosities compared to the value inferred for AT2017gfo at 10.4 days of $log_{10}L_{\rm{bol}} = 39.939\pm0.316 $ \citep[e.g.][]{Smartt.etal:17}, we find a range of $log_{10}L_{\rm{bol}} = $ 39.694 -- 39.961. This implies that we are not inherently lacking energy, but rather that a significant portion of the flux lies elsewhere, particularly at $\lambda \lesssim 1.2\,\mu$m in our models. Previous studies found significant blue emission in the SED in the range of 5 -- 20 days when using homogeneous composition, 1D NLTE simulations \citep[][]{Pognan.etal:23}, due to decreasing optical depths with time, and relatively high temperatures producing blue photons that are readily able to escape. It is also likely that none of our models correspond particularly well to AT2017gfo in terms of composition, ejecta mass, energy deposition etc.

Focussing now on the shape of our models, we that they do not fully reproduce the 10\,d observation. For the specific part of the spectrum between $1.2 \leq \lambda \leq 2.4\, \mu$m, the $f_{\rm{dyn}} = 0.5$ model is the closest (albeit too faint), decently reproducing the plateau at 1.6 -- 1.8 $\mu$m, but still lacking flux past $\sim$ 2.0 $\mu$m. As we decrease the contribution from the dynamical ejecta, we find that our SED shapes deviate even further,  with a notable trough at 1.6 $\mu$m appearing in lanthanide-poor models, which is not found in the observed data. These models additionally require greater flux scaling, following from their smaller energy deposition (see Fig. \ref{fig:dep_compare}).

The 2.1~$\mu$m emission in the $f_{\rm{dyn}}= 0.5$ model is shown in the left panel of Fig. \ref{fig:f05_massvar10d}, where we see that it is a blend of Te\,\textsc{iii}, as well as the lanthanides Nd\,\textsc{ii} and Ce\,\textsc{iii}, while the emission at 1.6 $\mu$m is a blend of many lanthanides, of which again Nd\,\textsc{ii} and Ce\,\textsc{iii} dominate. The 2.1 $\mu$m feature which is seen from 7.5 to 10.4 days in AT2017gfo has been attributed to forbidden emission from Te\,\textsc{iii} \citep[][]{Hotokezaka.etal:23,Gillanders.etal:24}. However, we find that the extent to which it dominates the flux emission in at 2.1 $\mu$m is quite broad, from 18 -- 44 per-cent at 10 days depending on model (see Fig. \ref{fig:TeIII_fracs}). We also find emission from diverse other species at this epoch across all compositions, such that the resultant feature is always blended. This is particularly significant in the heavier mass models, where the emitting species are found to be lanthanides and/or first r-process peak elements, depending on the value of $f_{\rm{dyn}}$.

The sub-dominant contribution of Te\,\textsc{iii} at 10 days is due to three main reasons: the suppression of forbidden emission due to higher density allowing for more efficient collisional deexcitation, higher density allowing more efficient recombination favouring neutral and singly ionized species in the ionisation structure, and the stronger emission from other species by E1 transitions, particularly in the lanthanide rich models, that overwhelm the Te\,\textsc{iii} emission. We expect these effects to be more significant at earlier times, and therefore relevant to the 7.4 -- 9.4 day epochs of AT2017gfo, as well as in ejecta profiles with more centrally concentrated mass. 

We also note that the peak emission from Te\,\textsc{iii} in our model at 10~d is blueshifted slightly, such that the resultant peak, including the lanthanide contributions, is closer to $\lambda \sim 2.0\, \mu$m, as is seen clearly in Fig. \ref{fig:AT2017gfo}. This shift corresponds to a velocity of $\sim 0.05$c, i.e. our inner boundary. Should the observed feature arise from a single emission line with $\lambda_0 = 2.10\, \mu$m, a lack of blueshifted centroid combined with relatively large Doppler broadening of $v \sim 0.05$c implies not only that the ejecta is optically thin more inwards than 0.05c \citep[e.g. 0.02c][]{Jerkstrand.etal:25}, but also that the distribution of Te\,\textsc{iii} in velocity space still reaches higher velocities in significant amounts. 

Another possibility is that the observed feature is actually a blend of two or more emission lines, with fits to the 10\,d spectrum suggesting two Gaussian profiles with $\lambda_0 = 2.05, 2.14 \mu$m respectively \citep[][]{Gillanders.etal:24}. From our models, we find that other species emitting at these wavelengths are typically Kr\,\textsc{iii} and diverse lanthanide species, though only the former arises from a single emission line which could yield a Gaussian profile. The relative strength of the Kr\,\textsc{iii} line is weak in our models, though this may not be the case with more up to date atomic data, while Se\,\textsc{iv} may additionally emit significantly at these wavelengths in more energetic ejecta \citep[][]{Jerkstrand.etal:25}. 

The plateau at 1.6 -- 1.8 $\mu$m appears to be well reproduced within our models as long as sufficient amounts of lanthanides are present, minimally $X_{\rm{La}} \gtrsim 0.0027$. The missing 2.3 $\mu$m emission may be indicative that we have omitted an important species in our selected composition, though we note that previous analyses of AT2017gfo have not considered this bump to be a feature \citep[e.g.][]{Gillanders.etal:24}. 

Given the above analysis, we support the suggestion that Te\,\textsc{iii} may be emitting significantly at 2.1 $\mu$m in the 10 day spectrum of AT2017gfo. However, we do not find that it is the sole species emitting there, and also do not systematically find that it dominates the emitted flux at 10 days. We find that other species provide significant flux, particularly Kr\,\textsc{ii}, Kr\,\textsc{iii}, Zr\,\textsc{ii} and/or lanthanide species such as Nd\,\textsc{ii} and Ce\,\textsc{iii}. Generally, finding the `perfect' model for AT2017gfo may require better constraints on the overall composition of the ejecta, with recent works making efforts towards this goal \citep[e.g.][]{Vieira.etal:23a,Vieira.etal:24,Vieira.etal:26}, as well as usage of more reliable atomic data for all relevant processes.

\subsubsection{AT2023vfi}
\label{subsubsec:AT2023vfi}

For AT2023vfi, we consider the 29 day spectrum\footnote{Reduction from \cite{Gillanders.Smartt:25}}, which was dominated by the GRB afterglow below $\sim$ 1.9 $\mu$m, and was otherwise well fit by a Blackbody-like continuum with $T_{\rm{BB}} \sim 660$~K and $v_{\rm{phot}} \sim 0.08$c, and three emission lines overlain \citep[][]{Levan.etal:24,Gillanders.Smartt:25}. The emission at 2.1 $\mu$m was best fit by the combination of two blended Gaussians, centred at $\lambda_0 = 2.02\, \mu$m and $\lambda_0 = 2.19\,\mu$m, with full-width half-max (FWHM) velocities of $v_{\rm{FWHM}} \approx 0.06,0.11$c respectively. The emission at $\lambda_0 =$ 4.4~$\mu$m was well fit by a single Gaussian equally with $v_{\rm{FWHM}} = 0.11$c.

It is immediately apparent from Section \ref{sec:spectra} that our models do not produce a Blackbody-like continuum between 2 -- 5~$\mu$m. In order to make a more relevant comparison of our models to AT2023vfi, we subtract the Blackbody and GRB afterglow continua components from the observed spectra, using the fit parameters from \citet{Gillanders.Smartt:25}; in this sense we compare only the line emission. We note that this fit assumes a simple, single temperature Blackbody function, while the underlying reality may be multi-temperature with relativistic effects due to the fast expansion of ejecta \citep[e.g.][]{Sneppen.etal:23,Sadeh:25}, but we use the single temperature fit for simplicity. The nature of the Blackbody continuum in AT2023vfi is unclear and we therefore consider two cases.

Case A: we assume that the Blackbody emission does not arise from a real photosphere, and we remain agnostic as to the nature of the process and component generating this continuum. We then compare our spectra directly to the continuum-subtracted observation.

Case B: we assume that the thermal continuum arises from a real photosphere moving at a velocity of 0.08c. Since our models' inner boundary is located at 0.05c, we scale down our spectra's flux proportionally to the amount of mass that is faster than 0.08c, i.e. that would be above the photosphere. By doing so, we make an assumption that the overall power is directly proportional to mass, and that the SED and line shapes will not drastically change. Additionally, we assume that photons from the photosphere will not impact the conditions in the outer ejecta layers, which follows from our models being optically thin at the relevant wavelengths of $\lambda \geq 2\,\mu$m.

\begin{figure*}
    \centering
    \includegraphics[trim={0.4cm 0.cm 0.4cm 0.3cm},width=0.49\linewidth]{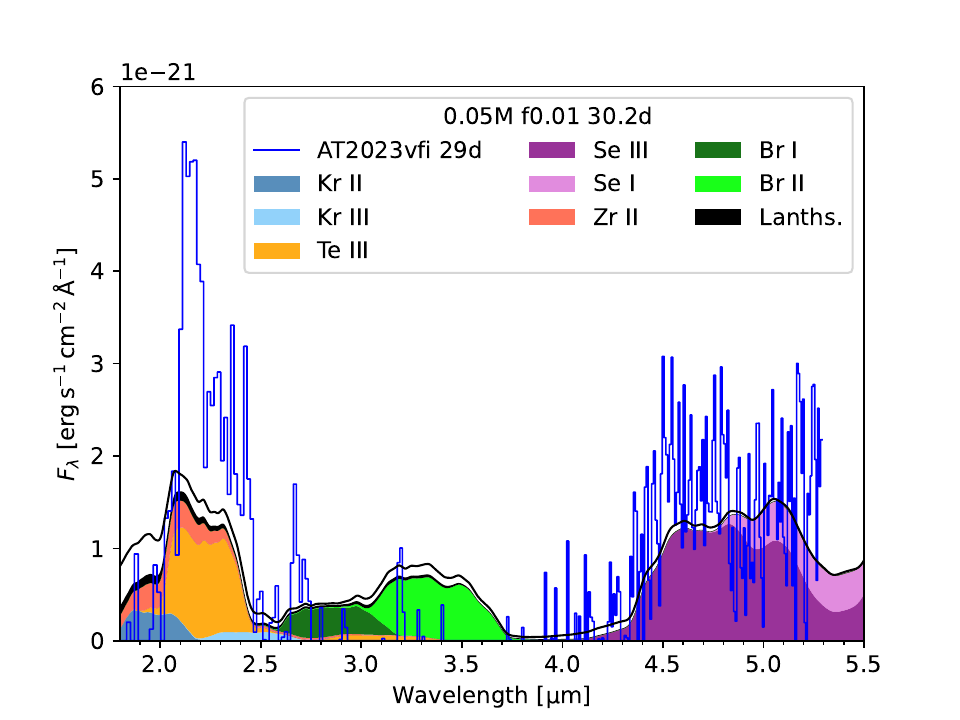}
    \includegraphics[trim={0.4cm 0.cm 0.4cm 0.3cm},width=0.49\linewidth]{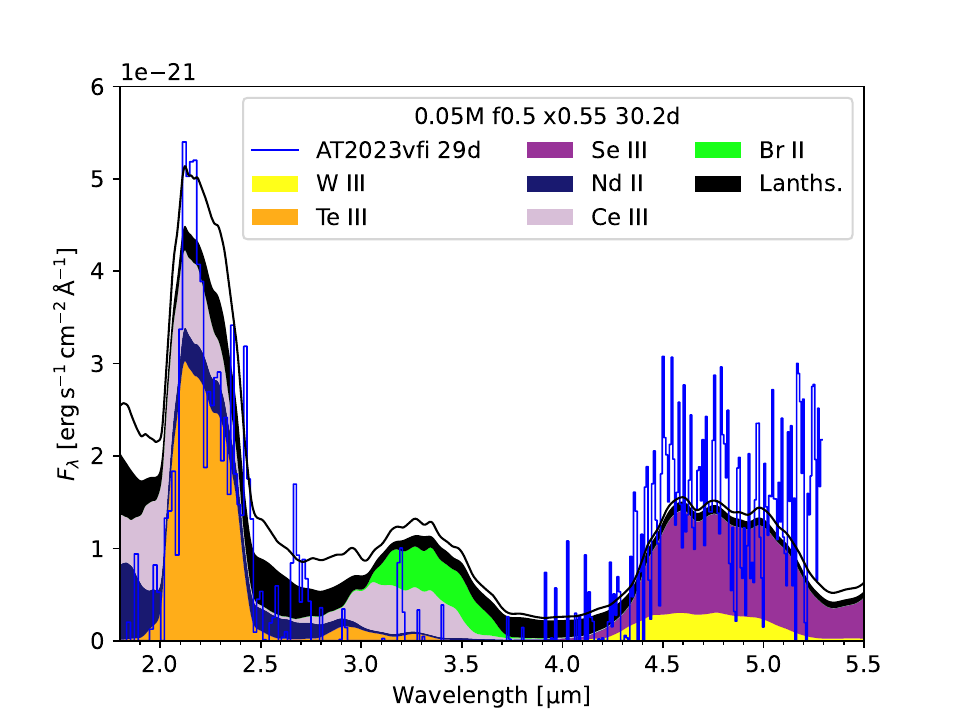}
    \caption{The models most closely matching the continuum-subtracted 29~d spectrum of AT2023vfi for case A (left panel) and case B (right panel) respectively. The model spectra have been shifted to $z = 0.065$, and the $f_{\rm{dyn}} = 0.5$ model has been scaled down correspondingly to the mass that is above the photosphere at $v_{\rm{phot}} = 0.08$c (see text, case B).}
    \label{fig:AT2023vfi_compare}
\end{figure*}

We show the results of our model comparison to the 29~d spectrum of AT2023vfi in Figure \ref{fig:AT2023vfi_compare}, where we have moved our models to $z = 0.065$ for direct comparison to the observed spectrum. For case A, we find that our lanthanide-poor $f_{\rm{dyn}} = 0.01$ model most closely matches the observation, while for case B, we find that our lanthanide-rich $f_{\rm{dyn}} = 0.5$ model is better.

From Figure \ref{fig:AT2023vfi_compare}, it is apparent that we do not fully reproduce the observed line emission. In the case of the lanthanide-poor model, the emission line at $\lambda = 4.6~\mu$m is relatively well reproduced by our models' Se\,\textsc{i} and Se\,\textsc{iii} emission. However, emission at $\sim$ 2.2~$\mu$m driven mainly by Te\,\textsc{iii} emission is too faint. In the case of the lanthanide-rich model, the 4.6~$\mu$m emission is once again relatively well reproduced, this time by a blend of Se\,\textsc{iii} and W\,\textsc{iii}, and we also find significant emission at 2.2~$\mu$m, though somewhat more than the observation and without reproducing the observed line shape. Instead we find blending of the Te\,\textsc{iii} line with a blanket of Nd\,\textsc{ii} and Ce\,\textsc{iii} lines which yield a structure resembling a single emission line. While no model reproduces the observed feature at 2.2~$\mu$m, the prominence of Te\,\textsc{iii} across all models suggests that it likely plays a significant role, consistent with previous analyses \citep[][]{Levan.etal:24,Gillanders.Smartt:25}.

Across all models, we find excess emission between 2.5 -- 4.0~$\mu$m, driven primarily by Br\,\textsc{i}, Br\,\textsc{ii}, and in the lanthanide-rich case additionally Ce\,\textsc{iii}. An absence of $_{35}$Br emission suggests that negligible amounts of it were synthesised in the ejecta of AT2023vfi, or alternatively, that it is ionised beyond the neutral and singly ionised stages. Br\,\textsc{iii} does not have any low-lying IR transitions, but Br\,\textsc{iv} does at $\lambda_0 = 3.1, 4.0\,\mu$m, which would again produce emission lines where observations do not show any, while it is unlikely that higher ionisation states would exist in significant abundances at these epochs. Therefore, it is more likely that little to no $_{35}$Br is present in the ejecta producing the other observed emission lines. Since $_{35}$Br is a first-peak element adjacent to $_{34}$Se in the periodic table, these are expected to be synthesised in similar conditions, therefore an interpretation of the above analysis is that $_{34}$Se, and other first-peak elements, are likewise sparse in the ejecta at this time. This then implies that the emission feature at 4.6 ~$\mu$m may be driven by W\,\textsc{iii}.

From \citet{McCann.etal:25}, a mass of between $\sim 5 \times 10^{-4}$ -- $10^{-3}\,\Msol$ of W\,\textsc{iii} is found to be necessary in order to produce the entire line luminosity of the 4.6\,$\mu$m feature in the 29~d spectrum of AT2023vfi, depending on ejecta electron density and temperature, while the $f_{\rm{dyn}} = 0.5$ model here only has $M_{\rm{WIII}} = 1.9 \times 10^{-4} \, \Msol$. As discussed in Section \ref{subsec:f05_spec}, having a total ejecta composition with relatively large masses of $_{74}$W and simultaneously little to no first peak elements requires significantly low $\rm{Y_e} \lesssim 0.2$ conditions. These are not met in the NR and hydrodynamical simulations used as input for the ejecta models here, which model a system with a long-lived NS remnant \citep[see figures 8 and 14 in][DD2-135M model]{Fujibayashi.etal:20b}, and we always find a greater mass-fraction of $_{34}$Se than $_{74}$W, even taking solely the dynamical component of the ejecta.

Based on the assumption that W\,\textsc{iii} is entirely responsible for the emission at 4.6~$\mu$m at 29~d, past works have suggested that AT2023vfi may be more compatible with a short-lived remnant merger scenario \citep[][]{McCann.etal:25}. Such a scenario could additionally favour the creation of 2nd to 3rd peak elements over 1st peak elements, in order to avoid excess $_{35}$Br emission as described above. However, low $\rm{Y_e}$ ejecta would additionally synthesise significant quantities of lanthanides, potentially leading to excess emission from Ce\,\textsc{iii} at $\sim 3.3~\mu$m as shown in Figure \ref{fig:AT2023vfi_compare}. Additionally, the inferred mass of AT2023vfi is on the order of 0.06$\,\Msol$ \citep[][]{Levan.etal:24}, in contrast to simulations predicting masses closer to $\sim 0.01\,\Msol$ \citep[e.g.][]{Fujibayashi.etal:23,Kawaguchi.etal:23}. NSBH mergers may be able to produce massive ejecta with significant low $\rm{Y_e}$ conditions \citep[see e.g.][for a review]{Kyutoku.etal:21}, though it remains to be seen if these are able reproduce the observed lightcurves \citep[see e.g. figure 13 of][]{Kawaguchi.etal:24}, while late time spectral models do not yet exist in literature.

Ultimately, it may be difficult to ever conclusively establish the merger scenario of AT2023vfi due to the absence of gravitational wave data precluding measurements of merger component masses, but further detailed modelling of the electromagnetic emission may allow constraints to be placed on ejecta parameters and thus favour certain scenarios over others. Based on the limited results from the models presented here, we suggest that BNS mergers with long-lived remnants are difficult to reconcile with the observed spectra of AT2023vfi.

Regardless of merger scenario and remnant lifetime, the origin of the Blackbody continuum observed in AT2023vfi is unknown, potentially attributed to lanthanide opacity \citep[][]{Levan.etal:24}, while classical graphite, silicate and metallic iron dust is ruled out \citep[][]{Arunchalam.etal:25}. We do not model dust here, but examine the possibility of creating an optically thick `photosphere' from r-process line opacity alone. This would require significant opacity in the IR, whereas our models are typically optically thin at these wavelengths at 30 days. Some actinides have been previously suggested to provide greater expansion opacity than lanthanides \citep[][]{Flors.etal:23}, though this does not seem to extend to redder wavelengths past 2~$\mu$m \citep[see e.g. fig. 5 of][]{Deprince.etal:25}, such that we do not expect the inclusion of actinides in the models to produce a continuum, given that our models have approximately $X_\mathrm{Ac} \sim 0.1 X_{\rm{La}}$ (see Fig. \ref{fig:dep_compare}). This also means that we do not necessarily require high actinide mass-fractions in the ejecta in order to have large opacities, as the lanthanides may prove sufficient. 

\begin{figure}
    \centering
    \includegraphics[trim={0.4cm 0.cm 0.4cm 0.3cm},width=1\linewidth]{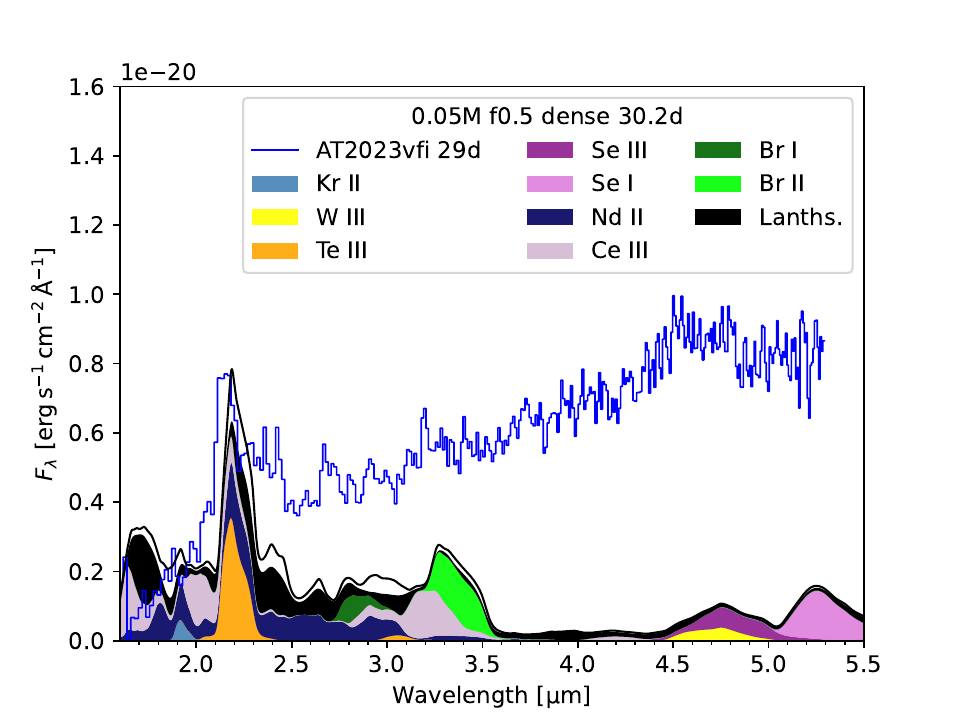}\\
    \caption{The modified $f_{\rm{dyn}} = 0.5$, $M_{\rm{ej}} = 0.05\,\Msol$ model with $v_{\rm{in}} = 0.02$c (see text) at 30\,d, compared to AT2023vfi at 29\,d. The model spectrum has been redshifted to $z = 0.065$ for comparison to the observation, and scaled such that the flux at 2.1$\mu$m approximately matches, for visualisation purposes.}
    \label{fig:f05_dense}
\end{figure}

We consider our most lanthanide rich model ($X_{\rm{La}} = 0.026$) with $M_{\rm{ej}} = 0.05\,\Msol$, which has the highest opacity in the IR, yet is nevertheless found to be optically thin past 1~$\mu$m aside from a few specific lines. In the 2.5 -- 5 $\mu$m range, we find only one optically thick Ce\,\textsc{iii} line at 3 $\mu$m (a low-lying E1 transition to the ground state), while the next closest are on the order of $\tau_{\rm{sob}} \lesssim 0.05$. Since for a given transition we have $\tau_{\rm{sob}} \propto n_l \propto M_{\rm{ion}}/V$, where $n_l$ is the number density of the lower level, proportional to the total ion mass $M_{\rm{ion}}$ within the ejecta divided by volume $V$, we would require a lanthanide abundance at least 20 times higher than the current composition in order for these lines to become optically thick, assuming all other conditions (e.g. temperature, ionisation structure, excitation structure, ejecta profile etc.) are held constant. Alternatively, a greater total ejecta mass by a factor of 20 for the same density profile and composition, or a compressed profile in order to reduce the volume by a factor of 20, would yield the same effect. 

In reality, any of the above changes would also impact energy deposition, temperature, ionisation and excitation structure solutions, and therefore the line optical depths themselves, making it difficult to calculate the exact amount of lanthanides needed to produce an optically thick continuum in NLTE conditions. Additionally, requiring a continuum that produces thermal emission at a low temperature of $\rm{T} \sim 660$~K may be difficult, as Planck mean opacities (note that this assumes LTE conditions) for the lanthanides are found to drop drastically below $\sim 2000$~K \citep[e.g.][]{Kasen.etal:13,Tanaka.etal:20,Deprince.etal:25}.

\begin{figure*}
    \centering
    \includegraphics[trim={0.4cm 0.cm 0.4cm 0.3cm},width=0.47\linewidth]{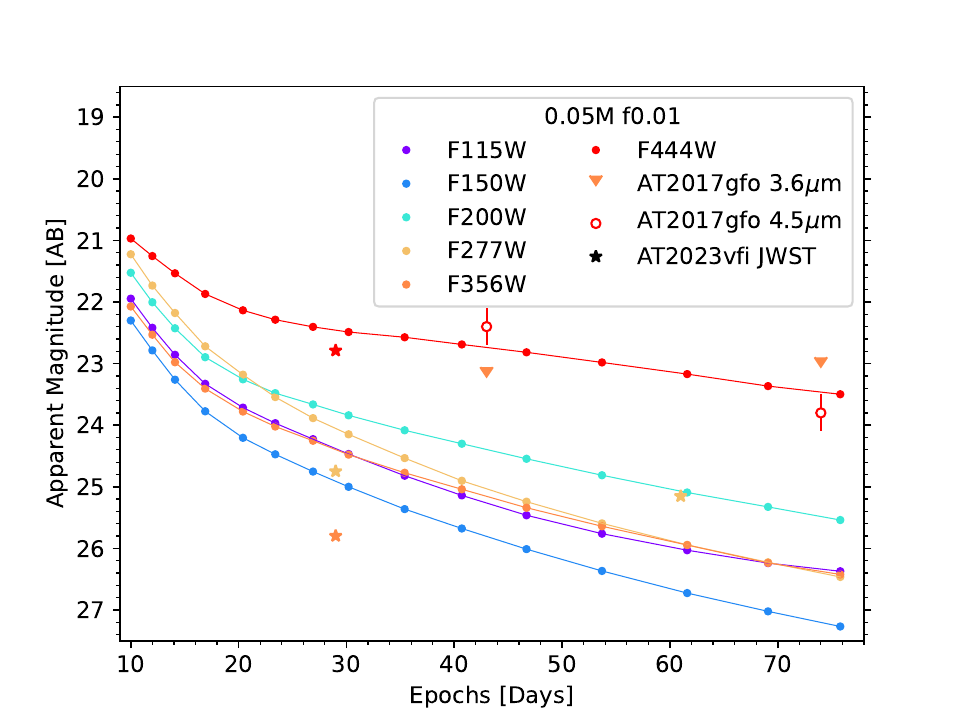}
    \includegraphics[trim={0.4cm 0.cm 0.4cm 0.3cm},width=0.47\linewidth]{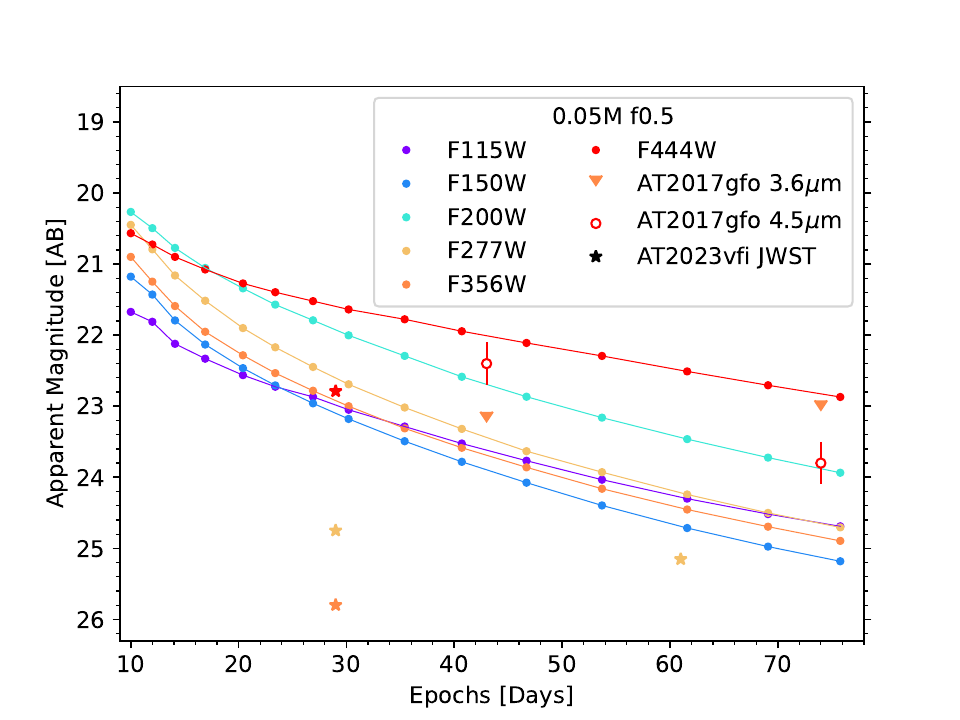}
    \includegraphics[trim={0.4cm 0.cm 0.4cm 0.3cm},width=0.47\linewidth]{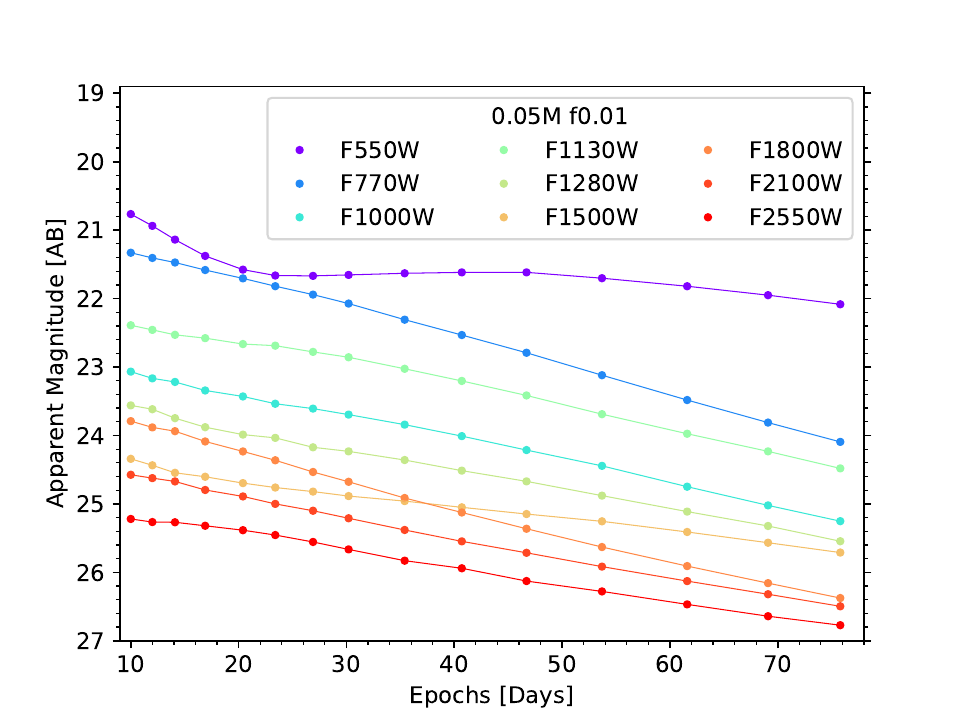}
    \includegraphics[trim={0.4cm 0.cm 0.4cm 0.3cm},width=0.47\linewidth]{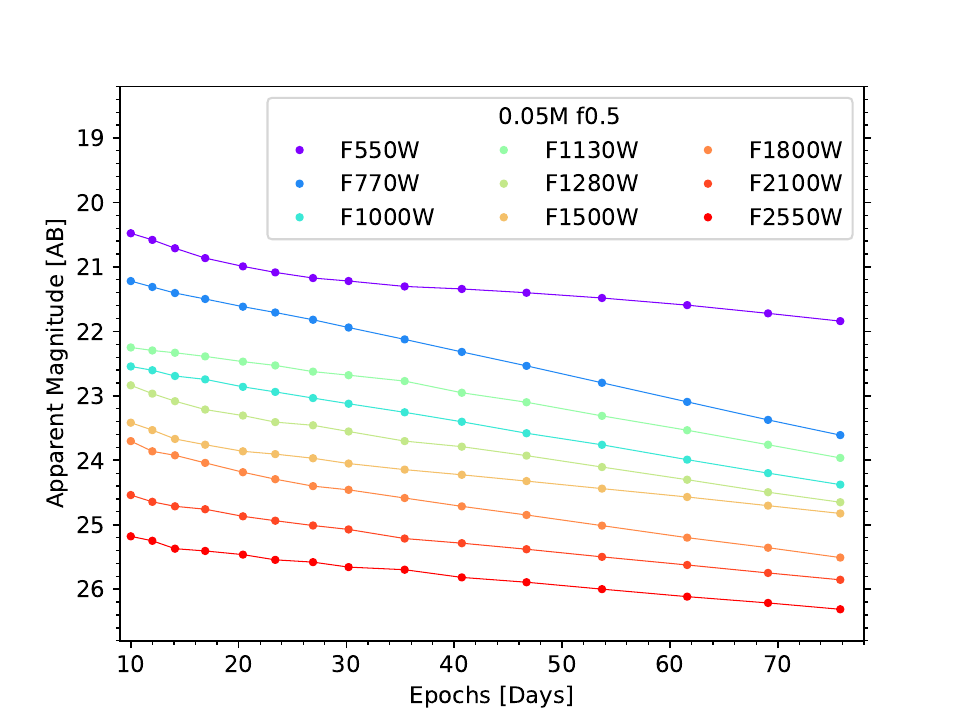}
    \caption{Broadband lightcurves in \textit{JWST} filters of the $f_{\rm{dyn}} = 0.01,0.5$ models with ejecta mass of $M_{\rm{ej}} = 0.05\, \Msol$. The top row shows the NIRcam bands and the bottom row the MIRI bands. The observed NIR photometry of AT2017gfo is included in the top row panels. Note that the 3.6\,$\mu$m points for AT2017gfo are upper limits. Synthetic photometry has been conducted on AT2023vfi the continuum-subtracted of AT2023vfi and scaled to be at a distance of 40~Mpc (see text). The colours of the markers for the synthetic photometry of AT2023vfi correspond to the relevant \textit{JWST} filter.}
    \label{fig:LCs}
\end{figure*}

While the observed continuum was well fit at 29\,d by a photosphere with velocity of $v= 0.08$c \citep[][]{Gillanders.Smartt:25}, and our models with $v_{\rm{in}} = 0.05$c are not optically thick at this time, we nevertheless explicitly examine the impact of our inner-boundary choice by taking the $f_{\rm{dyn}} = 0.5$ model with $M_{\rm{ej}} = 0.05\,\Msol$, removing the outer layer ($v = 0.25 - 0.3$c) and adding an inner layer between $v = 0.02 - 0.05$c, maintaining the $\rho \propto v^{-4}$ density profile. Since the outermost layer of the models at this time plays an extremely minor role in the spectral formation, we do not expect its removal to strongly impact the result. We run this model at 30\,d, the result of which is shown in Fig. \ref{fig:f05_dense}, where it is apparent that no continuum is formed.

The above calculation and test model, as well as results from previous studies on expansion opacities in LTE \citep[e.g.][]{Kasen.etal:13,Tanaka.etal:20,Deprince.etal:25}, suggest that the conditions required for line opacity alone to produce a photosphere emitting as a cool Blackbody at $\gtrsim$30~d are not easily achievable based on the the currently known range of possible ejecta properties, particularly if the photosphere is moving at a relatively fast velocity of $\sim 0.08$c. Though a dedicated study on this question should be conducted in order to yield a conclusive answer, we argue that other sources of opacity or thermal emission must be considered in order to produce such a cool, red continuum as observed in AT2023vfi, assuming this object originated from a BNS merger \citep[see][]{Arunchalam.etal:25}. 

\section{Photometry and lightcurves}
\label{sec:LCs}

Alongside the emergent spectra, we examine the photometric evolution of the models in terms of broadband lightcurves (LCs). We conduct synthetic photometry on the spectra using the \textit{JWST} NIRCam and MIRI wide filters. Since we consider our spectra from 1.2 $\mu$m onwards, we only use NIRCam filters with $\lambda_{\rm{ref}} > 1\, \mu$m, i.e. we start from the F115W filter. Following from the spectra, most of the LCs evolve similarly. In order to examine the range of broadband LCs, we therefore show the $f_{\rm{dyn}} = 0.01,0.5$ models in Fig. \ref{fig:LCs}, both with $M_{\rm{ej}} = 0.05\,\Msol$ as most relevant for comparison to the \textit{Spitzer} Infra-Red Array Camera (IRAC) 3.6 and 4.5 $\mu$m band observations of AT2017gfo \citep[][]{Villar.etal:18,Kasliwal.etal:22} and the \textit{JWST} NIRcam observations of AT2023vfi \citep[][]{Levan.etal:24}. Following the reasoning in Section \ref{subsubsec:AT2023vfi}, we conduct synthetic photometry on the continuum-subtracted spectra of AT2023vfi in order to compare the line emission only, since we do not include continuua in our models. The synthetic magnitudes are scaled to 40~Mpc for consistency with our models and AT2017gfo. Note that at 61~d we only include photometry from the F277W band, as the redder bands are measuring solely continuum emission which we subtract from the observed spectrum. 

\begin{table*}
    \centering
    \setlength\tabcolsep{0.4cm}
    \begin{tabular}{ccccc}
    \hline \hline
    Model & \multicolumn{2}{c}{40.7 days} & \multicolumn{2}{c}{75.8 days} \\
    $f_{\rm{dyn}}$     & 3.6$\mu$m & 4.5$\mu$m & 3.6$\mu$m & 4.5$\mu$m \\
    \hline 
    0                  & 25.17 & \textbf{22.57} & 26.56 & \textbf{23.71}\\
    0.01               & 25.13 & \textbf{22.51} & 26.54 & \textbf{23.32} \\
    0.05               & 24.94 & \textbf{22.39} & 26.37 & 23.21 \\
    0.1                & 24.74 & \textbf{22.27} & 26.17 & 23.10 \\
    0.5                & 23.64 & 21.80 & 24.95 & 22.72 \\
    \hline \hline
    \end{tabular}
    \caption{\textit{Spitzer} 3.6 $\mu$m and 4.5 $\mu$m band AB magnitudes of the models with $M_{\rm{ej}} = 0.05\,\rm{\Msol}$ at epochs closest to those of the observations of AT2017gfo. All of the model magnitudes in the 3.6 $\mu$m band are consistent with the non-detections of AT2017gfo. We mark magnitudes in the 4.5 $\mu$m band consistent with observations ($\pm 0.5$mags) in bold.} 
    \label{tab:Spitzer_mags}
\end{table*}

We begin by examining the broadband LCs of our $f_{\rm{dyn}} = 0.01$ model in the left hand-panels of Fig. \ref{fig:LCs}, which show the general broadband LC trends that are found in most parts of the parameter space. Notably, we have the F444W band as the brightest of the NIR bands across the entire timespan, corresponding to the important emission of Se\,\textsc{iii} at 4.5 $\mu$m. Similarly, the F550W band in the MIR is the brightest likewise due to the Se\,\textsc{iii} emission at 5.7 $\mu$m. This is followed by the F770W and F1000W bands, corresponding to the Ni\,\textsc{iii} features at both locations respectively, the redder of which is blended with Ru\,\textsc{ii}, Ru\,\textsc{iii} and other diverse species with smaller contributions, as explored in Section \ref{subsec:MIRspec}.

In the NIR bands, the F200W band is typically the second brightest after the F440W band, corresponding to the Te\,\textsc{iii} blended emission at 2.1 $\mu$m, with the F277W band sometimes being brighter at early times. This band is most affected by the early emission of Br\,\textsc{i}, and therefore drops quite steeply as this feature is lost with the ionisation of Br\,\textsc{i} to Br\,\textsc{ii}. The Br\,\textsc{ii} feature at 3.2 $\mu$m lies somewhat between the F277W and F356W features, such that both bands measure some flux from it. The F115W band initially follows the heavily blended emission at the bluest end of the model spectra (see Fig. \ref{fig:f005M001_NIRspec}), the flux levels of which drop rapidly in the first 20 - 30 days of the KN. However, we see a shallower slope in the LC from this point onwards, corresponding to the relatively constant emission from Y\,\textsc{ii} and Zr\,\textsc{ii} at $\sim 1.4\, \mu$m. 

In the MIR, we see that most bands typically follow a monotonically decreasing trend, aside from the F550W band, which shows a plateau or even faint brightening from $\sim$25~d onwards. This feature corresponds to the shift of ionisation structure from Se\,\textsc{i} to Se\,\textsc{iii}. Looking at top left panel of Fig. \ref{fig:f005spec_massvar}, we see that the 4 -- 6.5 $\mu$m range at 10 days is dominated by a single emission peak from Se\,\textsc{i}. This feature transitions into the usual double peaked Se\,\textsc{iii} emission feature in the 20 -- 30 days range, therefore yielding this plateau or re-brightening in the F550W band. This LC feature will not appear in models where Se\,\textsc{iii} is already dominant at 10\,d, e.g. in the models with less mass; instead a monotonic decrease like the other MIR bands is expected.

We now consider the more lanthanide-rich $f_{\rm{dyn}} = 0.5$ case shown in right-hand panels of Fig. \ref{fig:LCs}. We broadly recover a similar evolution for most bands that do not cover key spectral features as described in Section \ref{sec:spectra}, particularly for the 10 -- 30~$\mu$m bands as explored in Section \ref{subsec:MIRspec}. However, other differences in key bands are more noticeable. For instance, we find that the F444W band is not systematically the brightest at all epochs, particularly at early times, where we find that the F200W and F277W are slightly brighter. Considering the left-hand panel of Fig. \ref{fig:f05_massvar10d}, we see that the important lanthanide emission in the 1 -- 3 $\mu$m range of this model at 10 days far surpasses the emission of blended $_{34}$Se and $_{63}$Eu at $\sim 4.5\, \mu$m. In terms of certain NIR colours, for instance F440W - F200W, this implies that KN ejecta with a significant lanthanide-bearing dynamical component may in fact appear bluer at 10 days than models dominated more by post-merger ejecta, since NIR lanthanide contributions appear to be strongest in the 1 -- 3 $\mu$m range. 

We compare our broadband LCs to the NIR photometry of AT2017gfo \citep[][]{Villar.etal:18,Kasliwal.etal:22} and AT2023vfi in the top panels Fig. \ref{fig:LCs}. Starting with AT2023vfi, as seen from the spectra, the F444W band photometry of the $f_{\rm{dyn}} = 0.01$ model has the closest match to the 29~d values of AT2023vfi, though slightly brighter due to the greater flux on the blue side of the band at $\sim 4\,\mu$m (see Fig. \ref{fig:AT2023vfi_compare}). Conversely, the model is brighter in the F277W and F356W bands, due to the line emission from Br\,\textsc{i} and Br\,\textsc{ii}, while at 61~d, the model is dimmer in the F277W band due to relatively weak Doppler-broadened emission at 2.2~$\mu$m, which reaches the edge of the filter at 2.4~$\mu$m, compared to the observation. The $f_{\rm{dyn}} = 0.5$ model in the right-hand panel is consistently brighter than AT2023vfi due to its high power.

Considering AT2017gfo, we find similar ordering of magnitudes corresponding to the J, H and Ks bands in the 10 to 20 day range, notably that the Ks band is typically brighter than the J and H bands \citep[e.g.][]{Villar.etal:17}. It is more interesting to compare to the photometric observations of AT2017gfo by \textit{Spitzer} at 43 and 74 days. Two separate data reductions of these observations both yield non-detections in the 3.6 $\mu$m band down to $\sim 23$ AB magnitudes, and detections in the 4.5 $\mu$m band of $\sim 22.4$ and $23.8$ AB magnitudes respectively \citep[][]{Villar.etal:18,Kasliwal.etal:22}. We note that the first measurement in the 4.5 $\mu$m band has been reported as 22.9 AB mag \citep[][]{Villar.etal:18} and 21.9 AB mag \citep[][]{Kasliwal.etal:22}, and we consider the average of these two measurements. We compare the consistency of our models with $M_{\rm{ej}} = 0.05\,\Msol$ to these measurements in Table \ref{tab:Spitzer_mags}, taking magnitudes within $\pm 0.5$ mag to be consistent. 

We find that all models are consistent with the non-detections in the 3.6 $\mu$m bands at both epochs. However, only the most lanthanide-poor or even lanthanide-free model are consistent with the observed 4.5 $\mu$m band magnitudes, with more lanthanide-rich models being too bright at 75.8~d. It is perhaps surprising that the model which best reproduces the photometry is the fiducial, pure post-merger component case, not expected to correspond to the progenitor scenario of AT2017gfo, which is believed to have had a dynamical component and/or lanthanide-bearing component. However, given the simplicity of the ejecta models used here, as well as the uncertainties in NLTE modelling, the main result that should be taken away is that most of the models with $M_{\rm{ej}} = 0.05\, \Msol$ are consistent, so long as the post-merger ejecta component dominates, or alternatively, that the lanthanide mass fraction is $X_{\rm{La}} < 0.026$. Given that this contrasts with the closest spectrally matching model discussed in Section \ref{subsec:specobs_compare}, it is clear that AT2017gfo is not particularly well represented by any single model in our parameter space.

\section{Discussion and Conclusion}
\label{sec:discussion}

\subsection{Comparison to lanthanide-free models}
\label{subsec:companion_paper}

While this current paper has focused on the impact of lanthanides in the IR wavelengths of nebular phase KNe, we find that emission past $\lambda \sim 3.6\, \mu$m is typically dominated by first-peak species and $_{28}$Ni. Lanthanide-free models are used to study the nebular phase emission of these elements in depth in \citet{Jerkstrand.etal:25}. There, a solar abundance pattern restricted to $\rm{Z} = 31 - 40,52$, with trace abundances of $\rm{Z} = 26,28,30$ is considered, with usage of up to date dielectronic recombination rates for $_{34}$Se, $_{37}$Rb, $_{38}$Sr, $_{39}$Y, and $_{40}$Zr \citep[][]{Banerjee.etal:25}. Thermal collision strengths for the relevant low-lying forbidden lines in the chosen elements are also included when available, with all these lines being wavelength calibrated. 

Several key results obtained here are supported by similar findings. Notably, emission from Br\,\textsc{i} and Br\,\textsc{ii} is recovered in both cases centred at $\lambda \sim 3\,\mu$m, and Te\,\textsc{iii} is found to play a significant role at 2.1~$\mu$m, particularly at later times. In the MIR, the Ni\,\textsc{iii} 7.3 $\mu$m line is likewise recovered in both studies, as well as emission from $_{40}$Zr, and Br\,\textsc{ii}. Variations arising from different choices of ejecta profile, composition and atomic data usage are also present. In this study, we find a strong dominance of Se\,\textsc{i} and Se\,\textsc{iii} in the 4 -- 6 $\mu$m range, with the latter dominating at later times. However, Ge\,\textsc{i} is found to have an important emission line at 5.66~$\mu$m that blends, or even dominates over Se\,\textsc{iii} in cases of low-power, slow-moving ejecta \citep[model A of][]{Jerkstrand.etal:25}. High-power, fast-moving ejecta sees Ge\,\textsc{i} ionized out, and Se\,\textsc{iii} dominating as in the results found here. 

The relative strength of the Te\,\textsc{iii}, Kr\,\textsc{i} and Kr\,\textsc{iii} features around 2~$\mu$m also varies between models. While we find that Te\,\textsc{iii} consistently dominates, Kr\,\textsc{ii} may in fact be equally or more significant at early times in dense, low-power ejecta, and Kr\,\textsc{iii} may likewise play a greater role than that predicted by the models in this work. Additionally, Se\,\textsc{iv} is found to dominate emission at 2.29~$\mu$m in high-power, fast-moving ejecta at $t \geq 40$~d, while its contributions are found to be negligible in the models here. One must note, however, that the atomic data, recombination rates, and collision strengths for these particular species are all different between the models, on top of the ejecta models themselves. 

These variations highlight the complexity of the system in the nebular phase, showing that a full understanding of observed features will only come with a complete set of atomic data for all relevant processes, as well as strong constraints on ejecta properties. Despite this, the emergence of similar features arising from different underlying models, e.g. informed from Ab initio numerical simulations here cf. solar abundance pattern, suggest these results to be significant.

\subsection{Summary}
\label{subsec:summary}

Using simplified models informed from the outputs of numerical relativity and relativistic hydrodynamic simulations of a symmetric BNS merger, we employ NLTE radiative transfer simulations to explore a parameter space consisting of total ejecta mass and fraction of dynamical component. We evolve the models from 10 to 75 days post-merger in time-dependent mode with \textsc{sumo} \citep[][]{Jerkstrand.etal:11,Jerkstrand.etal:12,Pognan.etal:22a,Pognan.etal:25}, focusing on NIR and MIR emission in a range of 1.2 -- 30 $\mu$m, to investigate KN spectra in the context of existing and potential future \textit{JWST} observations. In particular we focus on the role of lanthanide species, complimenting current and previous works on nebular phase IR features \citep[][]{Hotokezaka.etal:22,Hotokezaka.etal:23,Gillanders.etal:24,Levan.etal:24,Gillanders.Smartt:25,Jerkstrand.etal:25}. 

We find that most of our models evolve in qualitatively similar ways across this timespan. We find a temperature and ionisation structure that follows predictions of time-dependent NLTE evolution: outer ejecta layers with lower density are more affected and experience cooling following the initial temperature increase predicted by steady-state calculations \citep[e.g.][]{Hotokezaka.etal:20,Hotokezaka.etal:21,Pognan.etal:22a}. The ionisation structure of these outer layers experiences a rapid freeze-out as early as 10 days, while the inner layers slowly become more ionized with time. This evolution of ionisation structure translates into slowly evolving spectra at late times, such that the same species emitting at 20 days are often also found at 75 days.

We find that most models spectrally evolve in similar fashion, though the lanthanide-rich $f_{\rm{dyn}} = 0.5$ model has more unique features driven by a markedly higher lanthanide mass fraction of $X_{\rm{La}} = 0.026$. We find several features common to most models. We recover Te\,\textsc{iii} emission at 2.1 $\mu$m across all epochs, though this feature is always blended, particularly at early times and in models with $M_{\rm{ej}} = 0.05\, \Msol$. We note more generally, that the emission at 2.1 $\mu$m is never found to be pure Te\,\textsc{iii}, with emission from Ce\,\textsc{iii}, Nd\,\textsc{ii}, Zr\,\textsc{ii}, Kr\,\textsc{ii} and Kr\,\textsc{iii} contributing to various degrees depending on model and epoch. Despite this, the persistent presence of Te\,\textsc{iii} emission at this wavelength across most epochs in our models supports previous claims of this species being at least partially responsible for the feature found in the spectra of both AT2017gfo \citep[][]{Hotokezaka.etal:23} and AT2023vfi \citep[][]{Levan.etal:24,Gillanders.Smartt:25}. 

We find a clear and persistent double-peaked Se\,\textsc{iii} feature in the 4 -- 6.4 $\mu$m range in the vast majority of our models across most epochs, noting that Se\,\textsc{iii} has been suggested as responsible for the 4.5 $\mu$m emission found in photometry of AT2017gfo \citep{Villar.etal:18,Hotokezaka.etal:22,Kasliwal.etal:22}, and in the spectra of AT2023vfi \citep[][]{Levan.etal:24,Gillanders.Smartt:25}. This double-peaked feature is initially absent at early times in the heaviest models with $M_{\rm{ej}} = 0.05\, \Msol$, where instead we recover a single Se\,\textsc{i} emission peak centred on 5 $\mu$m. In general, we expect a transition from a single-peaked Se\,\textsc{i} feature to a double-peaked Se\,\textsc{iii} feature, the time at which this occurs varying depending on model and atomic data, particularly recombination rate. W\,\textsc{iii} has also been suggested as a candidate for the emission at 4.5 $\mu$m \citep{Hotokezaka.etal:22}. However, we find W\,\textsc{iii} to be subdominant with respect to Se\,\textsc{iii} in every model at every epoch, largely due to a much smaller elemental abundance. 

We do not reproduce the Blackbody-like continuum as seen in AT2023vfi in any of our models. Considering the possibility of an optically thick `photosphere' creating this continuum, we argue that line opacity from lanthanides/actinides alone cannot produce such a feature in nebular phase KNe. To investigate this more quantitatively, we test a lanthanide-rich model with $v_{\rm{in}} = 0.02$c, finding that it is still optically thin in the NIR at 30\,d, and therefore does not produce an optically thick continuum. We therefore suggest that alternative sources of opacity which could yield an optically thick `photosphere', or otherwise different sources of low temperature, thermal emission, should be considered in future works. 

We compare our models to the line emission from AT2023vfi at 29~d by subtracting the Blackbody and GRB afterglow continua, considering a case where the thermal emission comes from a real photosphere, and a case where it is not produced by an optically thick photosphere, for which we remain agnostic as to the origin. We find that the 4.6~$\mu$m emission feature is relatively well reproduced by Se\,\textsc{i} and Se\,\textsc{iii}, or Se\,\textsc{iii} and W\,\textsc{iii} lines. However, other features do not match, and no model reproduces the double-peaked feature shape at 2.2~$\mu$m. We additionally find excess emission from $_{35}$Br lines in the 2.5 -- 4.0~$\mu$m range. Considering that this may indicate a lack of $_{35}$Br in the line-forming ejecta of AT2023vfi, we hypothesise that the ejecta may have had a general lack of first peak species, which would favour W\,\textsc{iii} as the main emitting species at 4.6~$\mu$m. While the merger scenario and nature of the remnant in AT2023vfi remains unclear, our models suggest that BNS mergers with long-lived remnants may be difficult to reconcile with the spectral observations.

Additionally, we make several new predictions for IR emission based on the results of our models. At late times ($t \gtrsim 30$ days) in models with $f_{\rm{dyn}} < 0.5$, we find Y\,\textsc{ii} emission at 1.39 $\mu$m, typically blended with Zr\,\textsc{ii} and traces of other species. The $f_{\rm{dyn}} = 0.5$ model instead has a significant blend of lanthanides, not all of which have been calibrated in these models, and so we make no claim in case of lanthanide-rich ejecta at this wavelength. At early times of $t = 10 \sim 30$ days, we find emission from Br\,\textsc{i}, Br\,\textsc{ii}, Ce\,\textsc{iii} and Nd\,\textsc{ii} in the 2.4 -- 3.6 $\mu$m range, with the relative strength of these contributions varying by model, mass, and epoch. Considering new recombination rates for Ce\,\textsc{iii} to Ce\,\textsc{ii}, we estimate that Ce\,\textsc{ii} may play a larger role, and expect its spectral impact to be greater than Ce\,\textsc{iii} due to a denser low-lying level structure. 

In the MIR, we predict Ni\,\textsc{iii} emission at 7.3 $\mu$m and 11 $\mu$m, both of which appear quasi-ubiquitously in our parameter space, though we note that the redder feature may in fact be blended with emission from Ge\,\textsc{i} \citep[][]{Jerkstrand.etal:25}. Although the flux of the 11\,$\mu$m feature is expected to be significantly smaller than that of the former, and despite blending with other species, most notably Ru\,\textsc{ii}, we find that it typically dominates the flux contribution, and therefore retain it as a potentially interesting feature to observe, at least as a way to confirm the Ni\,\textsc{iii} emission at 7.3 $\mu$m. The evolution of these features could potentially distinguish whether the dominating $_{28}$Ni isotopes are stable as predicted here, or unstable $^{56}$Ni as may be produced in the case of long-lived NS remnants \citep[][]{Jacobi.etal:25}. At $\sim$13 -- 14 $\mu$m, we additionally predict some emission mostly from Br\,\textsc{ii} and Zr\,\textsc{iii}. In the remainder of the 15 -- 30 $\mu$m range, we find some minor emission from diverse species including Zr\,\textsc{ii}, Mo\,\textsc{iii} and Fe\,\textsc{iii}, however the flux levels are markedly lower than other features, and we do not expect these to be easily observed. 

We also examine the broadband LC evolution of the models in the \textit{JWST} wide filters. We find that the F444W and F550W bands are usually the brightest due to the strong emission of Se\,\textsc{iii}. In the NIR, the F200W and F277W bands are usually the second brightest, with the former tracing Te\,\textsc{iii} emission at 2.1 $\mu$m, and the latter the blend of emission around 3 $\mu$m described above. In the MIR, the F770W band is the second brightest, as it tracks the Ni\,\textsc{iii} emission at 7.3 $\mu$m. We find that heavier models with $M_{\rm{ej}} = 0.05\, \Msol$ tend to have a plateau or even rebrightening in the F440W and F550W bands between 20 -- 30 days. This corresponds to the spectral shape of the models at these wavelengths changing from a single peak Se\,\textsc{i} feature centred at 5 $\mu$m, to the more common Se\,\textsc{iii} double-peak feature previously described. 

We find that our lanthanide rich $f_{\rm{dyn}} = 0.5$ model has initially different NIR colours than the other models. Notably, the F200W band is found to be as bright or even brighter than the F440W at 10 -- 20 days, due to the enhanced lanthanide emission at $\lambda \lesssim 3\, \mu$m. The general composition of KN ejecta parametrized by electron fraction $Y_e$ is often inferred by considering the colour and broadband LC evolution of the object, and by relating this to the average expansion opacity for given $Y_e$ \citep[e.g.][]{Tanaka.etal:20}. In this methodology, redder optical and NIR colours are usually taken to signify higher opacity material, typically associated to lanthanide-rich ejecta. We emphasise here that certain NIR colours may in fact appear bluer for more lanthanide-rich compositions, e.g. F440W - F200W. We therefore caution that while lanthanides do provide a significant amount of expansion opacity at NIR wavelengths, this does not appear to extend far past 2 -- 3 $\mu$m. From our models, we find that redder wavelengths are instead dominated by M1 emission from diverse first r-process peak elements, with potentially third peak elements playing a significant role in lower $Y_e$ ejecta where they may be synthesised in large quantities. 

While the results presented here inevitably depend on the ejecta model, choice of composition and atomic data, we find strong evidence showing that lanthanide impact is typically limited to $\lambda \lesssim 4\, \mu$m. This implies that MIR spectroscopy will be highly useful in probing first r-process peak species, particularly in cases where multiple lines of the same species are expected to be present, e.g. Ni\,\textsc{iii}, Se\,\textsc{iii}. In mildly bearing, or lanthanide-poor ejecta, observations in the NIR at later times can likewise probe `light' r-process species, while observations closer to $t \sim 10$d will instead measure lanthanide contributions. A full time-series of spectral observations will allow the separation of early time lanthanide features from more persistent, first-peak element features in the NIR. 

This current work and \citet{Jerkstrand.etal:25} are first steps into an exploration of the NIR and MIR emission of nebular phase KNe, conducted with NLTE radiative transfer simulations which self-consistently solve the temperature, excitation, and ionisation structures of the ejecta with the properties of the radiation field. While these studies provide an initial look into the possible features found there, further work exploring different merger scenarios, ejecta compositions and masses, as well as the generation of more complete atomic data required for NLTE modelling and usage of more realistic ejecta structure are still required in order to provide a complete picture of nebular phase, IR KN emission. 

\section*{Acknowledgements}

QP thanks James Gillanders for helpful discussion on the interpretation of AT2023vfi. JG thanks the Swedish Research Countil for the individual starting grant with contract no. 2020-05467. The radiative transfer simulations were performed on the Sakura and Raven clusters at the Max Planck Computational and Data Facility. 

\section*{Data Availability}

The data underlying this article are available upon reasonable request to the authors. Supplementary material may be viewed here: \url{https://zenodo.org/records/17017896}.



\bibliographystyle{mnras}
\bibliography{biblio} 

@PHDTHESIS{Axelrod:80,
       author = {{Axelrod}, T.~S.},
        title = "{Late Time Optical Spectra from the NICKEL(56) Model for Type i Supernovae.}",
       school = {California Univ., Santa Cruz.},
         year = 1980,
        month = jan,
       adsurl = {https://ui.adsabs.harvard.edu/abs/1980PhDT.........1A}
}

@ARTICLE{Jerkstrand.etal:11,
       author = {{Jerkstrand}, A. and {Fransson}, C. and {Kozma}, C.},
        title = "{The $^{44}$Ti-powered spectrum of SN 1987A}",
      journal = {\aap},
         year = 2011,
        month = jun,
       volume = {530},
          eid = {A45},
        pages = {A45},
          doi = {10.1051/0004-6361/201015937},
archivePrefix = {arXiv},
       eprint = {1103.3653},
 primaryClass = {astro-ph.HE},
       adsurl = {https://ui.adsabs.harvard.edu/abs/2011A&A...530A..45J}
}

@ARTICLE{Jerkstrand.etal:12,
       author = {{Jerkstrand}, A. and {Fransson}, C. and {Maguire}, K. and {Smartt}, S. and {Ergon}, M. and {Spyromilio}, J.},
        title = "{The progenitor mass of the Type IIP supernova SN 2004et from late-time spectral modeling}",
      journal = {\aap},
         year = 2012,
        month = oct,
       volume = {546},
          eid = {A28},
        pages = {A28},
          doi = {10.1051/0004-6361/201219528},
archivePrefix = {arXiv},
       eprint = {1208.2183},
 primaryClass = {astro-ph.HE},
       adsurl = {https://ui.adsabs.harvard.edu/abs/2012A&A...546A..28J}
}

@ARTICLE{Jerkstrand.etal:15,
       author = {{Jerkstrand}, A. and {Ergon}, M. and {Smartt}, S.~J. and {Fransson}, C. and {Sollerman}, J. and {Taubenberger}, S. and {Bersten}, M. and {Spyromilio}, J.},
        title = "{Late-time spectral line formation in Type IIb supernovae, with application to SN 1993J, SN 2008ax, and SN 2011dh}",
      journal = {\aap},
         year = 2015,
        month = jan,
       volume = {573},
          eid = {A12},
        pages = {A12},
          doi = {10.1051/0004-6361/201423983},
archivePrefix = {arXiv},
       eprint = {1408.0732},
 primaryClass = {astro-ph.HE},
       adsurl = {https://ui.adsabs.harvard.edu/abs/2015A&A...573A..12J}
}

@INCOLLECTION{Jerkstrand:17,
       author = {{Jerkstrand}, Anders},
        title = "{Spectra of Supernovae in the Nebular Phase}",
    booktitle = {Handbook of Supernovae},
         year = 2017,
       editor = {{Alsabti}, Athem W. and {Murdin}, Paul},
        pages = {795},
          doi = {10.1007/978-3-319-21846-5_29},
       adsurl = {https://ui.adsabs.harvard.edu/abs/2017hsn..book..795J}
}

@ARTICLE{Kiuchi.etal:24,
       author = {{Kiuchi}, Kenta and {Reboul-Salze}, Alexis and {Shibata}, Masaru and {Sekiguchi}, Yuichiro},
        title = "{A large-scale magnetic field produced by a solar-like dynamo in binary neutron star mergers}",
      journal = {Nature Astronomy},
         year = 2024,
        month = mar,
       volume = {8},
        pages = {298-307},
          doi = {10.1038/s41550-024-02194-y},
archivePrefix = {arXiv},
       eprint = {2306.15721},
 primaryClass = {astro-ph.HE},
       adsurl = {https://ui.adsabs.harvard.edu/abs/2024NatAs...8..298K}
}

@ARTICLE{Abbott.etal:17,
       author = {{Abbott}, B.~P. and {Abbott}, R. and {Abbott}, T.~D. and {Acernese}, F. and
         {Ackley}, K. and {Adams}, C. and {Adams}, T. and {et al.} },
        title = "{Multi-messenger Observations of a Binary Neutron Star Merger}",
      journal = {\apjl},
         year = "2017",
        month = "Oct",
       volume = {848},
       number = {2},
          eid = {L12},
        pages = {L12},
          doi = {10.3847/2041-8213/aa91c9},
archivePrefix = {arXiv},
       eprint = {1710.05833},
 primaryClass = {astro-ph.HE},
       adsurl = {https://ui.adsabs.harvard.edu/abs/2017ApJ...848L..12A}
}

@misc{Arunchalam.etal:25,
      title={GRB 230307A Formed No Dust or Was Not a Binary Neutron Star Merger}, 
      author={Prasiddha Arunachalam and Phillip Macias and Ryan. J. Foley},
      year={2025},
      eprint={2510.16121},
      archivePrefix={arXiv},
      primaryClass={astro-ph.HE},
      url={https://arxiv.org/abs/2510.16121}, 
}

@ARTICLE{Banerjee.etal:20,
       author = {{Banerjee}, Smaranika and {Tanaka}, Masaomi and {Kawaguchi}, Kyohei and {Kato}, Daiji and {Gaigalas}, Gediminas},
        title = "{Simulations of Early Kilonova Emission from Neutron Star Mergers}",
      journal = {\apj},
         year = 2020,
        month = sep,
       volume = {901},
       number = {1},
          eid = {29},
        pages = {29},
          doi = {10.3847/1538-4357/abae61},
archivePrefix = {arXiv},
       eprint = {2008.05495},
 primaryClass = {astro-ph.HE},
       adsurl = {https://ui.adsabs.harvard.edu/abs/2020ApJ...901...29B}
}

@ARTICLE{Banerjee.etal:22,
       author = {{Banerjee}, Smaranika and {Tanaka}, Masaomi and {Kato}, Daiji and {Gaigalas}, Gediminas and {Kawaguchi}, Kyohei and {Domoto}, Nanae},
        title = "{Opacity of the Highly Ionized Lanthanides and the Effect on the Early Kilonova}",
      journal = {\apj},
         year = 2022,
        month = aug,
       volume = {934},
       number = {2},
          eid = {117},
        pages = {117},
          doi = {10.3847/1538-4357/ac7565},
archivePrefix = {arXiv},
       eprint = {2204.06861},
 primaryClass = {astro-ph.HE},
       adsurl = {https://ui.adsabs.harvard.edu/abs/2022ApJ...934..117B}
}

@ARTICLE{Banerjee.etal:24,
       author = {{Banerjee}, Smaranika and {Tanaka}, Masaomi and {Kato}, Daiji and {Gaigalas}, Gediminas},
        title = "{Diversity of Early Kilonova with the Realistic Opacities of Highly Ionized Heavy Elements}",
      journal = {\apj},
         year = 2024,
        month = jun,
       volume = {968},
       number = {2},
          eid = {64},
        pages = {64},
          doi = {10.3847/1538-4357/ad4029},
archivePrefix = {arXiv},
       eprint = {2304.05810},
 primaryClass = {astro-ph.HE},
       adsurl = {https://ui.adsabs.harvard.edu/abs/2024ApJ...968...64B}
}

@ARTICLE{Banerjee.etal:25,
       author = {{Banerjee}, Smaranika and {Jerkstrand}, Anders and {Badnell}, Nigel and {Pognan}, Quentin and {Ferguson}, Niamh and {Grumer}, Jon},
        title = "{Nebular Spectra of Kilonovae with Detailed Recombination Rates. I. Light r-process Composition}",
      journal = {\apj},
         year = 2025,
        month = oct,
       volume = {992},
       number = {1},
          eid = {19},
        pages = {19},
          doi = {10.3847/1538-4357/adf6ba},
archivePrefix = {arXiv},
       eprint = {2501.18345},
 primaryClass = {astro-ph.HE},
       adsurl = {https://ui.adsabs.harvard.edu/abs/2025ApJ...992...19B}
}

@ARTICLE{Barnes.etal:16,
       author = {{Barnes}, Jennifer and {Kasen}, Daniel and {Wu}, Meng-Ru and {Mart{\'\i}nez-Pinedo}, Gabriel},
        title = "{Radioactivity and Thermalization in the Ejecta of Compact Object Mergers and Their Impact on Kilonova Light Curves}",
      journal = {\apj},
         year = 2016,
        month = oct,
       volume = {829},
       number = {2},
          eid = {110},
        pages = {110},
          doi = {10.3847/0004-637X/829/2/110},
archivePrefix = {arXiv},
       eprint = {1605.07218},
 primaryClass = {astro-ph.HE},
       adsurl = {https://ui.adsabs.harvard.edu/abs/2016ApJ...829..110B}
}

@ARTICLE{Brethauer.etal:26,
       author = {{Brethauer}, D. and {Kasen}, D. and {Margutti}, R. and {Chornock}, R.},
        title = "{Nonthermal Ionization of Kilonova Ejecta: Observable Impacts}",
      journal = {\apj},
         year = 2026,
        month = jan,
       volume = {996},
       number = {1},
          eid = {64},
        pages = {64},
          doi = {10.3847/1538-4357/ae1b8d},
archivePrefix = {arXiv},
       eprint = {2508.18364},
 primaryClass = {astro-ph.HE},
       adsurl = {https://ui.adsabs.harvard.edu/abs/2026ApJ...996...64B}
}

@ARTICLE{Carvajal.etal:22,
       author = {{Carvajal Gallego}, H. and {Berengut}, J.~C. and {Palmeri}, P. and {Quinet}, P.},
        title = "{Atomic data and opacity calculations in La V-X ions for the investigation of kilonova emission spectra}",
      journal = {\mnras},
         year = 2022,
        month = jun,
       volume = {513},
       number = {2},
        pages = {2302-2325},
          doi = {10.1093/mnras/stac1063},
       adsurl = {https://ui.adsabs.harvard.edu/abs/2022MNRAS.513.2302C}
}

@ARTICLE{Carvajal.etal:23,
       author = {{Carvajal Gallego}, H. and {Deprince}, J. and {Berengut}, J.~C. and {Palmeri}, P. and {Quinet}, P.},
        title = "{Opacity calculations in four to nine times ionized Pr, Nd, and Pm atoms for the spectral analysis of kilonovae}",
      journal = {\mnras},
         year = 2023,
        month = jan,
       volume = {518},
       number = {1},
        pages = {332-352},
          doi = {10.1093/mnras/stac3129},
       adsurl = {https://ui.adsabs.harvard.edu/abs/2023MNRAS.518..332C}
}

@ARTICLE{Cheong.etal:24,
       author = {{Cheong}, Patrick Chi-Kit and {Foucart}, Francois and {Duez}, Matthew D. and {Offermans}, Arthur and {Muhammed}, Nishad and {Chawhan}, Pavan},
        title = "{Energy-dependent and Energy-integrated Two-moment General-relativistic Neutrino Transport Simulations of a Hypermassive Neutron Star}",
      journal = {\apj},
         year = 2024,
        month = nov,
       volume = {975},
       number = {1},
          eid = {116},
        pages = {116},
          doi = {10.3847/1538-4357/ad7825},
archivePrefix = {arXiv},
       eprint = {2407.16017},
 primaryClass = {astro-ph.HE},
       adsurl = {https://ui.adsabs.harvard.edu/abs/2024ApJ...975..116C}
}

@ARTICLE{Cowperthwaite.etal:17,
       author = {{Cowperthwaite}, P.~S. and {Berger}, E. and {Villar}, V.~A. and {Metzger}, B.~D. and {Nicholl}, M. and {Chornock}, R. and {Blanchard}, P.~K. and {Fong}, W. and {Margutti}, R. and {Soares-Santos}, M. and {Alexander}, K.~D. and {Allam}, S. and {Annis}, J. and {Brout}, D. and {Brown}, D.~A. and {Butler}, R.~E. and {Chen}, H. -Y. and {Diehl}, H.~T. and {Doctor}, Z. and {Drout}, M.~R. and {Eftekhari}, T. and {Farr}, B. and {Finley}, D.~A. and {Foley}, R.~J. and {Frieman}, J.~A. and {Fryer}, C.~L. and {Garc{\'\i}a-Bellido}, J. and {Gill}, M.~S.~S. and {Guillochon}, J. and {Herner}, K. and {Holz}, D.~E. and {Kasen}, D. and {Kessler}, R. and {Marriner}, J. and {Matheson}, T. and {Neilsen}, E.~H., Jr. and {Quataert}, E. and {Palmese}, A. and {Rest}, A. and {Sako}, M. and {Scolnic}, D.~M. and {Smith}, N. and {Tucker}, D.~L. and {Williams}, P.~K.~G. and {Balbinot}, E. and {Carlin}, J.~L. and {Cook}, E.~R. and {Durret}, F. and {Li}, T.~S. and {Lopes}, P.~A.~A. and {Louren{\c{c}}o}, A.~C.~C. and {Marshall}, J.~L. and {Medina}, G.~E. and {Muir}, J. and {Mu{\~n}oz}, R.~R. and {Sauseda}, M. and {Schlegel}, D.~J. and {Secco}, L.~F. and {Vivas}, A.~K. and {Wester}, W. and {Zenteno}, A. and {Zhang}, Y. and {Abbott}, T.~M.~C. and {Banerji}, M. and {Bechtol}, K. and {Benoit-L{\'e}vy}, A. and {Bertin}, E. and {Buckley-Geer}, E. and {Burke}, D.~L. and {Capozzi}, D. and {Carnero Rosell}, A. and {Carrasco Kind}, M. and {Castander}, F.~J. and {Crocce}, M. and {Cunha}, C.~E. and {D'Andrea}, C.~B. and {da Costa}, L.~N. and {Davis}, C. and {DePoy}, D.~L. and {Desai}, S. and {Dietrich}, J.~P. and {Drlica-Wagner}, A. and {Eifler}, T.~F. and {Evrard}, A.~E. and {Fernandez}, E. and {Flaugher}, B. and {Fosalba}, P. and {Gaztanaga}, E. and {Gerdes}, D.~W. and {Giannantonio}, T. and {Goldstein}, D.~A. and {Gruen}, D. and {Gruendl}, R.~A. and {Gutierrez}, G. and {Honscheid}, K. and {Jain}, B. and {James}, D.~J. and {Jeltema}, T. and {Johnson}, M.~W.~G. and {Johnson}, M.~D. and {Kent}, S. and {Krause}, E. and {Kron}, R. and {Kuehn}, K. and {Nuropatkin}, N. and {Lahav}, O. and {Lima}, M. and {Lin}, H. and {Maia}, M.~A.~G. and {March}, M. and {Martini}, P. and {McMahon}, R.~G. and {Menanteau}, F. and {Miller}, C.~J. and {Miquel}, R. and {Mohr}, J.~J. and {Neilsen}, E. and {Nichol}, R.~C. and {Ogando}, R.~L.~C. and {Plazas}, A.~A. and {Roe}, N. and {Romer}, A.~K. and {Roodman}, A. and {Rykoff}, E.~S. and {Sanchez}, E. and {Scarpine}, V. and {Schindler}, R. and {Schubnell}, M. and {Sevilla-Noarbe}, I. and {Smith}, M. and {Smith}, R.~C. and {Sobreira}, F. and {Suchyta}, E. and {Swanson}, M.~E.~C. and {Tarle}, G. and {Thomas}, D. and {Thomas}, R.~C. and {Troxel}, M.~A. and {Vikram}, V. and {Walker}, A.~R. and {Wechsler}, R.~H. and {Weller}, J. and {Yanny}, B. and {Zuntz}, J.},
        title = "{The Electromagnetic Counterpart of the Binary Neutron Star Merger LIGO/Virgo GW170817. II. UV, Optical, and Near-infrared Light Curves and Comparison to Kilonova Models}",
      journal = {\apjl},
         year = 2017,
        month = oct,
       volume = {848},
       number = {2},
          eid = {L17},
        pages = {L17},
          doi = {10.3847/2041-8213/aa8fc7},
archivePrefix = {arXiv},
       eprint = {1710.05840},
 primaryClass = {astro-ph.HE},
       adsurl = {https://ui.adsabs.harvard.edu/abs/2017ApJ...848L..17C}
}

@ARTICLE{Curtis.etal:24,
       author = {{Curtis}, Sanjana and {Bosch}, Pablo and {M{\"o}sta}, Philipp and {Radice}, David and {Bernuzzi}, Sebastiano and {Perego}, Albino and {Haas}, Roland and {Schnetter}, Erik},
        title = "{Magnetized Outflows from Short-lived Neutron Star Merger Remnants Can Produce a Blue Kilonova}",
      journal = {\apjl},
         year = 2024,
        month = jan,
       volume = {961},
       number = {1},
          eid = {L26},
        pages = {L26},
          doi = {10.3847/2041-8213/ad0fe1},
archivePrefix = {arXiv},
       eprint = {2305.07738},
 primaryClass = {astro-ph.HE},
       adsurl = {https://ui.adsabs.harvard.edu/abs/2024ApJ...961L..26C}
}

@ARTICLE{Domoto.etal:21,
       author = {{Domoto}, Nanae and {Tanaka}, Masaomi and {Wanajo}, Shinya and {Kawaguchi}, Kyohei},
        title = "{Signatures of r-process Elements in Kilonova Spectra}",
      journal = {\apj},
         year = 2021,
        month = may,
       volume = {913},
       number = {1},
          eid = {26},
        pages = {26},
          doi = {10.3847/1538-4357/abf358},
archivePrefix = {arXiv},
       eprint = {2103.15284},
 primaryClass = {astro-ph.HE},
       adsurl = {https://ui.adsabs.harvard.edu/abs/2021ApJ...913...26D}
}

@ARTICLE{Domoto.etal:22,
       author = {{Domoto}, Nanae and {Tanaka}, Masaomi and {Kato}, Daiji and {Kawaguchi}, Kyohei and {Hotokezaka}, Kenta and {Wanajo}, Shinya},
        title = "{Lanthanide Features in Near-infrared Spectra of Kilonovae}",
      journal = {\apj},
         year = 2022,
        month = nov,
       volume = {939},
       number = {1},
          eid = {8},
        pages = {8},
          doi = {10.3847/1538-4357/ac8c36},
archivePrefix = {arXiv},
       eprint = {2206.04232},
 primaryClass = {astro-ph.HE},
       adsurl = {https://ui.adsabs.harvard.edu/abs/2022ApJ...939....8D}
}

@ARTICLE{Eichler.etal:89,
       author = {{Eichler}, David and {Livio}, Mario and {Piran}, Tsvi and
         {Schramm}, David N.},
        title = "{Nucleosynthesis, neutrino bursts and {\ensuremath{\gamma}}-rays from coalescing neutron stars}",
      journal = {\nat},
         year = "1989",
        month = "Jul",
       volume = {340},
       number = {6229},
        pages = {126-128},
          doi = {10.1038/340126a0},
       adsurl = {https://ui.adsabs.harvard.edu/abs/1989Natur.340..126E}
}

@ARTICLE{Flors.etal:23,
       author = {{Fl{\"o}rs}, A. and {Silva}, R.~F. and {Deprince}, J. and {Carvajal Gallego}, H. and {Leck}, G. and {Shingles}, L.~J. and {Mart{\'\i}nez-Pinedo}, G. and {Sampaio}, J.~M. and {Amaro}, P. and {Marques}, J.~P. and {Goriely}, S. and {Quinet}, P. and {Palmeri}, P. and {Godefroid}, M.},
        title = "{Opacities of singly and doubly ionized neodymium and uranium for kilonova emission modeling}",
      journal = {\mnras},
         year = 2023,
        month = sep,
       volume = {524},
       number = {2},
        pages = {3083-3101},
          doi = {10.1093/mnras/stad2053},
archivePrefix = {arXiv},
       eprint = {2302.01780},
 primaryClass = {astro-ph.HE},
       adsurl = {https://ui.adsabs.harvard.edu/abs/2023MNRAS.524.3083F}
}

@ARTICLE{Freiburghaus.etal:99,
       author = {{Freiburghaus}, C. and {Rosswog}, S. and {Thielemann}, F. -K.},
        title = "{R-Process in Neutron Star Mergers}",
      journal = {\apjl},
         year = "1999",
        month = "Nov",
       volume = {525},
       number = {2},
        pages = {L121-L124},
          doi = {10.1086/312343},
       adsurl = {https://ui.adsabs.harvard.edu/abs/1999ApJ...525L.121F}
}

@ARTICLE{Fujibayashi.etal:18,
       author = {{Fujibayashi}, Sho and {Kiuchi}, Kenta and {Nishimura}, Nobuya and {Sekiguchi}, Yuichiro and {Shibata}, Masaru},
        title = "{Mass Ejection from the Remnant of a Binary Neutron Star Merger: Viscous-radiation Hydrodynamics Study}",
      journal = {\apj},
         year = 2018,
        month = jun,
       volume = {860},
       number = {1},
          eid = {64},
        pages = {64},
          doi = {10.3847/1538-4357/aabafd},
archivePrefix = {arXiv},
       eprint = {1711.02093},
 primaryClass = {astro-ph.HE},
       adsurl = {https://ui.adsabs.harvard.edu/abs/2018ApJ...860...64F}
}

@ARTICLE{Fujibayashi.etal:20a,
       author = {{Fujibayashi}, Sho and {Shibata}, Masaru and {Wanajo}, Shinya and {Kiuchi}, Kenta and {Kyutoku}, Koutarou and {Sekiguchi}, Yuichiro},
        title = "{Mass ejection from disks surrounding a low-mass black hole: Viscous neutrino-radiation hydrodynamics simulation in full general relativity}",
      journal = {\prd},
         year = 2020,
        month = apr,
       volume = {101},
       number = {8},
          eid = {083029},
        pages = {083029},
          doi = {10.1103/PhysRevD.101.083029},
archivePrefix = {arXiv},
       eprint = {2001.04467},
 primaryClass = {astro-ph.HE},
       adsurl = {https://ui.adsabs.harvard.edu/abs/2020PhRvD.101h3029F}
}

@ARTICLE{Fujibayashi.etal:20b,
       author = {{Fujibayashi}, Sho and {Wanajo}, Shinya and {Kiuchi}, Kenta and {Kyutoku}, Koutarou and {Sekiguchi}, Yuichiro and {Shibata}, Masaru},
        title = "{Postmerger Mass Ejection of Low-mass Binary Neutron Stars}",
      journal = {\apj},
         year = 2020,
        month = oct,
       volume = {901},
       number = {2},
          eid = {122},
        pages = {122},
          doi = {10.3847/1538-4357/abafc2},
archivePrefix = {arXiv},
       eprint = {2007.00474},
 primaryClass = {astro-ph.HE},
       adsurl = {https://ui.adsabs.harvard.edu/abs/2020ApJ...901..122F}
}

@ARTICLE{Fujibayashi.etal:23,
       author = {{Fujibayashi}, Sho and {Kiuchi}, Kenta and {Wanajo}, Shinya and {Kyutoku}, Koutarou and {Sekiguchi}, Yuichiro and {Shibata}, Masaru},
        title = "{Comprehensive Study of Mass Ejection and Nucleosynthesis in Binary Neutron Star Mergers Leaving Short-lived Massive Neutron Stars}",
      journal = {\apj},
         year = 2023,
        month = jan,
       volume = {942},
       number = {1},
          eid = {39},
        pages = {39},
          doi = {10.3847/1538-4357/ac9ce0},
archivePrefix = {arXiv},
       eprint = {2205.05557},
 primaryClass = {astro-ph.HE},
       adsurl = {https://ui.adsabs.harvard.edu/abs/2023ApJ...942...39F}
}

@ARTICLE{Gillanders.etal:21,
       author = {{Gillanders}, J.~H. and {McCann}, M. and {Sim}, S.~A. and {Smartt}, S.~J. and {Ballance}, C.~P.},
        title = "{Constraints on the presence of platinum and gold in the spectra of the kilonova AT2017gfo}",
      journal = {\mnras},
         year = 2021,
        month = sep,
       volume = {506},
       number = {3},
        pages = {3560-3577},
          doi = {10.1093/mnras/stab1861},
archivePrefix = {arXiv},
       eprint = {2101.08271},
 primaryClass = {astro-ph.HE},
       adsurl = {https://ui.adsabs.harvard.edu/abs/2021MNRAS.506.3560G}
}

@ARTICLE{Gillanders.etal:22,
       author = {{Gillanders}, J.~H. and {Smartt}, S.~J. and {Sim}, S.~A. and {Bauswein}, A. and {Goriely}, S.},
        title = "{Modelling the spectra of the kilonova AT2017gfo - I. The photospheric epochs}",
      journal = {\mnras},
         year = 2022,
        month = sep,
       volume = {515},
       number = {1},
        pages = {631-651},
          doi = {10.1093/mnras/stac1258},
archivePrefix = {arXiv},
       eprint = {2202.01786},
 primaryClass = {astro-ph.HE},
       adsurl = {https://ui.adsabs.harvard.edu/abs/2022MNRAS.515..631G}
}

@ARTICLE{Gillanders.etal:24,
       author = {{Gillanders}, J.~H. and {Sim}, S.~A. and {Smartt}, S.~J. and {Goriely}, S. and {Bauswein}, A.},
        title = "{Modelling the spectra of the kilonova AT2017gfo - II. Beyond the photospheric epochs}",
      journal = {\mnras},
         year = 2024,
        month = apr,
       volume = {529},
       number = {3},
        pages = {2918-2945},
          doi = {10.1093/mnras/stad3688},
archivePrefix = {arXiv},
       eprint = {2306.15055},
 primaryClass = {astro-ph.HE},
       adsurl = {https://ui.adsabs.harvard.edu/abs/2024MNRAS.529.2918G}
}

@ARTICLE{Gillanders.Smartt:25,
       author = {{Gillanders}, J.~H. and {Smartt}, S.~J.},
        title = "{Analysis of the JWST spectra of the kilonova AT 2023vfi accompanying GRB 230307A}",
      journal = {\mnras},
         year = 2025,
        month = feb,
          doi = {10.1093/mnras/staf287},
archivePrefix = {arXiv},
       eprint = {2408.11093},
 primaryClass = {astro-ph.HE},
       adsurl = {https://ui.adsabs.harvard.edu/abs/2025MNRAS.tmp..272G}
}

@ARTICLE{Hayashi.etal:22,
       author = {{Hayashi}, Kota and {Fujibayashi}, Sho and {Kiuchi}, Kenta and {Kyutoku}, Koutarou and {Sekiguchi}, Yuichiro and {Shibata}, Masaru},
        title = "{General-relativistic neutrino-radiation magnetohydrodynamic simulation of seconds-long black hole-neutron star mergers}",
      journal = {\prd},
         year = 2022,
        month = jul,
       volume = {106},
       number = {2},
          eid = {023008},
        pages = {023008},
          doi = {10.1103/PhysRevD.106.023008},
archivePrefix = {arXiv},
       eprint = {2111.04621},
 primaryClass = {astro-ph.HE},
       adsurl = {https://ui.adsabs.harvard.edu/abs/2022PhRvD.106b3008H}
}

@ARTICLE{Hotokezaka.etal:20,
       author = {{Hotokezaka}, Kenta and {Nakar}, Ehud},
        title = "{Radioactive Heating Rate of r-process Elements and Macronova Light Curve}",
      journal = {\apj},
         year = 2020,
        month = mar,
       volume = {891},
       number = {2},
          eid = {152},
        pages = {152},
          doi = {10.3847/1538-4357/ab6a98},
archivePrefix = {arXiv},
       eprint = {1909.02581},
 primaryClass = {astro-ph.HE},
       adsurl = {https://ui.adsabs.harvard.edu/abs/2020ApJ...891..152H}
}

@ARTICLE{Hotokezaka.etal:21,
       author = {{Hotokezaka}, Kenta and {Tanaka}, Masaomi and {Kato}, Daiji and {Gaigalas}, Gediminas},
        title = "{Nebular emission from lanthanide-rich ejecta of neutron star merger}",
      journal = {\mnras},
         year = 2021,
        month = oct,
       volume = {506},
       number = {4},
        pages = {5863-5877},
          doi = {10.1093/mnras/stab1975},
archivePrefix = {arXiv},
       eprint = {2102.07879},
 primaryClass = {astro-ph.HE},
       adsurl = {https://ui.adsabs.harvard.edu/abs/2021MNRAS.506.5863H}
}

@ARTICLE{Hotokezaka.etal:22,
       author = {{Hotokezaka}, Kenta and {Tanaka}, Masaomi and {Kato}, Daiji and {Gaigalas}, Gediminas},
        title = "{Tungsten versus Selenium as a potential source of kilonova nebular emission observed by Spitzer}",
      journal = {\mnras},
         year = 2022,
        month = sep,
       volume = {515},
       number = {1},
        pages = {L89-L93},
          doi = {10.1093/mnrasl/slac071},
archivePrefix = {arXiv},
       eprint = {2204.00737},
 primaryClass = {astro-ph.HE},
       adsurl = {https://ui.adsabs.harvard.edu/abs/2022MNRAS.515L..89H}
}

@ARTICLE{Hotokezaka.etal:23,
       author = {{Hotokezaka}, Kenta and {Tanaka}, Masaomi and {Kato}, Daiji and {Gaigalas}, Gediminas},
        title = "{Tellurium emission line in kilonova AT 2017gfo}",
      journal = {\mnras},
         year = 2023,
        month = nov,
       volume = {526},
       number = {1},
        pages = {L155-L159},
          doi = {10.1093/mnrasl/slad128},
archivePrefix = {arXiv},
       eprint = {2307.00988},
 primaryClass = {astro-ph.HE},
       adsurl = {https://ui.adsabs.harvard.edu/abs/2023MNRAS.526L.155H}
}

@ARTICLE{Jacobi.etal:25,
       author = {{Jacobi}, Maximilian and {Magistrelli}, Fabio and {Loffredo}, Eleonora and {Ricigliano}, Giacomo and {Chiesa}, Leonardo and {Bernuzzi}, Sebastiano and {Perego}, Albino and {Arcones}, Almudena},
        title = "{$^{56}$Ni Production in Long-lived Binary Neutron Star Merger Remnants}",
      journal = {\apjl},
         year = 2026,
        month = mar,
       volume = {999},
       number = {1},
          eid = {L16},
        pages = {L16},
          doi = {10.3847/2041-8213/ae4104},
       adsurl = {https://ui.adsabs.harvard.edu/abs/2026ApJ...999L..16J}
}

@ARTICLE{Jerkstrand.etal:25,
       author = {{Jerkstrand}, Anders and {Pognan}, Quentin and {Banerjee}, Smaranika and {Sterling}, Nicholas and {Grumer}, Jon and {Ferguson}, Niamh and {Butler}, Keith and {Gillanders}, James and {Smartt}, Stephen and {Kawaguchi}, Kyohei and {Vilagos}, Blanka},
        title = "{Infrared spectral signatures of light r-process elements in kilonovae}",
      journal = {arXiv e-prints},
         year = 2025,
        month = oct,
          eid = {arXiv:2510.12410},
        pages = {arXiv:2510.12410},
          doi = {10.48550/arXiv.2510.12410},
archivePrefix = {arXiv},
       eprint = {2510.12410},
 primaryClass = {astro-ph.SR},
       adsurl = {https://ui.adsabs.harvard.edu/abs/2025arXiv251012410J}
}

@ARTICLE{Just.etal:23,
       author = {{Just}, O. and {Vijayan}, V. and {Xiong}, Z. and {Goriely}, S. and {Soultanis}, T. and {Bauswein}, A. and {Guilet}, J. and {Janka}, H. -Th. and {Mart{\'\i}nez-Pinedo}, G.},
        title = "{End-to-end Kilonova Models of Neutron Star Mergers with Delayed Black Hole Formation}",
      journal = {\apjl},
         year = 2023,
        month = jul,
       volume = {951},
       number = {1},
          eid = {L12},
        pages = {L12},
          doi = {10.3847/2041-8213/acdad2},
archivePrefix = {arXiv},
       eprint = {2302.10928},
 primaryClass = {astro-ph.HE},
       adsurl = {https://ui.adsabs.harvard.edu/abs/2023ApJ...951L..12J}
}

@ARTICLE{Kasen.etal:13,
       author = {{Kasen}, Daniel and {Badnell}, N.~R. and {Barnes}, Jennifer},
        title = "{Opacities and Spectra of the r-process Ejecta from Neutron Star Mergers}",
      journal = {\apj},
         year = 2013,
        month = sep,
       volume = {774},
       number = {1},
          eid = {25},
        pages = {25},
          doi = {10.1088/0004-637X/774/1/25},
archivePrefix = {arXiv},
       eprint = {1303.5788},
 primaryClass = {astro-ph.HE},
       adsurl = {https://ui.adsabs.harvard.edu/abs/2013ApJ...774...25K}
}

@ARTICLE{Kasen.etal:17,
       author = {{Kasen}, Daniel and {Metzger}, Brian and {Barnes}, Jennifer and {Quataert}, Eliot and {Ramirez-Ruiz}, Enrico},
        title = "{Origin of the heavy elements in binary neutron-star mergers from a gravitational-wave event}",
      journal = {\nat},
         year = 2017,
        month = nov,
       volume = {551},
       number = {7678},
        pages = {80-84},
          doi = {10.1038/nature24453},
archivePrefix = {arXiv},
       eprint = {1710.05463},
 primaryClass = {astro-ph.HE},
       adsurl = {https://ui.adsabs.harvard.edu/abs/2017Natur.551...80K}
}

@ARTICLE{Kasen.Barnes:19,
       author = {{Kasen}, Daniel and {Barnes}, Jennifer},
        title = "{Radioactive Heating and Late Time Kilonova Light Curves}",
      journal = {\apj},
         year = 2019,
        month = may,
       volume = {876},
       number = {2},
          eid = {128},
        pages = {128},
          doi = {10.3847/1538-4357/ab06c2},
archivePrefix = {arXiv},
       eprint = {1807.03319},
 primaryClass = {astro-ph.HE},
       adsurl = {https://ui.adsabs.harvard.edu/abs/2019ApJ...876..128K}
}

@ARTICLE{Kasliwal.etal:22,
       author = {{Kasliwal}, Mansi M. and {Kasen}, Daniel and {Lau}, Ryan M. and {Perley}, Daniel A. and {Rosswog}, Stephan and {Ofek}, Eran O. and {Hotokezaka}, Kenta and {Chary}, Ranga-Ram and {Sollerman}, Jesper and {Goobar}, Ariel and {Kaplan}, David L.},
        title = "{Spitzer mid-infrared detections of neutron star merger GW170817 suggests synthesis of the heaviest elements}",
      journal = {\mnras},
         year = 2022,
        month = feb,
       volume = {510},
       number = {1},
        pages = {L7-L12},
          doi = {10.1093/mnrasl/slz007},
archivePrefix = {arXiv},
       eprint = {1812.08708},
 primaryClass = {astro-ph.HE},
       adsurl = {https://ui.adsabs.harvard.edu/abs/2022MNRAS.510L...7K}
}

@ARTICLE{Kawaguchi.etal:18,
       author = {{Kawaguchi}, Kyohei and {Shibata}, Masaru and {Tanaka}, Masaomi},
        title = "{Radiative Transfer Simulation for the Optical and Near-infrared Electromagnetic Counterparts to GW170817}",
      journal = {\apjl},
         year = 2018,
        month = oct,
       volume = {865},
       number = {2},
          eid = {L21},
        pages = {L21},
          doi = {10.3847/2041-8213/aade02},
archivePrefix = {arXiv},
       eprint = {1806.04088},
 primaryClass = {astro-ph.HE},
       adsurl = {https://ui.adsabs.harvard.edu/abs/2018ApJ...865L..21K}
}

@ARTICLE{Kawaguchi.etal:21,
       author = {{Kawaguchi}, Kyohei and {Fujibayashi}, Sho and {Shibata}, Masaru and {Tanaka}, Masaomi and {Wanajo}, Shinya},
        title = "{A Low-mass Binary Neutron Star: Long-term Ejecta Evolution and Kilonovae with Weak Blue Emission}",
      journal = {\apj},
         year = 2021,
        month = jun,
       volume = {913},
       number = {2},
          eid = {100},
        pages = {100},
          doi = {10.3847/1538-4357/abf3bc},
archivePrefix = {arXiv},
       eprint = {2012.14711},
 primaryClass = {astro-ph.HE},
       adsurl = {https://ui.adsabs.harvard.edu/abs/2021ApJ...913..100K}
}

@ARTICLE{Kawaguchi.etal:22,
       author = {{Kawaguchi}, Kyohei and {Fujibayashi}, Sho and {Hotokezaka}, Kenta and {Shibata}, Masaru and {Wanajo}, Shinya},
        title = "{Electromagnetic Counterparts of Binary-neutron-star Mergers Leading to a Strongly Magnetized Long-lived Remnant Neutron Star}",
      journal = {\apj},
         year = 2022,
        month = jul,
       volume = {933},
       number = {1},
          eid = {22},
        pages = {22},
          doi = {10.3847/1538-4357/ac6ef7},
archivePrefix = {arXiv},
       eprint = {2202.13149},
 primaryClass = {astro-ph.HE},
       adsurl = {https://ui.adsabs.harvard.edu/abs/2022ApJ...933...22K}
}

@ARTICLE{Kawaguchi.etal:23,
       author = {{Kawaguchi}, Kyohei and {Fujibayashi}, Sho and {Domoto}, Nanae and {Kiuchi}, Kenta and {Shibata}, Masaru and {Wanajo}, Shinya},
        title = "{Kilonovae of binary neutron star mergers leading to short-lived remnant neutron star formation}",
      journal = {\mnras},
         year = 2023,
        month = nov,
       volume = {525},
       number = {3},
        pages = {3384-3398},
          doi = {10.1093/mnras/stad2430},
archivePrefix = {arXiv},
       eprint = {2306.06961},
 primaryClass = {astro-ph.HE},
       adsurl = {https://ui.adsabs.harvard.edu/abs/2023MNRAS.525.3384K}
}

@ARTICLE{Kawaguchi.etal:24,
       author = {{Kawaguchi}, Kyohei and {Domoto}, Nanae and {Fujibayashi}, Sho and {Hamidani}, Hamid and {Hayashi}, Kota and {Shibata}, Masaru and {Tanaka}, Masaomi and {Wanajo}, Shinya},
        title = "{Three dimensional end-to-end simulation for kilonova emission from a black-hole neutron-star merger}",
      journal = {\mnras},
         year = 2024,
        month = nov,
          doi = {10.1093/mnras/stae2594},
archivePrefix = {arXiv},
       eprint = {2404.15027},
 primaryClass = {astro-ph.HE},
       adsurl = {https://ui.adsabs.harvard.edu/abs/2024MNRAS.tmp.2516K}
}

@INPROCEEDINGS{Kurucz:18,
       author = {{Kurucz}, R.~L.},
        title = "{Including All the Lines: Data Releases for Spectra and Opacities through 2017}",
    booktitle = {Workshop on Astrophysical Opacities},
         year = 2018,
       series = {Astronomical Society of the Pacific Conference Series},
       volume = {515},
        month = aug,
        pages = {47},
       adsurl = {https://ui.adsabs.harvard.edu/abs/2018ASPC..515...47K}
}

@ARTICLE{Kyutoku.etal:21,
       author = {{Kyutoku}, Koutarou and {Shibata}, Masaru and {Taniguchi}, Keisuke},
        title = "{Coalescence of black hole{\textendash}neutron star binaries}",
      journal = {Living Reviews in Relativity},
         year = 2021,
        month = dec,
       volume = {24},
       number = {1},
          eid = {5},
        pages = {5},
          doi = {10.1007/s41114-021-00033-4},
archivePrefix = {arXiv},
       eprint = {2110.06218},
 primaryClass = {astro-ph.HE},
       adsurl = {https://ui.adsabs.harvard.edu/abs/2021LRR....24....5K}
}

@ARTICLE{Lattimer.Schramm:74,
       author = {{Lattimer}, J.~M. and {Schramm}, D.~N.},
        title = "{Black-Hole-Neutron-Star Collisions}",
      journal = {\apjl},
         year = "1974",
        month = "Sep",
       volume = {192},
        pages = {L145},
          doi = {10.1086/181612},
       adsurl = {https://ui.adsabs.harvard.edu/abs/1974ApJ...192L.145L}
}

@ARTICLE{Lattimer.Schramm:76,
       author = {{Lattimer}, J.~M. and {Schramm}, D.~N.},
        title = "{The tidal disruption of neutron stars by black holes in close binaries.}",
      journal = {\apj},
         year = 1976,
        month = dec,
       volume = {210},
        pages = {549-567},
          doi = {10.1086/154860},
       adsurl = {https://ui.adsabs.harvard.edu/abs/1976ApJ...210..549L}
}

@ARTICLE{Levan.etal:24,
       author = {{Levan}, Andrew J. and {Gompertz}, Benjamin P. and {Salafia}, Om Sharan and {Bulla}, Mattia and {Burns}, Eric and {Hotokezaka}, Kenta and {Izzo}, Luca and {Lamb}, Gavin P. and {Malesani}, Daniele B. and {Oates}, Samantha R. and {Ravasio}, Maria Edvige and {Rouco Escorial}, Alicia and {Schneider}, Benjamin and {Sarin}, Nikhil and {Schulze}, Steve and {Tanvir}, Nial R. and {Ackley}, Kendall and {Anderson}, Gemma and {Brammer}, Gabriel B. and {Christensen}, Lise and {Dhillon}, Vikram S. and {Evans}, Phil A. and {Fausnaugh}, Michael and {Fong}, Wen-fai and {Fruchter}, Andrew S. and {Fryer}, Chris and {Fynbo}, Johan P.~U. and {Gaspari}, Nicola and {Heintz}, Kasper E. and {Hjorth}, Jens and {Kennea}, Jamie A. and {Kennedy}, Mark R. and {Laskar}, Tanmoy and {Leloudas}, Giorgos and {Mandel}, Ilya and {Martin-Carrillo}, Antonio and {Metzger}, Brian D. and {Nicholl}, Matt and {Nugent}, Anya and {Palmerio}, Jesse T. and {Pugliese}, Giovanna and {Rastinejad}, Jillian and {Rhodes}, Lauren and {Rossi}, Andrea and {Saccardi}, Andrea and {Smartt}, Stephen J. and {Stevance}, Heloise F. and {Tohuvavohu}, Aaron and {van der Horst}, Alexander and {Vergani}, Susanna D. and {Watson}, Darach and {Barclay}, Thomas and {Bhirombhakdi}, Kornpob and {Breedt}, Elm{\'e} and {Breeveld}, Alice A. and {Brown}, Alexander J. and {Campana}, Sergio and {Chrimes}, Ashley A. and {D'Avanzo}, Paolo and {D'Elia}, Valerio and {De Pasquale}, Massimiliano and {Dyer}, Martin J. and {Galloway}, Duncan K. and {Garbutt}, James A. and {Green}, Matthew J. and {Hartmann}, Dieter H. and {Jakobsson}, P{\'a}ll and {Kerry}, Paul and {Kouveliotou}, Chryssa and {Langeroodi}, Danial and {Le Floc'h}, Emeric and {Leung}, James K. and {Littlefair}, Stuart P. and {Munday}, James and {O'Brien}, Paul and {Parsons}, Steven G. and {Pelisoli}, Ingrid and {Sahman}, David I. and {Salvaterra}, Ruben and {Sbarufatti}, Boris and {Steeghs}, Danny and {Tagliaferri}, Gianpiero and {Th{\"o}ne}, Christina C. and {de Ugarte Postigo}, Antonio and {Kann}, David Alexander},
        title = "{Heavy-element production in a compact object merger observed by JWST}",
      journal = {\nat},
         year = 2024,
        month = feb,
       volume = {626},
       number = {8000},
        pages = {737-741},
          doi = {10.1038/s41586-023-06759-1},
archivePrefix = {arXiv},
       eprint = {2307.02098},
 primaryClass = {astro-ph.HE},
       adsurl = {https://ui.adsabs.harvard.edu/abs/2024Natur.626..737L}
}

@ARTICLE{Li.Paczynski:98,
       author = {{Li}, Li-Xin and {Paczy{\'n}ski}, Bohdan},
        title = "{Transient Events from Neutron Star Mergers}",
      journal = {\apjl},
         year = "1998",
        month = "Nov",
       volume = {507},
       number = {1},
        pages = {L59-L62},
          doi = {10.1086/311680},
archivePrefix = {arXiv},
       eprint = {astro-ph/9807272},
 primaryClass = {astro-ph},
       adsurl = {https://ui.adsabs.harvard.edu/abs/1998ApJ...507L..59L}
}

@ARTICLE{Metzger.etal:10,
       author = {{Metzger}, B.~D. and {Mart{\'\i}nez-Pinedo}, G. and {Darbha}, S. and {Quataert}, E. and {Arcones}, A. and {Kasen}, D. and {Thomas}, R. and {Nugent}, P. and {Panov}, I.~V. and {Zinner}, N.~T.},
        title = "{Electromagnetic counterparts of compact object mergers powered by the radioactive decay of r-process nuclei}",
      journal = {\mnras},
         year = 2010,
        month = aug,
       volume = {406},
       number = {4},
        pages = {2650-2662},
          doi = {10.1111/j.1365-2966.2010.16864.x},
archivePrefix = {arXiv},
       eprint = {1001.5029},
 primaryClass = {astro-ph.HE},
       adsurl = {https://ui.adsabs.harvard.edu/abs/2010MNRAS.406.2650M}
}

@ARTICLE{Perego.etal:17,
       author = {{Perego}, Albino and {Radice}, David and {Bernuzzi}, Sebastiano},
        title = "{AT 2017gfo: An Anisotropic and Three-component Kilonova Counterpart of GW170817}",
      journal = {\apjl},
         year = 2017,
        month = dec,
       volume = {850},
       number = {2},
          eid = {L37},
        pages = {L37},
          doi = {10.3847/2041-8213/aa9ab9},
archivePrefix = {arXiv},
       eprint = {1711.03982},
 primaryClass = {astro-ph.HE},
       adsurl = {https://ui.adsabs.harvard.edu/abs/2017ApJ...850L..37P}
}

@ARTICLE{Pian.etal:17,
       author = {{Pian}, E. and {D'Avanzo}, P. and {Benetti}, S. and {Branchesi}, M. and {Brocato}, E. and {Campana}, S. and {Cappellaro}, E. and {Covino}, S. and {D'Elia}, V. and {Fynbo}, J.~P.~U. and {Getman}, F. and {Ghirlanda}, G. and {Ghisellini}, G. and {Grado}, A. and {Greco}, G. and {Hjorth}, J. and {Kouveliotou}, C. and {Levan}, A. and {Limatola}, L. and {Malesani}, D. and {Mazzali}, P.~A. and {Melandri}, A. and {M{\o}ller}, P. and {Nicastro}, L. and {Palazzi}, E. and {Piranomonte}, S. and {Rossi}, A. and {Salafia}, O.~S. and {Selsing}, J. and {Stratta}, G. and {Tanaka}, M. and {Tanvir}, N.~R. and {Tomasella}, L. and {Watson}, D. and {Yang}, S. and {Amati}, L. and {Antonelli}, L.~A. and {Ascenzi}, S. and {Bernardini}, M.~G. and {Bo{\"e}r}, M. and {Bufano}, F. and {Bulgarelli}, A. and {Capaccioli}, M. and {Casella}, P. and {Castro-Tirado}, A.~J. and {Chassande-Mottin}, E. and {Ciolfi}, R. and {Copperwheat}, C.~M. and {Dadina}, M. and {De Cesare}, G. and {di Paola}, A. and {Fan}, Y.~Z. and {Gendre}, B. and {Giuffrida}, G. and {Giunta}, A. and {Hunt}, L.~K. and {Israel}, G.~L. and {Jin}, Z. -P. and {Kasliwal}, M.~M. and {Klose}, S. and {Lisi}, M. and {Longo}, F. and {Maiorano}, E. and {Mapelli}, M. and {Masetti}, N. and {Nava}, L. and {Patricelli}, B. and {Perley}, D. and {Pescalli}, A. and {Piran}, T. and {Possenti}, A. and {Pulone}, L. and {Razzano}, M. and {Salvaterra}, R. and {Schipani}, P. and {Spera}, M. and {Stamerra}, A. and {Stella}, L. and {Tagliaferri}, G. and {Testa}, V. and {Troja}, E. and {Turatto}, M. and {Vergani}, S.~D. and {Vergani}, D.},
        title = "{Spectroscopic identification of r-process nucleosynthesis in a double neutron-star merger}",
      journal = {\nat},
         year = 2017,
        month = nov,
       volume = {551},
       number = {7678},
        pages = {67-70},
          doi = {10.1038/nature24298},
archivePrefix = {arXiv},
       eprint = {1710.05858},
 primaryClass = {astro-ph.HE},
       adsurl = {https://ui.adsabs.harvard.edu/abs/2017Natur.551...67P}
}

@ARTICLE{Pognan.etal:22a,
       author = {{Pognan}, Quentin and {Jerkstrand}, Anders and {Grumer}, Jon},
        title = "{On the validity of steady-state for nebular phase kilonovae}",
      journal = {\mnras},
         year = 2022,
        month = mar,
       volume = {510},
       number = {3},
        pages = {3806-3837},
          doi = {10.1093/mnras/stab3674},
archivePrefix = {arXiv},
       eprint = {2112.07484},
 primaryClass = {astro-ph.HE},
       adsurl = {https://ui.adsabs.harvard.edu/abs/2022MNRAS.510.3806P}
}

@ARTICLE{Pognan.etal:22b,
       author = {{Pognan}, Quentin and {Jerkstrand}, Anders and {Grumer}, Jon},
        title = "{NLTE effects on kilonova expansion opacities}",
      journal = {\mnras},
         year = 2022,
        month = jul,
       volume = {513},
       number = {4},
        pages = {5174-5197},
          doi = {10.1093/mnras/stac1253},
archivePrefix = {arXiv},
       eprint = {2202.09245},
 primaryClass = {astro-ph.HE},
       adsurl = {https://ui.adsabs.harvard.edu/abs/2022MNRAS.513.5174P}
}

@ARTICLE{Pognan.etal:23,
       author = {{Pognan}, Quentin and {Grumer}, Jon and {Jerkstrand}, Anders and {Wanajo}, Shinya},
        title = "{NLTE spectra of kilonovae}",
      journal = {\mnras},
         year = 2023,
        month = dec,
       volume = {526},
       number = {4},
        pages = {5220-5248},
          doi = {10.1093/mnras/stad3106},
archivePrefix = {arXiv},
       eprint = {2309.01134},
 primaryClass = {astro-ph.HE},
       adsurl = {https://ui.adsabs.harvard.edu/abs/2023MNRAS.526.5220P}
}

@ARTICLE{Pognan.etal:25,
       author = {{Pognan}, Quentin and {Wu}, Meng-Ru and {Mart{\'\i}nez-Pinedo}, Gabriel and {da Silva}, Ricardo Ferreira and {Jerkstrand}, Anders and {Grumer}, Jon and {Fl{\"o}rs}, Andreas},
        title = "{Actinide signatures in low electron fraction kilonova ejecta}",
      journal = {\mnras},
         year = 2025,
        month = jan,
       volume = {536},
       number = {3},
        pages = {2973-2992},
          doi = {10.1093/mnras/stae2778},
archivePrefix = {arXiv},
       eprint = {2409.16210},
 primaryClass = {astro-ph.HE},
       adsurl = {https://ui.adsabs.harvard.edu/abs/2025MNRAS.536.2973P}
}

@ARTICLE{Rosswog.etal:99,
       author = {{Rosswog}, S. and {Liebend{\"o}rfer}, M. and {Thielemann}, F. -K. and
         {Davies}, M.~B. and {Benz}, W. and {Piran}, T.},
        title = "{Mass ejection in neutron star mergers}",
      journal = {\aap},
         year = "1999",
        month = "Jan",
       volume = {341},
        pages = {499-526},
archivePrefix = {arXiv},
       eprint = {astro-ph/9811367},
 primaryClass = {astro-ph},
       adsurl = {https://ui.adsabs.harvard.edu/abs/1999A&A...341..499R}
}

@ARTICLE{Rosswog.etal:13,
       author = {{Rosswog}, S. and {Piran}, T. and {Nakar}, E.},
        title = "{The multimessenger picture of compact object encounters: binary mergers versus dynamical collisions}",
      journal = {\mnras},
         year = 2013,
        month = apr,
       volume = {430},
       number = {4},
        pages = {2585-2604},
          doi = {10.1093/mnras/sts708},
archivePrefix = {arXiv},
       eprint = {1204.6240},
 primaryClass = {astro-ph.HE},
       adsurl = {https://ui.adsabs.harvard.edu/abs/2013MNRAS.430.2585R}
}

@ARTICLE{Sadeh:25,
       author = {{Sadeh}, Gilad},
        title = "{Inferring Kilonova Ejecta Photospheric Properties from Early Blackbody Spectra}",
      journal = {\apj},
         year = 2025,
        month = jul,
       volume = {988},
       number = {1},
          eid = {130},
        pages = {130},
          doi = {10.3847/1538-4357/adecee},
archivePrefix = {arXiv},
       eprint = {2503.04700},
 primaryClass = {astro-ph.HE},
       adsurl = {https://ui.adsabs.harvard.edu/abs/2025ApJ...988..130S}
}

@ARTICLE{Shibata.Hotokezaka:19,
       author = {{Shibata}, Masaru and {Hotokezaka}, Kenta},
        title = "{Merger and Mass Ejection of Neutron Star Binaries}",
      journal = {Annual Review of Nuclear and Particle Science},
         year = 2019,
        month = oct,
       volume = {69},
        pages = {41-64},
          doi = {10.1146/annurev-nucl-101918-023625},
archivePrefix = {arXiv},
       eprint = {1908.02350},
 primaryClass = {astro-ph.HE},
       adsurl = {https://ui.adsabs.harvard.edu/abs/2019ARNPS..69...41S}
}

@ARTICLE{Sippens.etal:25,
       author = {{Sippens Groenewegen}, Lieke and {Curtis}, Sanjana and {M{\"o}sta}, Philipp and {Kasen}, Daniel and {Brethauer}, Daniel},
        title = "{2D end-to-end modelling of kilonovae from binary neutron star merger remnants}",
      journal = {\mnras},
         year = 2025,
        month = nov,
       volume = {543},
       number = {3},
        pages = {2836-2854},
          doi = {10.1093/mnras/staf1529},
archivePrefix = {arXiv},
       eprint = {2508.00062},
 primaryClass = {astro-ph.HE},
       adsurl = {https://ui.adsabs.harvard.edu/abs/2025MNRAS.543.2836S}
}

@ARTICLE{Smartt.etal:17,
       author = {{Smartt}, S.~J. and {Chen}, T. -W. and {Jerkstrand}, A. and {Coughlin}, M. and {Kankare}, E. and {Sim}, S.~A. and {Fraser}, M. and {Inserra}, C. and {Maguire}, K. and {Chambers}, K.~C. and {Huber}, M.~E. and {Kr{\"u}hler}, T. and {Leloudas}, G. and {Magee}, M. and {Shingles}, L.~J. and {Smith}, K.~W. and {Young}, D.~R. and {Tonry}, J. and {Kotak}, R. and {Gal-Yam}, A. and {Lyman}, J.~D. and {Homan}, D.~S. and {Agliozzo}, C. and {Anderson}, J.~P. and {Angus}, C.~R. and {Ashall}, C. and {Barbarino}, C. and {Bauer}, F.~E. and {Berton}, M. and {Botticella}, M.~T. and {Bulla}, M. and {Bulger}, J. and {Cannizzaro}, G. and {Cano}, Z. and {Cartier}, R. and {Cikota}, A. and {Clark}, P. and {De Cia}, A. and {Della Valle}, M. and {Denneau}, L. and {Dennefeld}, M. and {Dessart}, L. and {Dimitriadis}, G. and {Elias-Rosa}, N. and {Firth}, R.~E. and {Flewelling}, H. and {Fl{\"o}rs}, A. and {Franckowiak}, A. and {Frohmaier}, C. and {Galbany}, L. and {Gonz{\'a}lez-Gait{\'a}n}, S. and {Greiner}, J. and {Gromadzki}, M. and {Guelbenzu}, A. Nicuesa and {Guti{\'e}rrez}, C.~P. and {Hamanowicz}, A. and {Hanlon}, L. and {Harmanen}, J. and {Heintz}, K.~E. and {Heinze}, A. and {Hernandez}, M. -S. and {Hodgkin}, S.~T. and {Hook}, I.~M. and {Izzo}, L. and {James}, P.~A. and {Jonker}, P.~G. and {Kerzendorf}, W.~E. and {Klose}, S. and {Kostrzewa-Rutkowska}, Z. and {Kowalski}, M. and {Kromer}, M. and {Kuncarayakti}, H. and {Lawrence}, A. and {Lowe}, T.~B. and {Magnier}, E.~A. and {Manulis}, I. and {Martin-Carrillo}, A. and {Mattila}, S. and {McBrien}, O. and {M{\"u}ller}, A. and {Nordin}, J. and {O'Neill}, D. and {Onori}, F. and {Palmerio}, J.~T. and {Pastorello}, A. and {Patat}, F. and {Pignata}, G. and {Podsiadlowski}, Ph. and {Pumo}, M.~L. and {Prentice}, S.~J. and {Rau}, A. and {Razza}, A. and {Rest}, A. and {Reynolds}, T. and {Roy}, R. and {Ruiter}, A.~J. and {Rybicki}, K.~A. and {Salmon}, L. and {Schady}, P. and {Schultz}, A.~S.~B. and {Schweyer}, T. and {Seitenzahl}, I.~R. and {Smith}, M. and {Sollerman}, J. and {Stalder}, B. and {Stubbs}, C.~W. and {Sullivan}, M. and {Szegedi}, H. and {Taddia}, F. and {Taubenberger}, S. and {Terreran}, G. and {van Soelen}, B. and {Vos}, J. and {Wainscoat}, R.~J. and {Walton}, N.~A. and {Waters}, C. and {Weiland}, H. and {Willman}, M. and {Wiseman}, P. and {Wright}, D.~E. and {Wyrzykowski}, {\L}. and {Yaron}, O.},
        title = "{A kilonova as the electromagnetic counterpart to a gravitational-wave source}",
      journal = {\nat},
         year = 2017,
        month = nov,
       volume = {551},
       number = {7678},
        pages = {75-79},
          doi = {10.1038/nature24303},
archivePrefix = {arXiv},
       eprint = {1710.05841},
 primaryClass = {astro-ph.HE},
       adsurl = {https://ui.adsabs.harvard.edu/abs/2017Natur.551...75S}
}

@ARTICLE{Sneppen.etal:23,
       author = {{Sneppen}, Albert and {Watson}, Darach and {Bauswein}, Andreas and {Just}, Oliver and {Kotak}, Rubina and {Nakar}, Ehud and {Poznanski}, Dovi and {Sim}, Stuart},
        title = "{Spherical symmetry in the kilonova AT2017gfo/GW170817}",
      journal = {\nat},
         year = 2023,
        month = feb,
       volume = {614},
       number = {7948},
        pages = {436-439},
          doi = {10.1038/s41586-022-05616-x},
archivePrefix = {arXiv},
       eprint = {2302.06621},
 primaryClass = {astro-ph.HE},
       adsurl = {https://ui.adsabs.harvard.edu/abs/2023Natur.614..436S}
}

@ARTICLE{Sneppen.Watson:23,
       author = {{Sneppen}, Albert and {Watson}, Darach},
        title = "{Discovery of a 760 nm P Cygni line in AT2017gfo: Identification of yttrium in the kilonova photosphere}",
      journal = {\aap},
         year = 2023,
        month = jul,
       volume = {675},
          eid = {A194},
        pages = {A194},
          doi = {10.1051/0004-6361/202346421},
archivePrefix = {arXiv},
       eprint = {2306.14942},
 primaryClass = {astro-ph.HE},
       adsurl = {https://ui.adsabs.harvard.edu/abs/2023A&A...675A.194S}
}

@ARTICLE{Sneppen.etal:24a,
       author = {{Sneppen}, Albert and {Watson}, Darach and {Gillanders}, James H. and {Heintz}, Kasper E.},
        title = "{Rapid kilonova evolution: Recombination and reverberation effects}",
      journal = {\aap},
         year = 2024,
        month = aug,
       volume = {688},
          eid = {A95},
        pages = {A95},
          doi = {10.1051/0004-6361/202348758},
       adsurl = {https://ui.adsabs.harvard.edu/abs/2024A&A...688A..95S}
}

@ARTICLE{Sneppen.etal:24b,
       author = {{Sneppen}, Albert and {Watson}, Darach and {Damgaard}, Rasmus and {Heintz}, Kasper E. and {Vieira}, Nicholas and {V{\"a}is{\"a}nen}, Petri and {Mahoro}, Antoine},
        title = "{Emergence hour-by-hour of r-process features in the kilonova AT2017gfo}",
      journal = {\aap},
         year = 2024,
        month = oct,
       volume = {690},
          eid = {A398},
        pages = {A398},
          doi = {10.1051/0004-6361/202450317},
archivePrefix = {arXiv},
       eprint = {2404.08730},
 primaryClass = {astro-ph.HE},
       adsurl = {https://ui.adsabs.harvard.edu/abs/2024A&A...690A.398S}
}

@ARTICLE{Sneppen.etal:25,
       author = {{Sneppen}, Albert and {Just}, Oliver and {Bauswein}, Andreas and {Damgaard}, Rasmus and {Watson}, Darach and {Shingles}, Luke J. and {Collins}, Christine E. and {Sim}, Stuart A. and {Xiong}, Zewei and {Martinez-Pinedo}, Gabriel and {Soultanis}, Theodoros and {Vijayan}, Vimal},
        title = "{Helium as an Indicator of the Neutron-Star Merger Remnant Lifetime and its Potential for Equation of State Constraints}",
      journal = {arXiv e-prints},
         year = 2024,
        month = nov,
          eid = {arXiv:2411.03427},
        pages = {arXiv:2411.03427},
          doi = {10.48550/arXiv.2411.03427},
archivePrefix = {arXiv},
       eprint = {2411.03427},
 primaryClass = {astro-ph.HE},
       adsurl = {https://ui.adsabs.harvard.edu/abs/2024arXiv241103427S}
}

@ARTICLE{Singh.etal:25,
       author = {{Singh}, S. and {Harman}, Z. and {Keitel}, C.~H.},
        title = "{Dielectronic recombination studies of ions relevant to kilonovae and nonlocal thermodynamic equilibrium plasma}",
      journal = {\aap},
         year = 2025,
        month = aug,
       volume = {700},
          eid = {A110},
        pages = {A110},
          doi = {10.1051/0004-6361/202555098},
archivePrefix = {arXiv},
       eprint = {2504.06639},
 primaryClass = {astro-ph.HE},
       adsurl = {https://ui.adsabs.harvard.edu/abs/2025A&A...700A.110S}
}

@ARTICLE{Symbalisty.Schramm:82,
       author = {{Symbalisty}, E. and {Schramm}, D.~N.},
        title = "{Neutron Star Collisions and the r-Process}",
      journal = {\aplett},
         year = 1982,
        month = jan,
       volume = {22},
        pages = {143},
       adsurl = {https://ui.adsabs.harvard.edu/abs/1982ApL....22..143S}
}

@ARTICLE{Tanaka.etal:20,
       author = {{Tanaka}, Masaomi and {Kato}, Daiji and {Gaigalas}, Gediminas and {Kawaguchi}, Kyohei},
        title = "{Systematic opacity calculations for kilonovae}",
      journal = {\mnras},
         year = 2020,
        month = aug,
       volume = {496},
       number = {2},
        pages = {1369-1392},
          doi = {10.1093/mnras/staa1576},
archivePrefix = {arXiv},
       eprint = {1906.08914},
 primaryClass = {astro-ph.HE},
       adsurl = {https://ui.adsabs.harvard.edu/abs/2020MNRAS.496.1369T}
}

@ARTICLE{Tarumi.etal:23,
       author = {{Tarumi}, Yuta and {Hotokezaka}, Kenta and {Domoto}, Nanae and {Tanaka}, Masaomi},
        title = "{Non-LTE analysis for Helium and Strontium lines in the kilonova AT2017gfo}",
      journal = {arXiv e-prints},
         year = 2023,
        month = feb,
          eid = {arXiv:2302.13061},
        pages = {arXiv:2302.13061},
          doi = {10.48550/arXiv.2302.13061},
archivePrefix = {arXiv},
       eprint = {2302.13061},
 primaryClass = {astro-ph.HE},
       adsurl = {https://ui.adsabs.harvard.edu/abs/2023arXiv230213061T}
}

@ARTICLE{Vieira.etal:23a,
       author = {{Vieira}, Nicholas and {Ruan}, John J. and {Haggard}, Daryl and {Ford}, Nicole and {Drout}, Maria R. and {Fern{\'a}ndez}, Rodrigo and {Badnell}, N.~R.},
        title = "{Spectroscopic r-Process Abundance Retrieval for Kilonovae. I. The Inferred Abundance Pattern of Early Emission from GW170817}",
      journal = {\apj},
         year = 2023,
        month = feb,
       volume = {944},
       number = {2},
          eid = {123},
        pages = {123},
          doi = {10.3847/1538-4357/acae72},
archivePrefix = {arXiv},
       eprint = {2209.06951},
 primaryClass = {astro-ph.HE},
       adsurl = {https://ui.adsabs.harvard.edu/abs/2023ApJ...944..123V}
}

@ARTICLE{Vieira.etal:24,
       author = {{Vieira}, Nicholas and {Ruan}, John J. and {Haggard}, Daryl and {Ford}, Nicole M. and {Drout}, Maria R. and {Fern{\'a}ndez}, Rodrigo},
        title = "{Spectroscopic r-process Abundance Retrieval for Kilonovae. II. Lanthanides in the Inferred Abundance Patterns of Multicomponent Ejecta from the GW170817 Kilonova}",
      journal = {\apj},
         year = 2024,
        month = feb,
       volume = {962},
       number = {1},
          eid = {33},
        pages = {33},
          doi = {10.3847/1538-4357/ad1193},
archivePrefix = {arXiv},
       eprint = {2308.16796},
 primaryClass = {astro-ph.HE},
       adsurl = {https://ui.adsabs.harvard.edu/abs/2024ApJ...962...33V}
}

@ARTICLE{Vieira.etal:26,
       author = {{Vieira}, Nicholas and {Ruan}, John J. and {Haggard}, Daryl and {Drout}, Maria R. and {Fern{\'a}ndez}, Rodrigo},
        title = "{Spectroscopic r-process Abundance Retrieval for Kilonovae. III. Linking Spectral and Light-curve Modeling of the GW170817 Kilonova}",
      journal = {\apj},
         year = 2026,
        month = jan,
       volume = {997},
       number = {1},
          eid = {92},
        pages = {92},
          doi = {10.3847/1538-4357/ae1fd5},
archivePrefix = {arXiv},
       eprint = {2504.10696},
 primaryClass = {astro-ph.HE},
       adsurl = {https://ui.adsabs.harvard.edu/abs/2026ApJ...997...92V}
}

@ARTICLE{Villar.etal:17,
       author = {{Villar}, V.~A. and {Guillochon}, J. and {Berger}, E. and {Metzger}, B.~D. and {Cowperthwaite}, P.~S. and {Nicholl}, M. and {Alexander}, K.~D. and {Blanchard}, P.~K. and {Chornock}, R. and {Eftekhari}, T. and {Fong}, W. and {Margutti}, R. and {Williams}, P.~K.~G.},
        title = "{The Combined Ultraviolet, Optical, and Near-infrared Light Curves of the Kilonova Associated with the Binary Neutron Star Merger GW170817: Unified Data Set, Analytic Models, and Physical Implications}",
      journal = {\apjl},
         year = 2017,
        month = dec,
       volume = {851},
       number = {1},
          eid = {L21},
        pages = {L21},
          doi = {10.3847/2041-8213/aa9c84},
archivePrefix = {arXiv},
       eprint = {1710.11576},
 primaryClass = {astro-ph.HE},
       adsurl = {https://ui.adsabs.harvard.edu/abs/2017ApJ...851L..21V}
}

@ARTICLE{Villar.etal:18,
       author = {{Villar}, V.~A. and {Cowperthwaite}, P.~S. and {Berger}, E. and {Blanchard}, P.~K. and {Gomez}, S. and {Alexander}, K.~D. and {Margutti}, R. and {Chornock}, R. and {Eftekhari}, T. and {Fazio}, G.~G. and {Guillochon}, J. and {Hora}, J.~L. and {Nicholl}, M. and {Williams}, P.~K.~G.},
        title = "{Spitzer Space Telescope Infrared Observations of the Binary Neutron Star Merger GW170817}",
      journal = {\apjl},
         year = 2018,
        month = jul,
       volume = {862},
       number = {1},
          eid = {L11},
        pages = {L11},
          doi = {10.3847/2041-8213/aad281},
archivePrefix = {arXiv},
       eprint = {1805.08192},
 primaryClass = {astro-ph.HE},
       adsurl = {https://ui.adsabs.harvard.edu/abs/2018ApJ...862L..11V}
}

@ARTICLE{Wanajo:18,
       author = {{Wanajo}, Shinya},
        title = "{Physical Conditions for the r-process. I. Radioactive Energy Sources of Kilonovae}",
      journal = {\apj},
         year = 2018,
        month = nov,
       volume = {868},
       number = {1},
          eid = {65},
        pages = {65},
          doi = {10.3847/1538-4357/aae0f2},
archivePrefix = {arXiv},
       eprint = {1808.03763},
 primaryClass = {astro-ph.HE},
       adsurl = {https://ui.adsabs.harvard.edu/abs/2018ApJ...868...65W}
}

@ARTICLE{Watson.etal:2019,
       author = {{Watson}, Darach and {Hansen}, Camilla J. and {Selsing}, Jonatan and {Koch}, Andreas and {Malesani}, Daniele B. and {Andersen}, Anja C. and {Fynbo}, Johan P.~U. and {Arcones}, Almudena and {Bauswein}, Andreas and {Covino}, Stefano and {Grado}, Aniello and {Heintz}, Kasper E. and {Hunt}, Leslie and {Kouveliotou}, Chryssa and {Leloudas}, Giorgos and {Levan}, Andrew J. and {Mazzali}, Paolo and {Pian}, Elena},
        title = "{Identification of strontium in the merger of two neutron stars}",
      journal = {\nat},
         year = 2019,
        month = oct,
       volume = {574},
       number = {7779},
        pages = {497-500},
          doi = {10.1038/s41586-019-1676-3},
archivePrefix = {arXiv},
       eprint = {1910.10510},
 primaryClass = {astro-ph.HE},
       adsurl = {https://ui.adsabs.harvard.edu/abs/2019Natur.574..497W}
}

@ARTICLE{Waxman.etal:18,
       author = {{Waxman}, Eli and {Ofek}, Eran O. and {Kushnir}, Doron and {Gal-Yam}, Avishay},
        title = "{Constraints on the ejecta of the GW170817 neutron star merger from its electromagnetic emission}",
      journal = {\mnras},
         year = 2018,
        month = dec,
       volume = {481},
       number = {3},
        pages = {3423-3441},
          doi = {10.1093/mnras/sty2441},
archivePrefix = {arXiv},
       eprint = {1711.09638},
 primaryClass = {astro-ph.HE},
       adsurl = {https://ui.adsabs.harvard.edu/abs/2018MNRAS.481.3423W}
}

@ARTICLE{Waxman.etal:19,
       author = {{Waxman}, Eli and {Ofek}, Eran O. and {Kushnir}, Doron},
        title = "{Late-time Kilonova Light Curves and Implications to GW170817}",
      journal = {\apj},
         year = 2019,
        month = jun,
       volume = {878},
       number = {2},
          eid = {93},
        pages = {93},
          doi = {10.3847/1538-4357/ab1f71},
archivePrefix = {arXiv},
       eprint = {1902.01197},
 primaryClass = {astro-ph.HE},
       adsurl = {https://ui.adsabs.harvard.edu/abs/2019ApJ...878...93W}
}

@ARTICLE{Banik.etal:14,
       author = {{Banik}, Sarmistha and {Hempel}, Matthias and {Bandyopadhyay}, Debades},
        title = "{New Hyperon Equations of State for Supernovae and Neutron Stars in Density-dependent Hadron Field Theory}",
      journal = {\apjs},
         year = 2014,
        month = oct,
       volume = {214},
       number = {2},
          eid = {22},
        pages = {22},
          doi = {10.1088/0067-0049/214/2/22},
archivePrefix = {arXiv},
       eprint = {1404.6173},
 primaryClass = {astro-ph.HE},
       adsurl = {https://ui.adsabs.harvard.edu/abs/2014ApJS..214...22B}
}

@ARTICLE{Bromley.etal:23,
       author = {{Bromley}, S.~J. and {McCann}, M. and {Loch}, S.~D. and {Ballance}, C.~P.},
        title = "{Electron-impact Excitation of Pt I-III: The Importance of Metastables and Collision Processes in Neutron Star Merger and Laboratory Plasmas}",
      journal = {\apjs},
         year = 2023,
        month = sep,
       volume = {268},
       number = {1},
          eid = {22},
        pages = {22},
          doi = {10.3847/1538-4365/ace5a1},
       adsurl = {https://ui.adsabs.harvard.edu/abs/2023ApJS..268...22B}
}

@BOOK{Cowan_atom81,
       author = {{Cowan}, Robert D.},
        title = "{The theory of atomic structure and spectra}",
         year = 1981,
       adsurl = {https://ui.adsabs.harvard.edu/abs/1981tass.book.....C}
}

@ARTICLE{Deprince.etal:25,
       author = {{Deprince}, J. and {Wagle}, G. and {Ben Nasr}, S. and {Carvajal Gallego}, H. and {Godefroid}, M. and {Goriely}, S. and {Just}, O. and {Palmeri}, P. and {Quinet}, P. and {Van Eck}, S.},
        title = "{Kilonova ejecta opacity inferred from new large-scale HFR atomic calculations in all elements between Ca (Z = 20) and Lr (Z = 103)}",
      journal = {\aap},
         year = 2025,
        month = apr,
       volume = {696},
          eid = {A32},
        pages = {A32},
          doi = {10.1051/0004-6361/202452967},
archivePrefix = {arXiv},
       eprint = {2412.16688},
 primaryClass = {astro-ph.HE},
       adsurl = {https://ui.adsabs.harvard.edu/abs/2025A&A...696A..32D}
}

@ARTICLE{Dunleavy.etal:22,
       author = {{Dunleavy}, N.~L. and {Ballance}, C.~P. and {Ramsbottom}, C.~A. and {Johnson}, C.~A. and {Loch}, S.~D. and {Ennis}, D.~A.},
        title = "{A Dirac R-matrix calculation for the electron-impact excitation of W$^{+}$}",
      journal = {Journal of Physics B Atomic Molecular Physics},
         year = 2022,
        month = sep,
       volume = {55},
       number = {17},
          eid = {175002},
        pages = {175002},
          doi = {10.1088/1361-6455/ac8089},
       adsurl = {https://ui.adsabs.harvard.edu/abs/2022JPhB...55q5002D}
}

@ARTICLE{Gaigalas.etal:20,
       author = {{Gaigalas}, Gediminas and {Rynkun}, Pavel and {Rad{\v{z}}i{\={u}}t{\.{e}}}, Laima and {Kato}, Daiji and {Tanaka}, Masaomi and {J{\"o}nsson}, P.},
        title = "{Energy Level Structure and Transition Data of Er$^{2+}$}",
      journal = {\apjs},
         year = 2020,
        month = may,
       volume = {248},
       number = {1},
          eid = {13},
        pages = {13},
          doi = {10.3847/1538-4365/ab881a},
archivePrefix = {arXiv},
       eprint = {2002.07615},
 primaryClass = {physics.atom-ph},
       adsurl = {https://ui.adsabs.harvard.edu/abs/2020ApJS..248...13G}
}

@ARTICLE{Gu:08,
       author = {{Gu}, M.~F.},
        title = "{The flexible atomic code}",
      journal = {Canadian Journal of Physics},
         year = 2008,
        month = jan,
       volume = {86},
       number = {5},
        pages = {675-689},
          doi = {10.1139/P07-197},
       adsurl = {https://ui.adsabs.harvard.edu/abs/2008CaJPh..86..675G}
}

@ARTICLE{Hummer.Storey:98,
       author = {{Hummer}, D.~G. and {Storey}, P.~J.},
        title = "{Recombination of helium-like ions - I. Photoionization cross-sections and total recombination and cooling coefficients for atomic helium}",
      journal = {\mnras},
         year = 1998,
        month = jul,
       volume = {297},
       number = {4},
        pages = {1073-1078},
          doi = {10.1046/j.1365-8711.1998.2970041073.x},
       adsurl = {https://ui.adsabs.harvard.edu/abs/1998MNRAS.297.1073H}
}

@Misc{NIST_ASD,
author = {A.~Kramida and {Yu.~Ralchenko} and
J.~Reader and {and NIST ASD Team}},
HOWPUBLISHED = {{NIST Atomic Spectra Database
(ver. 5.8), [Online]. Available:
{\tt{https://physics.nist.gov/asd}} [2021, September 22].
National Institute of Standards and Technology,
Gaithersburg, MD.}},
year = {2020},
}

@ARTICLE{Lotz:67,
       author = {{Lotz}, Wolfgang},
        title = "{An empirical formula for the electron-impact ionization cross-section}",
      journal = {Zeitschrift fur Physik},
         year = 1967,
        month = apr,
       volume = {206},
       number = {2},
        pages = {205-211},
          doi = {10.1007/BF01325928},
       adsurl = {https://ui.adsabs.harvard.edu/abs/1967ZPhy..206..205L}
}

@ARTICLE{McCann.etal:24,
       author = {{McCann}, M. and {Ballance}, C.~P. and {Loch}, S.~D. and {Ennis}, D.~A.},
        title = "{Electron-impact excitation data for W$^{2+}$ in support of tungsten spectroscopy and re-deposition measurements for magnetically-confined plasmas}",
      journal = {Journal of Physics B Atomic Molecular Physics},
         year = 2024,
        month = dec,
       volume = {57},
       number = {23},
          eid = {235202},
        pages = {235202},
          doi = {10.1088/1361-6455/ad7cad},
       adsurl = {https://ui.adsabs.harvard.edu/abs/2024JPhB...57w5202M}
}

@ARTICLE{McCann.etal:25,
       author = {{McCann}, M. and {Mulholland}, L.~P. and {Xiong}, Z. and {Ramsbottom}, C.~A. and {Ballance}, C.~P. and {Just}, O. and {Bauswein}, A. and {Mart{\'\i}nez-Pinedo}, G. and {McNeill}, F. and {Sim}, S.~A.},
        title = "{Luminosity predictions for the first three ionisation stages of W, Pt and Au to probe potential sources of emission in kilonova}",
      journal = {\mnras},
         year = 2025,
        month = feb,
          doi = {10.1093/mnras/staf283},
archivePrefix = {arXiv},
       eprint = {2411.16476},
 primaryClass = {physics.atom-ph},
       adsurl = {https://ui.adsabs.harvard.edu/abs/2025MNRAS.tmp..261M}
}

@ARTICLE{Kato.etal:24,
       author = {{Kato}, Daiji and {Tanaka}, Masaomi and {Gaigalas}, Gediminas and {Kitovien{\.{e}}}, Laima and {Rynkun}, Pavel},
        title = "{Systematic opacity calculations for kilonovae - II. Improved atomic data for singly ionized lanthanides}",
      journal = {\mnras},
         year = 2024,
        month = dec,
       volume = {535},
       number = {3},
        pages = {2670-2686},
          doi = {10.1093/mnras/stae2504},
archivePrefix = {arXiv},
       eprint = {2501.13286},
 primaryClass = {astro-ph.HE},
       adsurl = {https://ui.adsabs.harvard.edu/abs/2024MNRAS.535.2670K}
}

@ARTICLE{Mulholland.etal:24a,
       author = {{Mulholland}, L.~P. and {McElroy}, N.~E. and {McNeill}, F.~L. and {Sim}, S.~A. and {Ballance}, C.~P. and {Ramsbottom}, C.~A.},
        title = "{New radiative and collisional atomic data for Sr II and Y II with application to Kilonova modelling}",
      journal = {\mnras},
         year = 2024,
        month = aug,
       volume = {532},
       number = {2},
        pages = {2289-2308},
          doi = {10.1093/mnras/stae1615},
archivePrefix = {arXiv},
       eprint = {2407.01398},
 primaryClass = {astro-ph.HE},
       adsurl = {https://ui.adsabs.harvard.edu/abs/2024MNRAS.532.2289M}
}

@ARTICLE{Mulholland.etal:24b,
       author = {{Mulholland}, L.~P. and {McNeill}, F. and {Sim}, S.~A. and {Ballance}, C.~P. and {Ramsbottom}, C.~A.},
        title = "{Collisional and radiative data for tellurium ions in kilonovae modelling and laboratory benchmarks}",
      journal = {\mnras},
         year = 2024,
        month = nov,
       volume = {534},
       number = {4},
        pages = {3423-3438},
          doi = {10.1093/mnras/stae2331},
archivePrefix = {arXiv},
       eprint = {2410.05958},
 primaryClass = {astro-ph.SR},
       adsurl = {https://ui.adsabs.harvard.edu/abs/2024MNRAS.534.3423M}
}

@ARTICLE{Nahar:96,
       author = {{Nahar}, Sultana N.},
        title = "{Total electron-ion recombination of Fe III}",
      journal = {\pra},
         year = 1996,
        month = apr,
       volume = {53},
       number = {4},
        pages = {2417-2424},
          doi = {10.1103/PhysRevA.53.2417},
       adsurl = {https://ui.adsabs.harvard.edu/abs/1996PhRvA..53.2417N}
}

@ARTICLE{Nahar:97a,
       author = {{Nahar}, Sultana N.},
        title = "{Electron-ion recombination of FetII}",
      journal = {\pra},
         year = 1997,
        month = mar,
       volume = {55},
       number = {3},
        pages = {1980-1987},
          doi = {10.1103/PhysRevA.55.1980},
       adsurl = {https://ui.adsabs.harvard.edu/abs/1997PhRvA..55.1980N}
}

@ARTICLE{Nahar.etal:97b,
       author = {{Nahar}, Sultana N. and {Bautista}, Manuel A. and {Pradhan}, Anil K.},
        title = "{Electron-Ion Recombination of Neutral Iron}",
      journal = {\apj},
         year = 1997,
        month = apr,
       volume = {479},
       number = {1},
        pages = {497-503},
          doi = {10.1086/303874},
       adsurl = {https://ui.adsabs.harvard.edu/abs/1997ApJ...479..497N}
}

@ARTICLE{Pequignot.Aldrovandi:86,
       author = {{Pequignot}, D. and {Aldrovandi}, S.~M.~V.},
        title = "{The ionization balance in HI regions.}",
      journal = {\aap},
         year = 1986,
        month = jun,
       volume = {161},
        pages = {169-176},
       adsurl = {https://ui.adsabs.harvard.edu/abs/1986A&A...161..169P}
}

@ARTICLE{Regemorter:62,
       author = {{van Regemorter}, Henri},
        title = "{Rate of Collisional Excitation in Stellar Atmospheres.}",
      journal = {\apj},
         year = 1962,
        month = nov,
       volume = {136},
        pages = {906},
          doi = {10.1086/147445},
       adsurl = {https://ui.adsabs.harvard.edu/abs/1962ApJ...136..906V}
}

@ARTICLE{Shull.Steenberg:82,
       author = {{Shull}, J.~M. and {van Steenberg}, M.},
        title = "{The ionization equilibrium of astrophysically abundant elements.}",
      journal = {\apjs},
         year = 1982,
        month = jan,
       volume = {48},
        pages = {95-107},
          doi = {10.1086/190769},
       adsurl = {https://ui.adsabs.harvard.edu/abs/1982ApJS...48...95S}
}

@ARTICLE{Smyth.etal:18,
       author = {{Smyth}, R.~T. and {Ballance}, C.~P. and {Ramsbottom}, C.~A. and {Johnson}, C.~A. and {Ennis}, D.~A. and {Loch}, S.~D.},
        title = "{Dirac R -matrix calculations for the electron-impact excitation of neutral tungsten providing noninvasive diagnostics for magnetic confinement fusion}",
      journal = {\pra},
         year = 2018,
        month = may,
       volume = {97},
       number = {5},
          eid = {052705},
        pages = {052705},
          doi = {10.1103/PhysRevA.97.052705},
archivePrefix = {arXiv},
       eprint = {1805.02757},
 primaryClass = {physics.atom-ph},
       adsurl = {https://ui.adsabs.harvard.edu/abs/2018PhRvA..97e2705S}
}

@ARTICLE{Sterling.etal:17,
       author = {{Sterling}, N.~C. and {Madonna}, S. and {Butler}, K. and {Garc{\'\i}a-Rojas}, J. and {Mashburn}, A.~L. and {Morisset}, C. and {Luridiana}, V. and {Roederer}, I.~U.},
        title = "{Identification of Near-infrared [Se III] and [Kr VI] Emission Lines in Planetary Nebulae}",
      journal = {\apj},
         year = 2017,
        month = may,
       volume = {840},
       number = {2},
          eid = {80},
        pages = {80},
          doi = {10.3847/1538-4357/aa6c28},
archivePrefix = {arXiv},
       eprint = {1704.00741},
 primaryClass = {astro-ph.SR},
       adsurl = {https://ui.adsabs.harvard.edu/abs/2017ApJ...840...80S}
}

@ARTICLE{Tauheed.Hala:12,
       author = {{Tauheed}, A. and {Hala}},
        title = "{Revised and extended analysis of doubly ionized selenium: Se III}",
      journal = {\physscr},
         year = 2012,
        month = feb,
       volume = {85},
       number = {2},
          eid = {025304},
        pages = {025304},
          doi = {10.1088/0031-8949/85/02/025304},
       adsurl = {https://ui.adsabs.harvard.edu/abs/2012PhyS...85b5304T}
}

@ARTICLE{Verner.Ferland:96,
       author = {{Verner}, D.~A. and {Ferland}, G.~J.},
        title = "{Atomic Data for Astrophysics. I. Radiative Recombination Rates for H-like, He-like, Li-like, and Na-like Ions over a Broad Range of Temperature}",
      journal = {\apjs},
         year = 1996,
        month = apr,
       volume = {103},
        pages = {467},
          doi = {10.1086/192284},
archivePrefix = {arXiv},
       eprint = {astro-ph/9509083},
 primaryClass = {astro-ph},
       adsurl = {https://ui.adsabs.harvard.edu/abs/1996ApJS..103..467V}
}




\appendix

\section{Atomic data}
\label{app:atomic_data}

In this appendix, we summarise the main changes to the atomic data in \textsc{sumo} made since the last publication \citep[][]{Pognan.etal:25}, and also describe the key differences in treatment of processes between light elements ($Z \leq 28$) included in the models, and r-process elements which have been previously described. Four light elements are included in the models: $_{2}$He, $_{22}$Ti, $_{26}$Fe, and $_{28}$Ni. The atomic level and line lists for these species come from various well known and verified sources, either from NIST \citep[][]{NIST_ASD}, or the Kurucz database \citep[][]{Kurucz:18}, with additional information such as collision strengths coming from diverse sources where available \citep[see details within][]{Jerkstrand.etal:11,Jerkstrand.etal:12,Jerkstrand.etal:15}. Cross-sections and rates for the various NLTE processes modelled within \textsc{sumo} for these species likewise originate from several sources, all detailed in the previously cited works. We describe in the next paragraphs the key differences between r-process species and these light elements that are most relevant to this study.

\begin{table*}
    \centering
    \setlength\tabcolsep{0.4cm}
    \begin{tabular}{cccc}
    \hline \hline
    Species & Number of levels calibrated & Calibrated up to E [$\mathrm{cm^{-1}}$] & Transitions with collision strengths \\
    \hline
    Se\,\textsc{i} & $44^a$ & 71659 & -- \\
    Se\,\textsc{iii} & $57^{a,b}$ & 198010 & $9^{c}$\\
    Br\,\textsc{i} & $50^a$ & 86279 & -- \\
    Br\,\textsc{ii} & $10^a$ & 100234 & -- \\
    Kr\,\textsc{ii} & $2^{a}$ & 5370 & -- \\
    Rb\,\textsc{i} & $17^{a}$ & 28689 & -- \\
    Sr\,\textsc{ii}* & $27^{d}$ & 78702 & $279^{d}$ \\
    Y\,\textsc{i} & $107^{a}$ & 42098 & -- \\
    Y\,\textsc{ii}* & $96^{d}$ & 72079 & $3700^d$ \\
    Y\,\textsc{iii} & $16^{a}$ & 123192 & -- \\
    Zr\,\textsc{i} & $81^{a}$ & 23489 & -- \\
    Zr\,\textsc{ii} & $4^a$ & 1322 & -- \\
    Zr\,\textsc{iii} & $3^a$ & 1486 & -- \\
    Zr\,\textsc{iv} & $14^a$ & 197930 & -- \\
    Pd\,\textsc{i} & $55^a$ & 56544 & -- \\
    Te\,\textsc{i}* & $30^{e}$ & 63610 & $435^e$ \\
    Te\,\textsc{ii}* & $50^e$ & 106119 & $1225^e$ \\
    Te\,\textsc{iii}* & $40^e$ & 142982 & $780^e$ \\
    Ce\,\textsc{iii} & $70^a$ & 90223 & -- \\
    Nd\,\textsc{ii} & $364^a$ & 30037 & -- \\
    Nd\,\textsc{iii} & $15^a$ & 16938 & -- \\
    Sm\,\textsc{i} & $153^a$ & 30931 & -- \\
    Sm\,\textsc{iii} & $17^a$ & 30048 & -- \\
    Dy\,\textsc{i} & $197^a$ & 30621 & -- \\
    Dy\,\textsc{ii} & $155^a$ & 30661 & -- \\
    Er\,\textsc{i} & $205^a$ & 30088 & -- \\
    Tm\,\textsc{i} & $220^a$ & 52003 & -- \\
    Yb\,\textsc{ii} & $67^a$ & 50468 & -- \\
    W\,\textsc{i}* & $250^f$ & 51123 & $31124^f$ \\ 
    W\,\textsc{ii}* & $450^g$ & 88848 & $101024^g$ \\
    W\,\textsc{iii}* & $300^h$ & 122124 & $44849^h$ \\
    Pt\,\textsc{i}* & $157^i$ & 72292 & $9415^i$ \\
    Pt\,\textsc{ii}* & $450^i$ & 146839 & $73415^i$ \\
    Pt\,\textsc{iii}* & $600^i$ & 192204 & $79409^i$ \\
    \hline \hline
    \end{tabular}
    \caption{Summary of calibrated atomic data used in this study. The column for transitions with collision strengths refers to Maxwellian averaged thermal collision strengths as described in Section \ref{subsec:RTsim}, with the number corresponding to the first n-transitions. Ions with an asterisk have had the original \textsc{fac} data replaced by r-matrix calculations from the corresponding source. The data sources are as follows: a:NIST \citet{NIST_ASD}, b:\citet{Tauheed.Hala:12}, c:\citet{Sterling.etal:17}, d:\citet{Mulholland.etal:24a}, e:\citet{Mulholland.etal:24b}, f:\citet{Smyth.etal:18}, g:\citet{Dunleavy.etal:22}, h: \citet{McCann.etal:24}, i:\citet{Bromley.etal:23}.} 
    \label{tab:atomic_data}
\end{table*}

A key difference between the light and r-process elements included in this study are the ionisation states that are included in the modelling. As the r-process elements all come from a homogeneous data set calculated by \textsc{fac}, data up to the third ionisation state is available. However, this is not the case for the light elements, with $_{2}$He naturally only being neutral and singly ionized, while the other three light species only have data included up to the doubly ionized stage. As their usage was originally intended for supernova studies, the relatively lower energy of nebular phase supernovae, alongside the high ionisation threshold of the doubly ionized species of $\sim 30~$eV meant that triply ionized species would only be present in trace amounts if at all. We find that this is not true in the KN case, as mentioned in the main text, and that triply ionized $_{22}$Ti, $_{26}$Fe, and $_{28}$Ni do appear in the ionisation structure calculation. Since no atomic data is included for these species, they do not actively emit or absorb, and are simply considered for the purpose of ionisation structure calculation. This leads to some inconsistency in the emergent spectrum, and we find that at least Ti\,\textsc{iv} and Ni\,\textsc{iv} have some relevant IR transitions that may have yielded features in the model spectra (described in Section \ref{sec:spectra}).

Another important difference to note, is that the light elements do not make use of the flat recombination rate of $10^{-11}\, \mathrm{cm^3 \, s^{-1}}$ applied to the r-process species, since data for the lighter elements is more readily available. Tabulated data for these rates comes from diverse sources e.g. \citep[He\,\textsc{i},][]{Hummer.Storey:98}, \citep[He\,\textsc{ii}][]{Verner.Ferland:96}, \citep[every Fe ion,][]{Nahar:96,Nahar:97a,Nahar.etal:97b}, while fitting formulae are also used \citep[Ti\,\textsc{i},][]{Pequignot.Aldrovandi:86} and \citep[Ni\,\textsc{i} and Ni\,\textsc{ii},][]{Shull.Steenberg:82}. When no data or particular fitting formula is employed, the recombination rates of light species are set equal to the identical ionisation state of $_{26}$Fe, i.e. Ti\,\textsc{ii}, Ti\,\textsc{iii}, and Ni\,\textsc{iii} are set to have the same recombination rates as Fe\,\textsc{ii} and Fe\,\textsc{iii} respectively. While this introduces some degree of inaccuracy, iron group elements tend to have very similar recombination rates \citep[][]{Jerkstrand.etal:11}, such that the error introduced from this approximation is likely far inferior to that of using a flat rate for the r-process species. 

Although the r-process species all have the same constant recombination rate, their ionisation structure may still differ due to the ionisation processes. It has been found that photoionisation (PI) and non-thermal collisional ionisation (NT) tend to dominate in KN conditions \citep[e.g.][]{Pognan.etal:22a}. For the former process, we use the hydrogenic cross-section and allow PI to occur from the first 50 levels of an atom. For the latter, we use the \citet{Lotz:67} cross-section, which mainly scales with the ground-state ionisation threshold. In this sense, elements with similar ionisation potentials and level structure will have similar ionisation structures. For instance, lanthanide species are typically singly to doubly ionised, while first peak species are more often found in neutral and singly ionised form for the same conditions. Inclusion of detailed recombination rates, both radiative and dielectronic, as well as accurate ionisation cross-sections for all processes and all r-process elements would be required to establish conclusively whether these trends are maintained with better data.

The information pertaining to updated r-process data is shown in Table \ref{tab:atomic_data}. Of the species shown there, ions with an asterisk next to their name indicate that the \textsc{fac} data has been entirely replaced by r-matrix calculations with thermal collision strengths (sources indicated in table caption). Other species have had the displayed number of levels following the method described in Section \ref{subsec:RTsim}.


\bsp	
\label{lastpage}
\end{document}